\documentclass[%
 aip,
 amsmath,amssymb,
preprint,%
]{revtex4-1}

\usepackage{graphicx}
\usepackage{dcolumn}
\usepackage{bm}

\usepackage[utf8]{inputenc}
\usepackage[T1]{fontenc}
\usepackage{mathptmx}
\usepackage{caption}
\usepackage{subcaption}

\usepackage{hhline}
\usepackage{multirow}
\usepackage{siunitx}
\usepackage{booktabs}
\usepackage{xcolor}
\usepackage{booktabs}

\begin{document}


\title[]{ {\small \it under consideration for publication in Physics of Fluids} \\ \vspace{8mm}
Effect of non-parallel mean flow on the acoustic spectrum of heated supersonic jets: explanation of `jet quietening'
}

\author{M. Z. Afsar}
 \affiliation{Department of Mechanical $\&$ Aerospace Engineering, University of Strathclyde, 75 Montrose Street, Glasgow, G41 1XJ, UK.}
  \email{mohammed.afsar@strath.ac.uk}
   \homepage{https://pureportal.strath.ac.uk/en/persons/mohammed-afsar.}
\author{A. Sescu}%
\affiliation{Department of Aerospace Engineering, Mississippi State University, Starkville, MS 39762,
USA
}%
\author{V. Sassanis}
\affiliation{%
Department of Aerospace Engineering, Mississippi State University, Starkville, MS 39762,
USA
}%

\date{\today}

\begin{abstract}

Noise measurements of heated axisymmetric jets at fixed supersonic acoustic Mach number indicate that the acoustic spectrum reduces when the temperature ratio increases. The `spectral quietening' effect has been observed both experimentally and computationally using Large Eddy Simulations (LES). It was explained by Afsar {\it et al}. (M. Z. Afsar and M. E. Goldstein $\&$ A. M. Fagan
AIAAJ., Vol. 49, p. 2522, 2011) through the cancellation introduced by enthalpy flux/momentum flux coupling term using the generalized acoustic analogy formulation. But the parallel flow assumption is known to give inaccurate predictions at high jet speeds. In this paper we therefore extend the non-parallel flow asymptotic theory of Goldstein {\it et al}. (M. E. Goldstein, A. Sescu $\&$ M. Z. Afsar, J. Fluid
Mech., Vol. 695, p. 199, 2012) for the vector Green's function of the adjoint linearized Euler equations (ALEE) in the analogy. Using a steady Reynolds Averaged Navier Stokes (RANS) calculation for the jet mean flow, we find that the coupling term propagator is positive-definite and asymptotically sub-dominant at low frequencies corresponding to the peak jet noise when non-parallel flow effects are taken into account and self-consistent approximations for the turbulence structure are made. The validity of the non-parallel flow-based acoustic analogy model is assessed at various observation angles by computing the overall sound pressure level (OASPL) and use this to suggest a more rational explanation of the quietening effect. In general, our noise predictions are in very good agreement with acoustic data beyond the peak frequency.

\end{abstract}

\maketitle

\section{\label{Intro} Introduction}
The peak sound of a high speed jet radiates at low frequencies (typically at Strouhal numbers, $St\sim0.2$) such that the Overall Sound Pressure Level (OASPL) is being greatest at a small polar observation angle, $\theta$, with respect to the jet centre-line, usually at $\theta = 30^\circ$  {\cite{Tanna77, BridgesWern}}. 
%
Although these trends are essentially the same for all acoustic Mach numbers in isothermal flows up to mild supersonic conditions, where the convection speed of the turbulence remains subsonic such that broad-band shock associated noise (see  Mayo {\it et al}. \cite{Mayo} and Kalyan $\&$ Karabasov \cite{KalKarab}) is basically negligible, there is an unexplained property of the acoustic spectrum with increasing jet temperature ratio, $TR = T_J/T_\infty >1.0$. 
That is, when an axisymmetric air jet at such mild supersonic speeds (i.e. turbulence convection Mach number, $M_c < 1$)  is heated at constant acoustic Mach number $Ma = U_J/c_\infty$ (where $U_J$ is the jet velocity and $c_\infty$ is the speed of sound at infinity) the jet noise reduces at all observation angles (Bodony $\&$ Lele\cite{BodLel}). 

The discovery of quieter supersonic jets by heating was first made in Hoch {\it et al}.\cite{Hoch} and later, more systematically reported by Tanna\cite{Tanna77}, for a round jet flow at $Ma = 1.47$. 
Fig. 7 in Tanna's experiments showed that the OASPL curve of the heated supersonic jet lies beneath the isothermal, or cold jet spectra, at all observation angles with the reduction in noise  greatest at small $\theta$.
Later experiments\cite{Seiner} were conducted to check the accuracy of earlier studies and to extend their applicability to a greater parameter range in $Ma$ $\&$ $TR$. 
Some of the pertinent literature in heated jet acoustics are Seiner {\it et al}. \cite{Seiner}, Lee $\&$ Bridges\cite{LeeBridge}, Bridges \cite{Bridges06},  Bhat\cite{Bhat}, Harper-Bourne\cite{HB07, HB09} and Tester $\&$ Morfey\cite{TestMorf}.
The Harper-Bourne\cite{HB07, HB09} and NASA experiments\cite{Bridges06,LeeBridge,BridgesWern} have both found similar generic behavior as Tanna\cite{Tanna77} and Tanna {\it et al}.\cite{Tanna75}; see for example, Fig. 27 and the discussion on p.21 in Harper-Bourne\cite{HB07}.
Similar spectral properties were found by Bodony $\&$ Lele\cite{BodLel} who performed Large-Eddy Simulations (LES) calculations of axisymmetric round jets at the Tanna set points. The quietening of the hot jet acoustic spectra was clearly apparent in the OASPL calculations in Fig. 9 (see also $\&$ p.249) of Bodony $\&$ Lele\cite{BodLel}.

More recent work in the acoustics of heated flows has looked at the structure of jet turbulence by measuring the local source convection velocities.
For example, Shea {\it et al}.'s\cite{Shea} data correlates the reduction in potential core length of a heated jet with increased jet temperature. Stuber {\it et al}.'s\cite{Stuber} measurements quantify the reduction in core length with an appropriate reduction in turbulence convection velocity (see Bridges $\&$ Wernet\cite{BridgesWern} also), which might indicate the weakness of the sound source in heated supersonic flows.
Although the $Ma$ and $TR$ of these studies are somewhat beyond our concerns (involving supersonic source convection effects), Liu {\it et al}. (\cite{Liu2016}, p.6) indicate that sound radiated by Mach waves is expected to enter at high frequencies at these speeds, hence the Shea/Stuber results\cite{Shea,Stuber} will be qualitatively applicable to the mild supersonic conditions under our consideration in which sound amplification due to turbulence/mean flow interaction (i.e. jet quietening effects) remain a low frequency phenomenon.

The aim of this paper is to develop a model for low frequency jet noise that gives a consistent physical explanation for the observed quietening of heated supersonic jets in addition to providing a `low-order' prediction scheme.
We use the acoustic analogy of Ref.\onlinecite{Gold03} as the basis of the model by extending a previously derived asymptotic theory for the vector Green's function of the adjoint linearized Euler equations (ALEE) operator (Eqs. 4.8--4.10 of Goldstein $\&$ Leib\cite{GanL}). 
The generalized analogy, owing to its first principle derivation, has laid the foundation for systematic analysis of jet noise to be performed without concern if the source terms (that are assumed known) or the propagator are defined consistently; these were the main issues relating to the Lighthill\cite{Lighthill} and Lilley\cite{Lilley} formulations. 
Goldstein\cite{Gold03} showed that the acoustic spectrum per unit volume can be expressed as a convolution product of a propagator and auto-covariance tensor of a generalized (rank-four) fluctuating stress tensor of a stationary random function that reduces to the fluctuating Reynolds stress in the absence of any temperature unsteadiness.
Although the auto-covariance tensor possesses $144$ components, not all of these are independent.
Symmetry between tensor suffixes, shows that it reduces (without approximation) to $63$ independent components, which is still too large for practical use. Thankfully, however, rational approximations such as axisymmetric turbulence further reduce this to a manageable number of $11$, only a few of which dominate the peak radiated sound.
The propagator tensor on the other hand is determined by the vector Green's function of the ALEE at $O(1)$ frequencies which itself is a function of the mean flow field for an, in general, arbitrary spreading jet.

Noise predictions using Ref.\onlinecite{Gold03} were successfully implemented by Goldstein $\&$ Leib\cite{GanL}, Karabasov {\it et al}.\cite{Karab2010} and Leib $\&$ Goldstein\cite{LG11} for various isothermal axisymmetric jets at a range of $Ma$.
The Reynolds stress auto-covariance tensor in these models was approximated to be kinematically axisymmetric and the functional form of the components that entered it was based on experiments and/or LES\cite{HB07,HB09, Karab2010}. More specifically the functional form was assumed to be proportional to the local RANS turbulent kinetic energy (TKE). The propagators were also determined using the RANS mean flow.
The above approaches still remain the current state-of-art for low-order prediction models based on the acoustic analogy models (see Ref. \onlinecite{GL18}). Note that while these models rely on accurate RANS mean flow solutions (which themselves are based on turbulence models) the centerline mean flow distribution is usually accurately predicted compared to particle-image velocimetry (PIV) data, with closest agreement being at the end of jet potential core (see Karabasov {\it et al}.\cite{Karab2010} and $\S.3$). Using LES to extract the mean flow/turbulence correlations would naturally be more computationally expensive. 
%

The Goldstein $\&$ Leib\cite{GanL} (G$\&$L) predictions were computed at $O(1)$ frequencies for a propagator tensor based on a weakly non-parallel mean flow. Non-parallelism appeared in the analysis at supersonic speeds and only affected the solution within a thin critical layer where the adjoint vector Green's function is singular for the locally parallel mean flow. G$\&$L constructed a uniformly valid composite solution for the adjoint Green's function, to eliminate the critical-layer singularity at a particular $\theta$. As $\theta \rightarrow 0$, the dominant contribution to the propagator in their calculations was due to the radial derivative of the Fourier transformed adjoint Green's function for the streamwise mass flux perturbation. This was also confirmed by Karabasov {\it et al}.'s\cite{Karab2010} numerical calculations and by Afsar\cite{Afs2010} all of whom showed that the acoustic efficiency of this term is raised to a dipole at low frequencies even though, intrinsically, it appears as a quadrupole when multiplied by the appropriate Reynolds stress auto-covariance component in the acoustic spectrum formula\cite{Gold75}.

Despite the predictive success of the models described above, there still remain several basic research problems involving the generalized acoustic analogy such as (a). how to model the components of the auto-covariance tensor that enter the prediction model? and (b). what is the appropriate base flow (i.e. uniform, parallel or weakly/strongly non-parallel) to determine the solution to the ALEE, and therefore the propagator tensor? 
We shall answer both of these questions in this paper when applied to the technologically relevant axisymmetric jet flow.
Previous attempts at explaining the quietening of heated supersonic axisymmetric jets using the generalized analogy approach showed that cancellation introduced by the enthalpy flux/momentum flux auto-covariance (coupling) term could potentially reduce the magnitude of the acoustic spectrum (see Afsar, Goldstein $\&$ Fagan\cite{AGF11} hereafter referred to as AGF). However, the AGF results were based on low frequency asymptotic estimates of the propagator in a parallel (non-spreading) mean flow. But even in the absence of jet heating, a parallel (i.e. a uni-directional transversely sheared) mean flow does not capture the correct amplification in sound at these frequencies for high subsonic jets (Karabasov {\it et al}.\cite{KBH13}), hence this approach cannot be expected to provide a consistent model for the reduction in sound with heating at supersonic jet speeds. 
%

In isothermal flows, Goldstein {\it et al}. (\cite{GSA12} – hereafter referred to as GSA) found that non-parallel flow effects can be properly accounted for in the propagator solution at low frequencies when the time-dependent Green's function evolves temporally at the same order as its spatial development.
Further extensions to the GSA theory represents an important area of research since the peak jet noise is observed at low frequencies (typically at Strouhal numbers, $St$, based on jet exit conditions of $St\sim 0.2$) and small $\theta$ (Karabasov {\it et al}.\cite{Karab2010}).
Reduced order models of this phenomenon will therefore support noise reduction efforts while remaining computationally inexpensive compared to the full numerical solution of the ALEE. 
%

The above distinguished scaling allows non-parallelism to enter the small $\theta$/low$St$ region of parameter space affecting the lowest order solution to the propagator everywhere in the jet.
GSA found a qualitatively similar spatial structure for the dominant dipole-like momentum flux (i.e. fluctuating Reynolds stress) associated propagator component relative to Karabasov \textit{et al}.'s\cite{KBH13} full numerical solution to the ALEE (see Fig. 16c in Karabasov {\it et al}. AIAA 2011-2929 or Fig. 12b in Karabasov {\it et al}.\cite{KBH13}). 
At the scaled frequency $\Omega = \omega/\epsilon = O(1)$, the `inner' region of the ALEE is governed by a single hyperbolic partial differential equation (PDE) when the streamwise mean flow is taken as one of the independent variables\cite{GSA12}. We show, however, that this transformation can be introduced prior to any asymptotic expansion of the ALEE. The inner equation then, essentially, follows at once after relatively straightforward dominant balance considerations are made.

The rest of the paper is organized as follows. In $\S$.\ref{S:2} we briefly summarize the generalized acoustic analogy for heated jets at moderate supersonic acoustic Mach numbers. In $\S$.\ref{S:3}, we expand the propagator tensor in a heated flow by showing that the same distinguished asymptotic limit of the ALEE derived in GSA must be applicable when $TR>1$.
%
The same inner equation is then valid in heated jets as for the isothermal case\cite{GSA12} but with the Favre-averaged mean square speed of sound determined by the Crocco-Busemann rather than the Crocco relation (which applies only at unity $TR$).
In $\S$.\ref{S:4} we use the propagator expansion to obtain an approximate formula for the peak acoustic spectrum in heated jets; this formula is used in $\S$.\ref{S:5} to understand what contribution the temperature-associated noise source terms play in the quietening of low frequency sound. 

We consider two axisymmetric jets at the same supersonic $Ma$ based on Bridges\cite{Bridges06} data set that were derived from the Tanna\cite{Tanna77} set points.
Previous authors have confirmed experimentally\cite{Tanna77, Bridges06} and computationally\cite{BodLel} that these jets possess a definite amount of spectral quietening.  
The conditions are summarized in Table \ref{tab:table1}.
The mean flow used in the study was obtained by a RANS calculation using {\sc{Fluent}}.
Our results are presented in $\S$.\ref{S:5} and discussed in $\S$.\ref{S:6}.
We find that a jet noise model based on an extended version of the GSA asymptotic theory gives accurate (i.e. within $1-2$ dB) sound predictions for the peak noise of the heated and isothermal supersonic jets under consideration. 
Non-parallel flow effects are seen to play a crucial role in correctly determining the sign and magnitude of the coupling and enthalpy flux terms within the acoustic spectrum.

\begin{table}
\caption{\label{tab:table1} axisymmetric jets used in this study. 
}
\begin{ruledtabular}
\begin{tabular}{lcr}
 Tanna\cite{Tanna77} set point  &  $Ma =U_J/c_\infty$ & $TR= T_J/T_\infty$ \\
\hline
SP90 & 1.48 & 1.0\\
SP49 & 1.48 & 2.7\\
\end{tabular}
\end{ruledtabular}
\end{table}

\section{Acoustic spectrum formula in heated flows}
\label{S:2}

\subsection{\label{sec2.1:GAA}Basic formalism of generalized analogy}
%
Consider a high speed axisymmetric air jet possessing a spatially evolving (i.e. non-parallel) mean flow with arbitrary temperature ratio convecting a localized region of turbulence. 
It is necessary to briefly summarize the acoustic spectrum formula derived in AGF to appropriately set the context of the asymptotic analysis of the ALEE developed in $\S.$\ref{S:3}.
Thus, let the (dimensional) pressure $p$, density $\rho$, enthalpy $h$, and speed of sound, $c$, satisfy the ideal gas law equation of state: $p = \rho c^2/\gamma$ where $\gamma$ denotes the specific heat ratio such that $h= c^2/(\gamma-1)$. 
The acoustic spectrum at the observation point, $\boldsymbol{x}$, is given by Fourier transform 
\begin{equation}
\label{eq:Iom}
I({\boldsymbol x}, \omega) 
\equiv
\frac{1}{2 \pi}
\int\limits_{-\infty}^{\infty}
e^{i \omega\tau}
\overline{
p^\prime({\boldsymbol x}, t)
p^\prime({\boldsymbol x}, t+\tau)
}
\,d\tau,
\end{equation}
of the far-field pressure auto-covariance, $\overline{
p^\prime({\boldsymbol x}, t)
p^\prime({\boldsymbol x}, t+\tau)
}$. The acoustic spectrum at ${\boldsymbol x} = (x_1, {\boldsymbol x}_T) = (x_1,x_2,x_3)$, due to a unit volume of turbulence at ${\boldsymbol y} = (y_1, {\boldsymbol y}_T) = (y_1,y_2,y_3)$, is given by
\begin{equation}
\label{eq:Iom2}
I({\boldsymbol x}; \omega) 
=
\int
\limits_{V_\infty({\boldsymbol y})}
I({\boldsymbol x}, {\boldsymbol y};\omega) 
\,d{\boldsymbol y},
\end{equation}
where, $V_\infty({\boldsymbol y})$, is the entire source region, $p^\prime ({\boldsymbol y}, \tau) \equiv p({\boldsymbol y}, \tau) -\bar{p}({\boldsymbol y})$, and over-bars are being used to denote time averages are defined as:
\begin{equation}
\label{eq:t_avg}
\bar{\bullet}({\boldsymbol x}) 
\equiv
\lim_{T\rightarrow\infty}
\frac{1}{2T}
\int\limits_{-T}^{T}
\bullet({\boldsymbol x}, t)
\,dt,
\end{equation}
where ${\bullet}$ in (\ref{eq:t_avg}) is a place holder for any fluid mechanical variable.

Goldstein $\&$ Leib \cite{GanL} showed that the integrand in (\ref{eq:Iom2}) can be determined by the exact integral solution,
\begin{equation}
\label{eq:Iom3}
\begin{split}
\frac{I({\boldsymbol x}, {\boldsymbol y};\omega)}{(2\pi)^2} 
=
&
\Gamma_{\lambda, j}
({\boldsymbol y}| {\boldsymbol x}; \omega)
\\
\times
&
\int
\limits_{V_\infty({\boldsymbol \eta})}
\Gamma{}_{\mu, l}^*
({\boldsymbol y} + {\boldsymbol \eta}| {\boldsymbol x}; \omega)
\mathcal{H}_{\lambda j \mu l}
({\boldsymbol y}, {\boldsymbol \eta}; \omega)
\,d{\boldsymbol \eta}.
\end{split}
\end{equation}
The asterisks in (\ref{eq:Iom3}) denotes complex conjugate and the Einstein summation convention is being used with the Greek tensor suffixes ranging $(\lambda,\mu)=(1,2,3,4)$ and Latin suffixes $(i,j,k,l)=(1,2,3)$ representing the components of a rectangular Cartesian co-ordinate system, where `$1$' is the streamwise direction and $(2,3)$ represent components in the transverse plane.
The mean flow is now contained within the propagator tensor
\begin{equation}
\label{eq:Prop}
\begin{split}
\Gamma_{\lambda, j}
({\boldsymbol y}| {\boldsymbol x}; \omega)
\equiv
&
\Lambda_{\lambda\sigma,j}
({\boldsymbol y})
G_\sigma
({\boldsymbol y}| {\boldsymbol x}; \omega)
\\
:=
&
\left(
\delta_{\lambda\sigma} 
\frac{\partial }{\partial y_j}
-
(\gamma-1)
\delta_{4\sigma}
\frac{\partial\tilde{v}_\lambda}{\partial y_j}
\right)
G_\sigma
({\boldsymbol y}| {\boldsymbol x}; \omega)
\end{split}
\end{equation}
that involves an inner tensor product in suffix $\sigma$, of operator $\Lambda_{\lambda\sigma,j}
({\boldsymbol y})$, that spans $(4\times4\times3)$ dimensions corresponding to suffixes $(\lambda,\sigma,j)$ where comma after $j$ indicates that this suffix belongs to a derivative, and the first four components of the Fourier transform
\begin{equation}
\label{eq:GFT}
{G_\sigma}
({\boldsymbol y}| {\boldsymbol x}; \omega)
=
\frac{1}{2 \pi}
\int\limits_{-\infty}^{\infty}
e^{i \omega(t-\tau)}
{ g}{}_{\sigma4}^{a}
({\boldsymbol y}, t - \tau| {\boldsymbol x})
\,d(t-\tau)
,
\end{equation}
of the five-dimensional adjoint vector Green's function, ${ g}{}_{\sigma4}^{a}
({\boldsymbol y}, \tau| {\boldsymbol x}, t)$, 
(with suffix $\sigma=1,2,3,4,5$) 
on the left hand sides of the five ALEE given by Eqs. (4.8)--(4.10) of G $\&$ L\cite{GanL} subject to the strict causality condition for the adjoint pressure-like Green's function,
${g}{}_{44}^{a}
({\boldsymbol y}, t - \tau| {\boldsymbol x})=0$ for 
$t<\tau$ when $|{\boldsymbol x}|\rightarrow \infty$. 
As frequently commented in previous papers\cite{Gold03,GanL,LG11}, (\ref{eq:Iom3}) and (\ref{eq:Prop}) above are completely general and apply to any localized turbulent flow, even in the presence of fixed solid surfaces whose boundaries are given by level curves $S({\boldsymbol y}) =  const.$ as long as ${g}{}_{\sigma 4}^{a}
({\boldsymbol y}, \tau| {\boldsymbol x}, t)$ is assumed to satisfy appropriate surface rigidity conditions $ \hat{n}_\sigma {g}{}_{\sigma 4}^{a}
({\boldsymbol y}, \tau| {\boldsymbol x}, t) = 0$ where $\hat{n}_\sigma=\{\hat{n}_i, 0, 0\}=\{\hat{n}_1,\hat{n}_2,\hat{n}_3, 0, 0\}$ denotes the unit normal to $S({\boldsymbol y})$.

The unit tensor, $\delta_{\lambda\sigma}$, that appears in both terms on the second line of (\ref{eq:Prop}) above is the symmetric four-dimensional Kronecker delta function and tilde refers to the Favre averaged quantity $\tilde{\bullet} = \overline{\rho \bullet}/\bar{\rho}$, so that the four-dimensional mean velocity vector in (\ref{eq:Prop}) is $\tilde{v}_\lambda = \{\tilde{v}_i,0\}, i=(1,2,3)$.
The 5th component of ${ G_\sigma}
({\boldsymbol y}| {\boldsymbol x}; \omega)$ -- the Fourier transform of adjoint Green's function for the continuity equation in the generalized analogy (defined by Eq. 2.9a in Goldstein\cite{Gold03}) -- does not enter the propagator formula, given by (\ref{eq:Prop}); it does, however, affect its solution through the linearized adjoint equations:
\begin{subequations}
\label{eq:GAA}
\begin{align}
-{D}_0 G_i + G_j \frac{\partial \tilde{v}_j}{\partial y_i}
-\widetilde{c^2}\frac{\partial G_4}{\partial y_i} +(\gamma-1)\tilde{X}_i G_4
-\frac{\partial G_5}{\partial y_i} & = 0 \\
-{D}_0 G_4 - \frac{\partial G_i}{\partial y_i}
+(\gamma-1)G_4 \frac{\partial\tilde{v}_i}{\partial y_i} & = \frac{\delta({\boldsymbol{x}-\boldsymbol{y}})}{2\pi} \\
-{D}_0 G_5 +\tilde{X}_i G_i & = 0, 
%
\end{align}
\end{subequations}
where ${D}_0 \equiv i\omega + \tilde{{\boldsymbol v}}({\boldsymbol y}).{\boldsymbol{\nabla}}$ is the convective derivative in which ${\boldsymbol{\nabla}}$ is the three-dimensional gradient operator and $i=(1,2,3)$.

Reciprocity (see pp. 878--886 of Morse and Feshbach\cite{Morse} and Eq. 4.7 in G $\&$ L) of the space-time Green's function demands that ${g}{}_{\sigma4}^{a}
({\boldsymbol y}, \tau| {\boldsymbol x}, t)={g}{}_{4 \sigma}
({\boldsymbol x}, t|{\boldsymbol y}, \tau)$. The ${\boldsymbol y}$ independent variable in (\ref{eq:GAA}) corresponds to the actual physical source point and ${\boldsymbol x}$, the observation point, which is taken as a parameter in the solution and located in the far field, $|{\boldsymbol x}|\rightarrow\infty$.
The coefficients that multiply the derivatives that act on ${G_\sigma}
({\boldsymbol y}| {\boldsymbol x}; \omega)$ in the system of equations given by (\ref{eq:GAA}) depend on mean flow field through: $\tilde{v}_i = (\tilde{v}_1, \tilde{v}_2, \tilde{v}_3)$,  the Favre-averaged speed of sound squared $\widetilde{c^2}({\boldsymbol y})\equiv {\gamma \bar{p}}/{\bar{\rho}}$ and $\boldsymbol{\tilde{X}}({\boldsymbol y}) = (\boldsymbol{\tilde{v}}.\boldsymbol{\nabla}) \boldsymbol{\tilde{v}}$, the mean flow advection vector.

The scripted tensor, $\mathcal{H}_{\lambda j \mu l} ({\boldsymbol y}, {\boldsymbol \eta}; \omega)$, in the acoustic spectrum formula, (\ref{eq:Iom3}), is related to the Fourier transform 
 \begin{equation}
\label{eq:HFT}
{H}_{\lambda j \mu l}
({\boldsymbol y}, {\boldsymbol \eta}; \omega)
=
\frac{1}{2 \pi}
\int\limits_{-\infty}^{\infty}
e^{i \omega\tau}
{R}_{\lambda j \mu l}
({\boldsymbol y}, {\boldsymbol \eta}; \tau)
\,d(\tau)
\end{equation}
of the generalized auto-covariance tensor,
\begin{equation}
\label{eq:Rijkl}
{R}_{\lambda j \mu l}
({\boldsymbol y}, {\boldsymbol \eta}; \tau)
\equiv
\lim_{T\rightarrow\infty}
\frac{1}{2T}
\int\limits_{-T}^{T}
e{}_{\lambda j}
({\boldsymbol y}, \tau)
e{}_{\mu l}
({\boldsymbol y} + {\boldsymbol \eta}, \tau+\tau_0)
\,d\tau_0,
\end{equation}
of the stationary random function, $e{}_{\lambda j}({\boldsymbol y}, \tau)= [\rho v{}_\lambda^\prime v{}_j^\prime -
\overline{\rho
v{}_\lambda^\prime
v{}_j^\prime}
]({\boldsymbol y}, \tau)
$, by the linear transformation $\mathcal{H}_{\lambda j \mu l}({\boldsymbol y}, {\boldsymbol \eta}; \omega):=\epsilon_{\lambda j \sigma m} H_{\sigma m \gamma n} ({\boldsymbol y}, {\boldsymbol \eta}; \omega) \epsilon_{\mu l \gamma n}$.
The four-dimensional vector, $v{}_\lambda^\prime$, denotes the perturbation, $v{}_\lambda^\prime ({\boldsymbol y},\tau) \equiv v{}_\lambda ({\boldsymbol y},\tau)- \tilde{v}{}_\lambda ({\boldsymbol y})$ in which $v{}_\lambda^\prime =v{}_i^\prime$ is the ordinary fluid velocity perturbation when suffix, $\lambda=i= (1,2,3)$, otherwise $v{}_\lambda^\prime =v{}_4^\prime$. 
The latter denotes $v{}_4^\prime :=(\gamma-1)(h^\prime + v{}^{\prime 2}/2)\equiv (c^2)^\prime +(\gamma-1)v{}^{\prime 2}/2$ where $h^\prime$ is the  fluctuating static enthalpy and $(c^2)^\prime$ is the fluctuations in the sound speed squared such that $v{}_4^\prime/(\gamma-1)$ represents the moving frame stagnation enthalpy fluctuation\cite{Gold03}. Since the above relation shows that $c^\prime = O(\sqrt{v{}^{\prime 2}})$, Eq. (30) in AGF allows the importance of $e{}_{4 l}$ compared to $e{}_{i l}$ to be quantified by the dimensionless ratio, $|e{}_{4 l}|/(c_\infty |e{}_{i l}|)$, that remains $O(\sqrt{v{}^{\prime 2}}/(c_\infty v{}^{\prime}))$. This can also be expressed as $|e{}_{4 l}|/(c_\infty |e{}_{i l}|)=O(M^{\prime})$ where $M^{\prime}=v{}^{\prime}/c_\infty$ is proportional to square root of the fluctuating temperature ratio $T^\prime/T_\infty$ when $Ma = O(1)$; 
this is obtained after linearizing the definition of the speed of sound and using Eq.(2.14) in G $\&$ L\cite{GanL}. Similar arguments were made for heated jets at low subsonic speeds by Morfey \textit{et al}. (Eq.22 of Ref. \onlinecite{Morf_etal_1978}) and Lilley (\cite{Lilley1996}, p.471).
The  above scaling is, however, slightly different to Ref.\onlinecite{GanL} (p.307f.). 
It implies that for isothermal (or, slightly cold) jets, where $T^{\prime} \approx 0$ and $|e{}_{4 l}|/(c_\infty |e{}_{i l}|) \rightarrow 0$. 
In other words, $v{}_4^\prime = o(h^\prime)$ when $T^\prime \approx 0$ so that the $(\lambda = \mu=4)$ component of 
${R}_{\lambda j \mu l}
$ can be set equal to zero.
For heated jets on the other hand, $|e{}_{4 l}|/(c_\infty |e{}_{i l}|) \neq 0$ and is expected to be fairly large, especially for $Ma>1$ (see Figs.1 $\&$ 2 in AGF and Fig. 20 in Sharma $\&$ Lele\cite{SharmLele}).

Comparing Eq. (5.12) to (5.13) in G $\&$ L\cite{GanL} and using appropriate outer products of unit tensors (see also sentence below Eq.\ref{eq:GFT}) in suffixes $(\lambda,j,\sigma, m)$ allows definition of the tensor $\epsilon_{\lambda j \sigma m}$ as, $\epsilon_{\lambda j \sigma m} \equiv \delta_{\lambda\sigma} \delta_{j m} - \delta_{\lambda j} \delta_{\sigma m}(\gamma-1)/2$ in the linear relation for $\mathcal{H}_{\lambda j \mu l}$.

\section{Including non-parallel flow effects at lowest order for non-unity temperature effects ($TR>1$) in (\ref{eq:Prop}) $\&$ (\ref{eq:GAA})}
\label{S:3}
GSA worked out an asymptotic approximation of the ALEE, given by the system in (\ref{eq:GAA}), for a slowly diverging jet flow at temporal frequencies of the order of the small jet spread rate, that is, at $\omega = O(\epsilon)$. 
This theory applied to a particular class of flow where Crocco’s relation\cite{Crocco} is valid.
We now show that this asymptotic expansion procedure can be naturally extended to heated jets and, interestingly, results in the same inner equation for the appropriate Green's function variable as in isothermal flow but now with the speed of sound determined via the temperature-dependent Crocco-Busseman formula.  
Eq.(5.21) (p.209 $\&$ \textit{f}.) in GSA showed that significant reduction in complexity of the lowest order inner equations took place when the streamwise mean flow component, $U$, was taken in place of the radial co-ordinate, $r$, as one of the independent variables.
But this transformation can be easily applied to the ALEE {\it at the outset}, prior to any asymptotic analysis. 
The advantage of this being that when the latter is utilized, in the form of method of multiple scales and matched asymptotic expansions (in that order), the basic inner equation for the more general heated flow results at once.

\subsection{\label{sec3.1:GAA}Transformation of (\ref{eq:GAA}) at $O(1)$ spread rate}

We non-dimensionalize the dependent and independent variables in the ALEE (\ref{eq:GAA}), in preparation for the asymptotic analysis of $\S$.{\ref{sec3.2:redux}}.
Therefore, let independent variables $({\boldsymbol y}, \tau)$ be normalized by the $O(1)$ characteristic length $D_J$ and time $D_J/U_J$ respectively where $U_J$ $\&$  $D_J$ are the mean velocity and nozzle exit diameter respectively. The fluid mechanical variables $({\boldsymbol \tilde{\boldsymbol v}},p,\rho)$ are then normalized by $U_J$, $\rho_J U{}_J^2$ and $\rho_J$ (nozzle exit density) respectively. Taking $({\boldsymbol e}_1, {\boldsymbol e}_r, {\boldsymbol e}_\phi)$ as an orthogonal triad of basis vectors in a cylindrical co-ordinate space, shows that the first three components of the vector ${\boldsymbol G}\equiv G_\sigma=(G_i, G_4,G_5)$, that is determined by (\ref{eq:GAA}a), can be expressed as a linear function of that basis 
by $G_j = (G_i {\boldsymbol e}_i){\boldsymbol e}_j = G_1 \delta_{j1} + G_r \delta_{jr} + G_\phi \delta_{j\phi}$ where ${G}_i = (G_1, G_r, G_\phi)$ are the respective components of ${G_i}$ along the $({\boldsymbol e}_1, {\boldsymbol e}_r, {\boldsymbol e}_\phi)$ directions. 
The mean flow field (commensurate with an axisymmetric jet) has components, ${\boldsymbol v} = (U, V_r)$ where, at this point, we leave the jet spread rate to be otherwise arbitrary at $\epsilon=O(1)$.

As GSA did, we take $U$ to be one of the independent variables of choice; i.e. under the one-to-one mapping $(y_1,r)$ $\rightarrow(y_1,U)$ where $r \equiv |{\boldsymbol y}_T|= \sqrt{y{}_2^2 + y{}_3^2}$. The co-ordinate surfaces $U(y_1,r) = const.$ and $y_1 = const.$ are such that ${\boldsymbol\nabla} U. {\boldsymbol\nabla} y_1 = 0$ at any fixed radial location, $r$, in the field space. Since the gradient operator
shows that ${\boldsymbol e}_1 \equiv  {\boldsymbol\nabla} y_1$ and ${\boldsymbol\nabla} U \equiv {\boldsymbol e}_1 {\partial U}/{\partial y_1} + {\boldsymbol e}_r {\partial U}/{\partial r}$, the definition of the partial derivative requires that ${\boldsymbol\nabla} U.{\boldsymbol\nabla} y_1 = {\partial U}/{\partial y_1 }=0$ in the transformed co-ordinate system. Using the fact that 
${ G_\sigma}
(y_1, r, \phi| {\boldsymbol x}; \omega)$ is implicitly related to $\tilde{G}_\sigma=\tilde{G}_\sigma (y_1, U,\phi|{\boldsymbol x}; \omega)$ via:
\begin{equation}
\label{G_implicit}
\tilde{G}_\sigma
(y_1, U(y_1,r),\phi|{\boldsymbol x}; \omega)
=  
{G}{}_\sigma
(y_1,r,\phi|{\boldsymbol x}; \omega),
\end{equation}
the orthogonality condition and the chain rule in $(y_1,U)$ co-ordinates similarly shows that the mean flow advection vector, ${X}_i=(X_1, X_r)$ in (\ref{eq:GAA}a--c), takes the slightly more general form
%
\begin{equation}
   \tilde{X}_1(y_1,U) = V_r \frac{\partial U}{\partial r}  \label{subX1}
    \hspace{0.5cm}
    \&
       \hspace{0.5cm}
  \tilde{X}_r (y_1, U)= \left(U\frac{\partial}{\partial y_1} + V_r \frac{\partial}{\partial r} \right) V_r. 
\end{equation}
compared to Eq.(5.15) in GSA since (\ref{subX1}) is now applicable at $\epsilon=O(1)$. The operator $D_0$, when acting on $\tilde{G}_\sigma
(y_1, U(y_1,r),\phi|{\boldsymbol x}; \omega)$ in (\ref{eq:GAA}a--c), can also be transformed as follows,
\begin{equation}
\label{D0_trans}
D_0
{G}_\sigma
(y_1,r)
=  
  \left( i\omega + U\frac{\partial}{\partial y_1} + V_r \frac{\partial}{\partial r}
  \right) {G}{}_\sigma
\equiv
\left(
\tilde{D}_0 + \tilde{X}_1 \frac{\partial}{\partial U}
\right)
\tilde{G}_\sigma (y_1, U),
   \end{equation}
where we have suppressed the remaining arguments in $G_\sigma$ and $\tilde{D}_0 \equiv i\omega + U{\partial}/{\partial y_1}$.

Since ${\partial U}/{\partial r} = ({\partial r}/{\partial U})^{-1}$
and the chain rule shows that ${\partial}/{\partial U} = ({\partial r}/{\partial U}){\partial}/{\partial r}$, the $i=r$ component of (\ref{eq:GAA}b) can be transformed to 
\begin{equation}
\label{G_1}
\tilde{G}_1
(y_1, U)
=  
\widetilde{c^2}
\frac{\partial \tilde{G}_4} {\partial U}
+
\frac{\partial \tilde{G}_5} {\partial U}
+
\tilde{S}_r
(y_1, U)
\end{equation}
where $\tilde{S}_r$, one component of the vector  $\tilde{S}_i = (\tilde{S}_1,\tilde{S}_r,\tilde{S}_5)$,
represents an $O(1)$ `left-over' term that acts to couple the various components of ${G}_\sigma=(G_i, G_4,G_5)$ in the ALEE, (\ref{eq:GAA}). As we shall see shortly, the retention of $\tilde{S}_i$, while being an exact consequence of the algebraic manipulation, prevents the solution to the lowest order asymptotic expansion of the Green's function variable $\tilde{\nu}$ (introduced below) from being governed by a hyperbolic PDE. 
This form of (\ref{G_1}) also generalizes Eq. (5.23) in GSA for an axisymmetric jet in which $\epsilon = O(1)$ where $\tilde{S}_r(y_1, U)$ is defined by

\begin{equation}
\label{S_r}
\tilde{S}_r
(y_1, U)
=  
\frac{\partial r} {\partial U}
\left[
\left(
D_0
-
\frac{\partial {V}_r} {\partial r}
\right)
\tilde{G}_r
-(\gamma-1)
\tilde{X}_r
\tilde{G}_4
\right]
(y_1, U),
\end{equation}
such that mean flow components $(U, V_r)$ in (\ref{subX1}), (\ref{D0_trans}) $\&$ (\ref{S_r}) are arbitrary (i.e. $O(1)$) at this point in the analysis. 
Eqs.(\ref{G_1}) and (\ref{S_r}) can now be used to generalize Eq.(5.26) in GSA. Inserting the second member of (\ref{D0_trans}) into (\ref{eq:GAA}c) and using (\ref{G_1}) $\&$ (\ref{S_r}) shows that (\ref{eq:GAA}c) can be transformed to 
\begin{equation}
\label{nu_eqn}
\tilde{D}_0 \tilde{\nu}
(y_1, U)
=  
\widetilde{c^2}
D_0 \tilde{G}_4
+
\tilde{S}_5
 (y_1, U)
\end{equation}
for the Green's function variable, $\tilde{\nu}= \tilde{\nu}
(y_1, U) \equiv \widetilde{c^2} \tilde{G}_4 + \tilde{G}_5$ when $\widetilde{c^2} = f (U)$ in which $f$ is an arbitrary function at this point but will be specified shortly to eliminate any `$\tilde{G}_4$ terms' appearing on the left side of (\ref{hyb_eqn1}). Here, $\tilde{S}_5 (y_1, U)$ is given by 
$
\tilde{S}_5
=
\tilde{X}_r
\tilde{G}_r
+\tilde{X}_1
\tilde{S}_r
$. 

To set about showing that any $\tilde{G}_4$ terms on the left hand side of the equation that governs $\tilde{\nu}
(y_1, U)$ vanish depending on the choice of $\widetilde{c^2} = f (U)$, we first integrate (\ref{G_1}) by parts to re-write its right hand side in terms of $\tilde{\nu}(y_1, U)$ and insert the result, (\ref{D0_trans}) $\&$ (\ref{nu_eqn}) into the $i=1$ component of (\ref{eq:GAA}a).
The latter can then be written in the following form
\begin{equation}
\begin{split}
\label{hyb_eqn1}
\frac{\partial }{\partial U}
\tilde{D}_0 \tilde{\nu}
-
\frac{1}{\widetilde{c^2}}
\frac{\partial \widetilde{c^2}}{\partial U}
\tilde{D}_0 \tilde{\nu}
+
\tilde{X}_1
\frac{\partial^2 \tilde{\nu}}{\partial U^2}
-
&
\tilde{X}_1
%
\left[
(\gamma-1)
+
\frac{\partial^2 \widetilde{c^2}}{\partial U^2}
\right]
\tilde{G}_4 
=
\\
& 
-\tilde{S}_1 
+
\left(
\frac{\tilde{S}_5}{\widetilde{c^2}}
+
D_0
\tilde{S}_r
\right) 
,
\end{split}
\end{equation}
where $\tilde{S}_1$ is defined below.

The pre-factor multiplying $\tilde{G}_4$ in (\ref{hyb_eqn1}) is identically zero when $\widetilde{c^2} = f(U)$ is assumed to satisfy Crocco`s relation\cite{GSA12, Crocco}, $\widetilde{c^2}(U)  = c{}^2_\infty - (\gamma-1)U^2/2$ (where $c{}_\infty$ is the speed of sound at infinity). 
This approximation assumes that the mean flow stagnation enthalpy is constant and although it is used as a first approximation to the static temperature (enthalpy) field in a compressible laminar boundary layer (White\cite{FWhite}, p. 579 and van Oudheusden\cite{vanOud}), Fig.$4$ in Afsar {\it et al}.\cite{Afsetal2016} showed that it remains within $2\%$ of the mean flow obtained from a steady RANS calculation of an unheated jet at $Ma=0.9$.

For a heated jet, on the other hand, it is more appropriate to use the Crocco-Buseman relation in which a flow of unity Prandtl number and zero streamwise pressure gradient\footnote{This will be consistent when the leading order of mean flow expansion, (\ref{eq:Meanflow_exp}), is considered in $\S.$\ref{S:3}} possesses a static enthalpy $\tilde{h} = \tilde{h}(U) = -U^2/2 +\tilde{c}_1 U + \tilde{c}_2$ (White\cite{FWhite}, p.579 $\&$ \textit{f}. and 627--628). Since the jet has zero flow in the outer region, evaluating constants $(\tilde{c}_1,\tilde{c}_2)$ gives (cf. Eq. 2.4c in Leeshaft {\it et al}.\cite{Leeshafft})
\begin{equation}
\label{CB_reln}
\widetilde{c^2} (U)
=
c{}^2_\infty + c{}^2_\infty(TR-1)U + \frac{\gamma-1}{2} U(1-U)
\end{equation}
and therefore that ${\partial^2 \widetilde{c^2}}/{\partial U^2} = -(\gamma-1)$.
Eq.(\ref{CB_reln}) implies that, even for a heated jet, the square brackets on the left side of (\ref{hyb_eqn1}) is identically eliminated. We verify the Crocco-Bussemann relation in $\S.$\ref{S:5} using the streamwise meanflow, $U$, obtained from a steady RANS solution of the heated and isothermal supersonic jets in Table \ref{tab:table1}. Note, however, that many researchers have used approximations of this type before and shown its similarity with mean temperature or density profiles of high speed turbulent jets. For example, Figs. 2.2 $\&$ 2.3 in Dahl\cite{Dahl} and p.36 of Fontaine {\it et al}.\cite{Fontaine} who showed that the effect of assuming the Crocco-Bussemann relation produced a slight change in the momentum thickness of initial shear layer at jet Mach numbers ranging: $(0.5-0.98)$. 

Owing to the fact that the square brackets in (\ref{hyb_eqn1}) vanish when using (\ref{CB_reln}) for $\widetilde{c^2}$, we can integrate by parts in (\ref{hyb_eqn1}) to show that the combined Green's function variable, $\tilde{\nu}(y_1,U)$, is determined by the following PDE:
\begin{equation}
\label{Hyp2}
\mathcal{L}
\tilde{\nu}
(y_1, U)
=
\mathcal{F}(\tilde{\boldsymbol S}),
\hspace{0.5cm}
\textnormal{for}
\hspace{0.25cm}
\epsilon = O(1),
\end{equation}
where 
\begin{equation}
\label{L_Hyp}
\mathcal{L}(y_1, U)
\equiv
\widetilde{c^2}
\frac{\partial }{\partial U}
\frac{1}{\widetilde{c^2}}
\tilde{D}_0 
+
\tilde{X}_1
\frac{\partial^2 }{\partial U^2},
\end{equation}
is a hyperbolic PDE operator and $\tilde{\boldsymbol S}= \{\tilde{ S}_1,\tilde{S}_r, \tilde{S}_5\}$, the vector of so-called `left over' terms enters via the functional $\mathcal{F}(\tilde{\boldsymbol S}) = (\delta_{i1} + \delta_{ir}D_0 -\delta_{ir}/\widetilde{c^2})S_i $ defined explicitly by,
\begin{equation}
\label{S_funcs}
\mathcal{F}(\tilde{\boldsymbol S})
=
\mathcal{F}(\tilde{S}_1,\tilde{S}_r, \tilde{S}_5)
:=
\tilde{S}_1
-
\left(
\frac{\tilde{S}_5}{\widetilde{c^2}}
+
D_0
\tilde{S}_r
\right). 
\end{equation}
The components of $\tilde{S}_i$ that enter $\mathcal{F}(\tilde{\boldsymbol S})$ are linearly related to the adjoint Green's function component for the radial momentum equation, $\tilde{G}_r$, and the 
mean flow component, $V_r$. The latter enters $\tilde{S}_i$ as a coefficient or derivative, for example as in,
\begin{equation}
\label{S_1}
\tilde{S}_1
(y_1, U)
=  
\frac{\partial V_r}{\partial y_1}
\tilde{G}_r
(y_1, U),
\end{equation}
or via $\tilde{X}_r$, in (\ref{subX1}), which is present in both the $\tilde{S}_r$ component, given by (\ref{S_r}), and $\tilde{S}_5$ defined below (\ref{nu_eqn}). 
%

To sum up thus far, the final equation we have derived, (\ref{Hyp2}), is simply a direct re-arrangement of Fourier transformed ALEE, (\ref{eq:GAA}a-c), where $\mathcal{F}(\tilde{\boldsymbol S} )$ is defined explicitly in (\ref{S_funcs}).
It is valid for an arbitrary axisymmetric jet flow with mean flow components, ${\boldsymbol v} = (U, V_r)$ at $O(1)$ jet spread rates, where the speed of sound is determined by Crocco-Busemann relation, (\ref{CB_reln}) and 
$\tilde{G}_\sigma=\tilde{G}_\sigma (y_1, U,\phi|{\boldsymbol x}; \omega)$ is the appropriate $O(1)$ frequency adjoint vector Green`s function solution to (\ref{eq:GAA}) with suffix $\sigma = 1,2,...5$.
The mapping of independent variables $(y_1,r)$ $\rightarrow(y_1,U)$ can, in principle, be used for any problem governed by a system of equations of the type given by (\ref{eq:GAA}).

Although we have reduced the total number of independent equations that need to be solved from $5$ in (\ref{eq:GAA}) down to $4$ in (\ref{Hyp2})--(\ref{S_funcs}), the Green's function problem is just as complex as the original ALEE system. 
This is because the functional, $\mathcal{F}(\tilde{\boldsymbol S})$, depends on the `leftover terms' through the vector, $\tilde{S}_i = \{\tilde{S}_1,\tilde{S}_r,\tilde{S}_5\}$, that appears on the right hand side of (\ref{Hyp2}) and which transforms it to a mixed PDE that requires the solution of $4$ coupled equations for $(\tilde{\nu},\tilde{G}_4,\tilde{G}_r, \tilde{G}_\phi)$ using (\ref{nu_eqn}), (\ref{Hyp2}) and $i=(r,\phi)$ components of (\ref{eq:GAA}a) when (\ref{G_1}) is substituted for $\tilde{G}_1$.

But the vector $\tilde{S}_i = (\tilde{S}_1,\tilde{S}_r,\tilde{S}_5)$ turns out to be asymptotically sub-dominant (i.e., negligible in comparison to the lowest order solution to Eq. \ref{Hyp2}) at low frequencies when $\omega = O(\epsilon)$ and under an appropriate distinguished scaling\cite{GSA12} for the $(r, \phi)$ components of  $\tilde{G}_\sigma (y_1, U,\phi|{\boldsymbol x}; \omega)$ that balances (\ref{eq:GAA}). 
In the next section we prove that $\mathcal{F}(\tilde{\boldsymbol S}) = o(1)$ in the limit as $\epsilon\rightarrow 0$ and how it necessarily shows that (\ref{Hyp2}) decouples into a homogeneous (i.e. right hand side equal to zero) hyperbolic PDE for the single dependent variable, $\tilde{\nu}$.
But even though $\mathcal{F}(\tilde{\boldsymbol S})$ will be shown to play a largely irrelevant role in the solution at the lowest order asymptotic expansion of $\tilde{\nu}$, our simplified derivation highlights the crucial role the dominant balance of $\tilde{G}_r$ plays in the elimination of the $\tilde{S}_i = \{\tilde{S}_1,\tilde{S}_r,\tilde{S}_5\}$ vector. The Green's function, $\tilde{G}_\phi$, is also important to this dominant balance calculation thanks to the $i=\phi$ component of (\ref{eq:GAA}a) and the adjoint energy equation, (\ref{eq:GAA}b), both of which were not explicitly used in deriving (\ref{Hyp2}).  
As opposed to GSA, our analysis applies in flows where $TR\neq1$.

\subsection{\label{sec3.2:redux}Reduction of (\ref{Hyp2}) to a single hyperbolic PDE at $\epsilon\ll O(1)$}

Experiments by Panchapasekan $\&$ Lumley\cite{Pancha} indicate (see p.101\textit{ff}. in Pope\cite{Pope}) that the jet spread rate, $\epsilon$, is virtually constant with Reynolds number in jets and nearly equal to $0.1$ at isothermal conditions.
It makes sense, therefore, if we allow the mean flow to vary over a slow streamwise length, $Y \equiv\epsilon y_1= O(1)$ (corresponding to a long physical scale, $y_1$, relative to an origin placed at the nozzle exit plane). In other words, we consider an axisymmetric jet that diverges with small spread rate inasmuch as $\epsilon \ll O(1)$.
Whence, the mean flow must expand according to (A.1--A.2) in G $\&$ L\cite{GanL}; viz.:
\begin{equation}
\label{eq:Meanflow_exp}
%
\begin{split}
{\tilde{v}_i} = & \{U(Y), V_r(Y, U)\} 
\\= &\begin{cases}
      U^{} +\epsilon U^{(1)}(Y, U)+ O(\epsilon^2),\hspace{0.75cm} i=1 \\
      \epsilon (V{}_r^{} + \epsilon V_r^{(2)})(Y, U) + O(\epsilon^3), \hspace{0.25cm} i=r 
            \end{cases} 
\end{split}
\end{equation}
when the lowest order expansion of $\widetilde{c^2}$ is determined by the Crocco-Busemann relation, (\ref{CB_reln}). We have not put superscripts on the lowest order mean flow components, that would otherwise appear as $(U^{(0)},V{}_r^{(1)})$ respectively; they will be taken as that computed by the RANS solution.
Moreover, at this order in $\epsilon$, ${\bar \rho}(Y, U) =  \bar{\rho}(U)$, ${\bar p}(Y, U) = const.$ and the mean flow advection vector, $X_i(\boldsymbol{y})$, that enters algebraically in the $\{\tilde{S}_r,\tilde{S}_5\}$ components of $\tilde{S}_i$, similarly expands as
\begin{equation}
\label{eq:X_exp}
\begin{split}
{\tilde{X}_i}=& \{\tilde{X}_1, \tilde{X}_r\}(Y, U)  \\=&\begin{cases}
      \epsilon \bar{X}{}_1^{}(Y, U) +\epsilon^2 \tilde{X}{}_1^{(2)}(Y, U)+ O(\epsilon^3),\hspace{0.25cm} i=1 \\
      \epsilon^2\bar{X}{}_r (Y, U) + O(\epsilon^3), \hspace{2.0cm} i=r 
            \end{cases} 
\end{split}
\end{equation}
where the leading terms are defined by, $ \bar{X}{}_1 \equiv \bar{X}{}_1^{(1)}  = V_r (\partial U/\partial r)$ and $\bar{X}{}_r \equiv \bar{X}{}_r^{(2)}=(U\partial/\partial Y + V_r \partial/\partial r) V_r$ for the streamwise and radial components respectively.
Hence, when measured from the jet centerline, the mean flow separates into an inner region, given by (\ref{eq:Meanflow_exp}) $\&$ (\ref{eq:X_exp}), where the inner radial co-ordinate is $r = O(1)$, and an outer region where this expansion break downs---at large radial locations (using inner variable, $r$) for which $R \equiv \epsilon r = O(1)$. 

But as discussed in $\S$.1, GSA show that the long $O(1/\epsilon)$ streamwise variation of non-parallel flow alters the leading order asymptotic structure of propagator  ${\Gamma}_{\lambda, j}({\boldsymbol y}| {\boldsymbol x}; \omega)$ everywhere in the flow at $Ma=O(1)$ when ${g}{}_{\sigma 4}^{a}
({\boldsymbol y}, \tau| {\boldsymbol x}, t)$ modulates in time under an appropriate slowly breathing asymptotic scaling. 
This happens at low frequencies when time variations are slow and (crucially) of the same order as the streamwise variations in the mean flow. Mathematically, ${ g}{}_{\sigma4}^{a}
({\boldsymbol y}, \tau| {\boldsymbol x}, t)$ depends on $\tau$ through re-scaled $O(1)$ time variable $\tilde{T}\equiv\epsilon\tau = O(1)$. 
In frequency space, the Strouhal number, $St$, is of the order of the jet spread rate, $\epsilon$, in the solution ${ G_\sigma}
({\boldsymbol y}| {\boldsymbol x}; \omega)$ of (\ref{eq:GAA}).
Hence, the distinguished asymptotic scaling in the latter occurs when $\epsilon \rightarrow 0$ and the scaled frequency, $\Omega \equiv \omega/\epsilon = O(1)$ is held fixed. It is only in this limit, where the solution to the ALEE becomes asymptotically disparate as $\epsilon \rightarrow 0$ and (just as Eqs. \ref{eq:Meanflow_exp} $\&$ \ref{eq:X_exp}) divides into an inner solution where $r = O(1)$ and an outer solution valid at $R \equiv \epsilon r = O(1)$. Note that if ${g}{}_{\sigma4}^{a}
({\boldsymbol y}, \tau| {\boldsymbol x}, t)$ were to depend on $\tau$ through $O(\epsilon^{-m})$ scaled times in such a manner that ${ G_\sigma}
({\boldsymbol y}| {\boldsymbol x}; \omega)$ depends on relatively $O(1)$ frequencies, $\omega = O(\epsilon^m)$, for exponent values $-N\leq m\leq 1$, it would confine non-parallel flow effects to supersonic speeds in the thin critical layer where ${ G_\sigma}
({\boldsymbol y}| {\boldsymbol x}; \omega)$ is otherwise singular and the mean flow can be reduced to a locally parallel flow away from this region at all frequencies of interest\cite{GanL}.

Inserting the above distinguished scaling into (\ref{D0_trans}) and using the mean flow expansion (\ref{eq:Meanflow_exp}) $\&$ (\ref{eq:X_exp}) shows that the latter operator acting on $\tilde{\nu}(Y, U)$ is given by
\begin{equation}
\label{D0nu_trans}
 \begin{split}
D_0
\tilde{\nu}
(y_1,U)
=  
&
\epsilon  \left( i\Omega + U\frac{\partial}{\partial Y} + V_r \frac{\partial}{\partial r}
  \right) \tilde{\nu}
\\
\equiv
&
\epsilon
\left(
\bar{D}_0 + \bar{X}_1 \frac{\partial}{\partial U}
\right)
\tilde{\nu}(Y, U),
  \end{split}
   \end{equation}
where $\bar{D}_0 \equiv i\Omega + U{\partial}/{\partial Y}$ at $\Omega=O(1)$. Eq. (\ref{D0nu_trans}), and the first line of (\ref{eq:X_exp}), shows that the left side of (\ref{Hyp2}) will be at most $O(\epsilon)$ when $\tilde{\nu}=O(1)$. The latter of which must be the case since the solution to $\tilde{\nu}$ in the outer region (see Eq.5.40 of GSA) expands in this manner. 
The right side of (\ref{Hyp2}) will then be $O(\epsilon^2)$ prior to considering the dominant balance of $\tilde{G}_r$ since, by (\ref{D0nu_trans}) $\&$ (\ref{Sr_exp}), the `leftover' terms on the right side of (\ref{Hyp2}) (defined by \ref{S_1}, \ref{S_r} $\&$ line below \ref{nu_eqn}) when substituted into (\ref{S_funcs}), expand to leading order as follows 
\begin{equation}
\label{Sterm_exp}
\mathcal{F}(\tilde{\boldsymbol S})
\rightarrow
\epsilon^2 \frac{\partial V_r}{\partial Y}
\tilde{G}_r
-\frac{1}{\widetilde{c^2}}
\epsilon
\left(
\epsilon\bar{X}{}_r
\tilde{G}_r
+ 
\bar{X}_1
\tilde{S}_r
\right)
-
\epsilon
\left(
\bar{D}_0 + \bar{X}_1 \frac{\partial}{\partial U}
\right)
\tilde{S}_r,
\end{equation}
where,
\begin{equation}
\begin{split}
\label{Sr_exp}
\tilde{S}_r
(y_1, U)
&
=  
\frac{\partial r} {\partial U}
\left(
D_0
-
\frac{\partial {V}{}_r^{}} {\partial r}
\right)
\tilde{G}_r
+o(1)
\\
\equiv
&
\epsilon
\frac{\partial r} {\partial U}
%
%
\left(
\bar{D}_0 + \bar{X}_1 \frac{\partial}{\partial U}
-
\frac{\partial {V}{}_r^{}} {\partial r}
\right)
\tilde{G}_r
(Y, U)
+
O(\epsilon^2).
\end{split}
\end{equation}
after again using (\ref{eq:Meanflow_exp}), (\ref{eq:X_exp}) $\&$  (\ref{D0nu_trans}).

Although an asymptotic expansion of $\tilde{G}_r$ starting as $\tilde{G}_r=O(1)$ would therefore cause $\mathcal{F}(\tilde{\boldsymbol S})$, on the right side of (\ref{Hyp2}), to drop out of the lowest order $\tilde{\nu}$--equation, this does not turn out to give the richest possible balance. The only other self-consistent asymptotic expansion of $\tilde{G}_r$, which turns out to also be the least degenerate solution to $\tilde{\nu}$ is given by the Fourier transform of Eq. (5.6) in GSA; namely, in the present formalism, one where $\tilde{G}_{(r,\phi)}=O(1/\epsilon)$ at leading order. 
However this would cause $\tilde{S}_r
(y_1, U)= O(1)$ in (\ref{Sterm_exp}) and (\ref{Sr_exp}) which would, on the face of it, balance the $O(\epsilon)$ expansion of the left hand side of (\ref{Hyp2}) after inserting (\ref{eq:X_exp}) $\&$(\ref{D0nu_trans}) in the latter.
But thankfully $\mathcal{F}(\tilde{\boldsymbol S})$ still drops out of (\ref{Hyp2}) because, whatever asymptotic expansion we take for $\tilde{G}_{(r,\phi)}$, both of these Green's functions must remain bounded on the jet axis.
That is, by considering the conditions across the surface $r=0$ in the $i=\phi$ component of (\ref{eq:GAA}a) and using $\boldsymbol{\nabla}.{\boldsymbol\tilde{v}} \sim D_0\tilde{G}_4 = O(\epsilon)$ in the adjoint energy equation, (\ref{eq:GAA}b), 
shows that $\tilde{G}_{(r,\phi)}= 0$ at lowest order in (\ref{eq:GAA}), (\ref{Hyp2}), (\ref{S_funcs}), (\ref{Sterm_exp}) $\&$ (\ref{Sr_exp}). 

GSA use an alternative explanation to restrict the modal expansion of the azimuthal Fourier transform of $\tilde{\nu}$; this, however, follows straightforwardly here because if $\tilde{G}_{\phi}= 0$ at lowest order, the $i=\phi$ component of (\ref{eq:GAA}a) recovers the fact that the lowest order (axisymmetric) solution $\tilde{\nu}$ to (\ref{Hyp2}) is independent of azimuthal angle $\phi$.  
In other words, the Fourier transform of $\tilde{\nu} (Y,U, \phi|X, \Phi;\Omega)$ in the difference, $(\Phi-\phi)$, is given by
%
 \begin{equation}
  \begin{split}
\label{nu_FT_azim}
\hat{\nu}{}^{ (n)} 
(Y, U) 
 = &
\frac{1}{2 \pi}
\int\limits_{-\infty}^{\infty}
\tilde{\nu} (Y, U| X,|\boldsymbol{x}_T|,\Phi-\phi; \Omega) 
e^{i n(\Phi-\phi)}
\,d(\Phi-\phi) \\
 \equiv &
\delta(n)
\tilde{\nu} (Y, U)\mid_{(\Phi- \phi)=0}, 
 \end{split}
\end{equation}
where $\delta(\bullet)$ is the Dirac delta function of argument $(\bullet)$ (we have suppressed repeated variable list in the $\tilde{\nu}$--solution). 
Using (\ref{eq:GFT}), the solution, $\bar{\nu} (Y,U)$, given by the scaled Fourier transform (note error in pre-factor of Eq. 5.8 in GSA); 
\begin{widetext}
 \begin{eqnarray}
\label{eq:Scaled_G}
\tilde{\nu}(Y,U)
&& \equiv 
\frac{\epsilon}{4\pi c{}_\infty^2 |{\boldsymbol x}|}
e^{i\Omega X/c_\infty}
\bar{\nu}
(Y, U| X,|\boldsymbol{x}_T|,0 ; \Omega)  \nonumber
\\
&& =
\frac{1}{2 \pi\epsilon}
\int\limits_{-\infty}^{\infty}
e^{i \Omega(\tilde{T}_0-\tilde{T})}
(\widetilde{c^2} \tilde{g}{}_{4 4} + \tilde{g}{}_{5 4}
)(Y, U| X,|\boldsymbol{x}_T|,0 ;\tilde{T}_0-\tilde{T}) 
\,d(\tilde{T}_0-\tilde{T}),
\end{eqnarray}
\end{widetext}
is now determined by (\ref{Hyp2}) when $\mathcal{F}(\tilde{\boldsymbol S}) = o(1)$ at arbitrary $\Omega=O(1)$ frequencies. Hence, setting the right hand side in (\ref{Hyp2}) equal to zero, shows that the lowest order term in the expansion $\nu(y_1, r) = \bar{\nu}(y_1,r) + \bar{\nu}^{(1)}(y_1,r) +...$ is given by the solution to
\begin{equation}
\label{Hyp3}
\mathcal{L}
\bar{\nu}
(Y, U)
\equiv
\widetilde{c^2}
\frac{\partial }{\partial U}
\left(
\frac{1}{\widetilde{c^2}}
\bar{D}_0 \bar{\nu}
\right)
+
\bar{X}_1
\frac{\partial^2 \bar{\nu}}{\partial U^2}
=
0,
\hspace{0.25cm}
\textnormal{for}
\hspace{0.25cm}
\epsilon \ll O(1),
\end{equation}
by the implicit function theorem where $\bar{\nu}
(Y, U)\equiv  \widetilde{c^2} \bar{G}{}_{4} + \bar{G}{}_{5}
$ is related to the zeroth-order azimuthal mode $\hat{\nu}{}^{ (0)} (Y, U)$ through the inverse Fourier transform of (\ref{nu_FT_azim}) in $(\Phi-\phi)$ where $(X,T_0) = \epsilon (x_1,t)$ are appropriate $O(1)$ slow variables for the observation field point $(x_1,t)$. 
Moreover, $Y = const.$ and ${dU}/{dY} = \bar{X}_1/U$ represent the characteristic curves (Garebedian\cite{Garab}, pp. 121-122) of (\ref{Hyp3}). The pre-factor of the second member on the first line of (\ref{eq:Scaled_G}) allows the outer boundary conditions in (\ref{BC1}) $\&$ (\ref{BC2}) for the scaled inner solution $\bar{\nu} (Y, U)$ to depend on the observation point, $\boldsymbol{x}$, only through $\theta$.
Eq. (\ref{Hyp3}) applies to jet flows with $TR>1$ and is identical to Eq. (5.31) in GSA but with $\widetilde{c^2}$ now determined by (\ref{CB_reln}).

The hyperbolic structure of (\ref{Hyp3}) shows that it is unnecessary to impose a downstream boundary condition. Fig. 1 in GSA indicates how `$\bar{\nu}$-waves' propagate to both left and right from the $U = 0$ boundary and that no boundary conditions are required on the $Y = 0$ and $Y \rightarrow \infty$ level curves (i.e. no inflow condition is necessary). Hence $\bar{\nu}(Y,U)$ is now uniquely determined by the outer boundary conditions (i.e., by matching to the inner limit of the outer solution using Van Dyke's rule\cite{VanDyke}) obtained from the (zero flow wave equation) solution to (\ref{Hyp3}) when $\bar{X}_1 = 0$. That is,
\begin{equation}
\label{BC1}
   \bar{\nu}(Y,0) = 
     -i\Omega c{}_\infty^2 e^{-i\Omega Y\cos\theta/c{}_\infty}, 
\end{equation}
\begin{equation}
\label{BC2}
   \frac{\partial\bar{\nu}}{\partial U}(Y,0) = 
     -i\Omega c{}_\infty \cos\theta e^{-i\Omega Y\cos\theta/c{}_\infty}, \\ 
\end{equation}
apply on the non-characteristic curve, $U = 0$, where $U \rightarrow 0$ corresponds to the outer limit, $r \rightarrow \infty$
and, $Y\geq 0$ (note the sign error in Eqs. 5.45 $\&$ 5.48 in GSA).
Eqs. (\ref{Hyp3}) -- (\ref{BC2}) show that the Green's function $\bar{\nu}(Y,U; \Omega)$ is independent of jet spread-rate, $\epsilon$, at lowest order after the numerical solution to (\ref{Hyp3}) is determined in $(Y,U)$ co-ordinates  at fixed scaled frequencies, $\Omega$. 
The matching conditions, (\ref{BC1}) $\&$ (\ref{BC2}), also show that any oscillatory behavior (which Eq. {\ref{Hyp3}} admits near the outer boundary $U\rightarrow 0$) of the form, $\bar{\nu} \sim U\Gamma e^{-i\Omega \ln U/E}$, where $\Gamma=\Gamma(Y)$ is an arbitrary function and $\bar{X}_1\rightarrow E(Y) U$ as $U\rightarrow0$, is entirely eliminated (see Eqs. 5.40 $\&$ 5.47 in GSA). 

Jet heating does not obviously effect the contribution the nozzle plays to the solution of (\ref{Hyp3}) at $\Omega = O(1)$ frequencies. 
GSA (p.207 $\&$ \textit{f}.) show that nozzle contribution to the outer boundary condition enters through the inner limit of a `scattered potential' function that satisfies the homogeneous two-dimensional Helmholtz equation on a half plane extending to upstream infinity at $O(r_J)$ distance from the jet centerline (Morse $\&$ Feshbach\cite{Morse}, p. 891). The axisymmetric mode of which behaves logarithmically.
But the inner solution $\bar{\nu}(Y,U)$ generates scattered waves (i.e. is induced by outer incoming waves) through the matching conditions, (\ref{BC1}) $\&$ (\ref{BC2}), and it cannot behave logarithmically as $r \rightarrow \infty$ when matched to that outer solution at any $TR$. 
Hence Van Dyke's rule\cite{VanDyke} shows that $\bar{\nu}(Y,0)$ will not behave like $\ln R$ as $R \rightarrow 0$ at $O(\Omega^0)$. Or, alternatively, $\bar{\nu}(Y,U)$ cannot behave as $\ln r \sim \ln(\ln (1/U))^{1/2}$ as $U\rightarrow 0$, where $r^2 \sim \ln(1/U)$ as $U\rightarrow0$ via Eq. (6.1) in GSA.
Therefore, any influence of the nozzle can also be entirely neglected in the solution to (\ref{Hyp3}) for all $TR\geq1$.

\subsection{\label{sec3.3:exp} Propagator expansion at $\Omega=O(1)$ frequencies}
Since the propagator, (\ref{eq:Prop}), depends on $\bar{G}_\sigma (Y,r| {\boldsymbol x}; \Omega)$ and the mean flow expansion (\ref{eq:Meanflow_exp}), its solution must also separate out into the same asymptotic regions as (\ref{eq:Meanflow_exp}) $\&$ (\ref{eq:X_exp}) and depend on scaled variable/parameter $(Y, \Omega)=O(1)$.
Hence, $\bar{\Gamma}_{\lambda, j} (y_1,r| {\boldsymbol x};\omega) =\bar{\Gamma}_{\lambda, j}
(Y,r| {\boldsymbol x};\Omega)$. 
Taking the gradient operator, ${\boldsymbol\nabla} \equiv \boldsymbol{e}_1 {\partial}/ {\partial y_1}+ \boldsymbol{e}_r {\partial}/ {\partial r} + {\boldsymbol e}_\phi {\partial}/ {r \partial \phi}$ of the lowest order mean flow vector, $\tilde{\boldsymbol{v}}({\boldsymbol y})$, in (\ref{eq:Meanflow_exp}) 
we can easily show that non-symmetric rank-two tensor, ${\partial \tilde{v}_\lambda}/{\partial y_j}$, in (\ref{eq:Prop}), where $\tilde{v}_\lambda \equiv \{ \tilde{v}_i, 0\}= \{ U, V_r, 0, 0\}$ at $\epsilon \ll O(1)$, possesses the following asymptotic expansion,
\begin{equation}
\label{Mean flow gradient exp}
\begin{split}
\frac{\partial\tilde{v}_\lambda}{\partial y_j}
(Y,r)
=
&
\delta_{\lambda 1}
\delta_{j r}
\frac{\partial U }{\partial r}
+
\epsilon
\delta_{\lambda 1}
\left(
\delta_{j r}
\frac{\partial U }{\partial Y}
+
\delta_{j r}
\frac{\partial V_r }{\partial r}
\right)
\\
+
&
\epsilon
\frac{V_r}{r}
\delta_{\lambda \phi}
\delta_{j \phi}
+ O(\epsilon^2),
\end{split}
\end{equation}
in $(Y,r, \phi)$ cylindrical co-ordinates using (\ref{eq:Meanflow_exp}) and ${\partial \boldsymbol{e}_r}/ {\partial \phi} = \boldsymbol{e}_\phi$ $\&$ ${\partial {\boldsymbol e}_\phi} /{ \partial \phi} = -\boldsymbol{e}_r$. 
Then, inserting (\ref{eq:Meanflow_exp}) and the lowest order scaled Green's function vector, $\bar{G}_\sigma (Y,r| {\boldsymbol x}; \Omega) = \bar{G}_1 \delta_{\sigma 1}+\bar{G}_4 \delta_{\sigma 4}$ into (\ref{eq:Prop}) shows that the latter propagator expands like,
\begin{equation}
\begin{split}
\label{Prop_Exp}
&
\bar{\Gamma}_{\lambda, j}
(Y,r| {\boldsymbol x};\Omega)
=
\delta_{\lambda 1}
\delta_{j r}
\left(
\frac{\partial \bar{G}_1}{\partial r}
-
(\gamma-1)
\frac{\partial U}{\partial r}
\bar{G}_4
\right)
+
\\ 
+
&
\delta_{\lambda 4}
\delta_{j r}
\frac{\partial \bar{G}_4}{\partial r}
+
\epsilon
\delta_{\lambda 1}
\delta_{j 1}
\left(
\frac{\partial \bar{G}_1}{\partial Y}
-
(\gamma-1)
\frac{\partial U}{\partial Y}
\bar{G}_4
\right)
+
\epsilon
\delta_{\lambda 4}
\delta_{j 1}
\frac{\partial \bar{G}_4}{\partial Y}
\\
-
&
\epsilon(\gamma-1)
\left(
\delta_{\lambda 1}
\delta_{j r}
\frac{\partial V_r}{\partial r}
+
\delta_{\lambda\phi}
\delta_{j\phi}
\frac{V_r}{r}
\right)
\bar{G}_4
+
O(\epsilon^2)
,
\end{split}
\end{equation}
in $(Y,r)$ co-ordinates at $\Omega=O(1)$ frequencies. $\bar{G}_\sigma (Y,r| {\boldsymbol x}; \Omega)$ is found in $(Y, U)$ co-ordinates using an equivalent re-scaling as (\ref{eq:Scaled_G}).
It is transformed back to $(Y=\epsilon y_1,r)$ co-ordinates for integration over $\boldsymbol{y}$ in (\ref{eq:Iom2}).
More specifically, since $\tilde{S}_i = \{\tilde{S}_1,\tilde{S}_r,\tilde{S}_5\}\equiv 0$ at lowest order, the solution to $\bar{\nu}(Y,U)$ found by solving (\ref{Hyp3}) allows $\bar{G}_4$ to be determined using (\ref{nu_eqn}) after re-scaling (\ref{D0nu_trans}) using (\ref{eq:Scaled_G}) and inserting the latter into the left hand side of (\ref{nu_eqn}).
$\bar{G}_1$ is then determined by substituting $\bar{G}_4$ into (\ref{G_1}) replacing $\bar{G}_{5}$ with $\bar{G}_{4}$  and $\bar{\nu}$ (see sentence below \ref{Hyp3}) 
where, again, both (\ref{G_1}) and (\ref{nu_eqn}) are interpreted in terms of the scaled Green's function variables via (\ref{eq:Scaled_G}) and use is made of the chain rule, $\partial \bar{G}_1/\partial r = (\partial U/\partial r)\partial \bar{G}_1/\partial U $.

\section{\label{S:4} Approximate formula for the peak jet noise in heated flows}

\subsection{\label{sec4.1:exp} WKB reduction of (\ref{eq:Iom3})}

The variation of the propagator $\Gamma{}_{\mu, l}^*
({\boldsymbol y} + {\boldsymbol \eta}| {\boldsymbol x}; \omega)
$  over ${\boldsymbol \eta}$ can be approximated by taking advantage of the scale disparity  between the mean flow and turbulence relative to the acoustic wavelength, $\lambda_\text{acoustic}$, in the correlation volume $V({\boldsymbol \eta})$ of integral in (\ref{eq:Iom2}).
In an asymptotic sense, the ALEE solution that determines $\Gamma{}_{\mu, l}^*$ in (\ref{eq:Prop}) will only contribute to integral over $O(|\boldsymbol{\eta}|)$ distances in (\ref{eq:Iom2}) 
when the mean flow length scales that determine the coefficients (and therefore solution structure) of (\ref{eq:GAA})
are of the same order as the turbulence correlation lengths in their respective directions.
This is because the latter propagator tensor, evaluated at $({\boldsymbol y} + {\boldsymbol \eta})$, multiplies $R_{\lambda j \mu l}$ 
in (\ref{eq:Iom2}).
At minimum, the critical variation in $\Gamma{}_{\mu, l}^*$ occurs at $k_\infty\gg 1$, thus allowing $\Gamma{}_{\mu, l}^*$ to be represented by a Wentzel-Kramers-Brillioun-Jeffreys (WKBJ) approximation inasmuch as $\Gamma{}_{\mu, l}^*
({\boldsymbol y} + {\boldsymbol \eta} | {\boldsymbol x}; \omega) \approx\Gamma{}_{\mu, l}^*({\boldsymbol y} | {\boldsymbol x}; \omega) e^{i {\boldsymbol k}.{\boldsymbol\eta}}$.

Inserting the above 
into (\ref{eq:Iom3}) therefore gives an algebraic formula for the acoustic spectrum:
\begin{equation}
\label{eq:IomWKB}
\frac{I({\boldsymbol x}, {\boldsymbol y};\omega)}{(2\pi)^2} 
\approx
\Gamma_{\lambda, j}
({\boldsymbol y}| {\boldsymbol x}; \omega)
\Gamma{}_{\mu, l}^*
({\boldsymbol y} | {\boldsymbol x}; \omega)
\Phi{}^*_{\lambda j \mu l}
({\boldsymbol y}, k_1, {\boldsymbol k}_T  ; \omega),
\end{equation}
where
\begin{equation}
\label{eq:Spec_Ten}
\Phi{}^*_{\lambda j \mu l}
({\boldsymbol y}, k_1, {\boldsymbol k}_T ; \omega)
:=
\int
\limits_{V_\infty({\boldsymbol \eta})}
\mathcal{H}_{\lambda j \mu l}
({\boldsymbol y}, {\boldsymbol \eta}; \omega)
e^{i {\boldsymbol k}.{\boldsymbol \eta} }
\,d{\boldsymbol \eta},
\end{equation}
such that the spectral tensor, $\Phi{}^*_{\lambda j \mu l}$, possesses two-pair symmetries, $\Phi_{ij kl}=\Phi_{ji kl}=\Phi_{ij lk}$ when $(\lambda, \mu)=(i,k)$ and one-pair symmetry, $\Phi{}^*_{ 4 j k l} = \Phi{}^*_{ 4 j l k} $ when $(\lambda,\mu)=4$.

The robustness of this approximation (Wundrow $\&$ Khavaran\cite{Wundrow}) at $\omega = O(1)$ allows it to remain valid for long, $\lambda_\text{acoustic} = O(1/\epsilon)$, wavelengths (of focus in this paper) when the propagators in (\ref{eq:IomWKB}) are determined at these frequencies and where  $\Gamma{}_{\mu, l}^*
({\boldsymbol y} + {\boldsymbol \eta}| {\boldsymbol x}; \omega)
$, varies slowly over $V({\boldsymbol \eta})$ relative to $\lambda_\text{acoustic}$.
It also implies that the amplitude and phase approximation for $\Gamma{}_{\mu, l}^*
({\boldsymbol y} | {\boldsymbol x}; \omega)$ are given by appropriate Taylor expansions (Eqs.B.2, B.3 $\&$ B.7 in AGF). 
The latter, phase function $S({\boldsymbol y} | {\boldsymbol x})$, is related to wavenumber vector by $\boldsymbol{k} = (k_1,\boldsymbol{k}_T) = k_\infty {\boldsymbol{\nabla} S}$ where, $k_\infty = \omega/c_\infty$, is the far-field wavenumber.
The transverse wave-number vector, $\boldsymbol{k}_T = (k_2,k_3)$, is then defined by the parallel flow Eikonal equation (cf. Eq. 13 in Durbin\cite{Durbin83}) $ |\boldsymbol{k}_T|^2/k{}_\infty^2 = |{\boldsymbol \nabla}_\perp S|^2 =({{c{}_\infty^2}/\widetilde{c^2}})(1-M(\boldsymbol{y}_T) \cos\theta)^2 - \cos^2\theta  $ when $k_1 = k_\infty \cos\theta$ as $|\boldsymbol{x}|\rightarrow \infty$ and $\boldsymbol{\nabla}_\perp$ is the gradient operator in transverse $(y_2, y_3)$ plane (Leib $\&$ Goldstein\cite{LG11}, Eq. 18).

\subsection{Generalizing the axisymmetric representation of ${R}_{\lambda j \mu l}$}
As mentioned in $\S.$\ref{Intro}, the tensor ${R}_{\lambda j \mu l}({\boldsymbol y},\eta_{_1}, \eta_{_\perp} ; \tau)$ possesses $144$ components $(3\times4\times3\times4)$, however, owing to its two pair symmetry property -- inasmuch as  ${R}_{i j k l} = {R}_{ j i k l} $ and ${R}_{i j k l} = {R}_{ i j l k}$ when $(\lambda, \mu) = (1,2,3)$ -- not all of these are independent.
AGF (see table 1 on p.2525 of their paper) show that $144$ reduces to $63$ independent components when these symmetries are taken into account and prior to any kinematic approximation, such as isotropy for example.

%
In this paper, we use an axisymmetric turbulence model that is a much more realistic kinematic representation for jets and which reduces the 63 components to a manageable number.
The approximation assumes that the transverse correlation lengths are small compared to that in the streamwise flow direction. 
This is a well founded assertion in jets (see, for example, Pokora $\&$ McGuirk's measurements\cite{Pokora} in Figs. 19-21 and also Fig.10 of their conference paper, AIAA 2008-3028).
AGF used Pokora $\&$ McGuirk's data to propose that ${R}_{\lambda j \mu l}({\boldsymbol y},\eta_{_1}, \eta_{_\perp} ; \tau)$ is an axisymmetric tensor where  $\eta_{_\perp}=|{\boldsymbol \eta}_{_\perp}|$ and ${\boldsymbol \eta}_{_\perp}= (\eta_{_2}, \eta_{_3})$.
The spectral equivalent of this (lemma's $3.1$ and $3.2$ in Afsar\cite{Afs2012}) requires that $\Phi{}^*_{\lambda j \mu l}({\boldsymbol y},k_{_1}, k{}_{_\perp}^2 ; \omega)$  is axisymmetric with the streamwise direction, $k_{_1}$, being the principle direction of invariance. 
The physical space approximation is consistent with experiments by Morris $\&$ Zaman\cite{M&Z} who show in their Fig. 15 that the transverse and azimuthal correlation lengths are virtually constant across range, $St=(0.01-1.0)$ for an isothermal axisymmetric jet.
Indeed, the axisymmetric approximation in the form used by AGF and Afsar\cite{Afs2012} was corroborated using LES data for high speed subsonic jet in Afsar {\it et al.}\cite{Afsetal2010} and also by AGF using turbulence correlations extracted via PIV data of incompressible water jet.

Since the momentum flux/enthalpy flux propagator term involves a spectral tensor component with odd number of suffixes (of type, $\Phi{}^*_{4 j k l}$), an obvious generalization of the axisymmetric representation of this term worked out in AGF (Eq. C.5) is to allow its {\it three-form} defined by $\Phi{}^*_{4 j k l}$ 
to remain invariant to proper rotations only thus taking into account any possible sign changes coming about by an improper rotation of axes. 
This is worked out in App. \ref{App:A}. Interestingly, since this tensor has one-pair symmetry ($\Phi{}^*_{ 4 j k l} = \Phi{}^*_{ 4 j l k} $), the general formula for $\Phi{}^*_{4 j k l}$ given by (\ref{eq:Grassman4}) reduces to Eq. (C.5) in AGF (i.e. $A_2 = A_3$ and $A_4 =0$ in \ref{eq:Grassman4}).

The $63$ independent components of the real-space tensor ${R}_{\lambda j \mu l}({\boldsymbol y},\eta_{_1}, \eta_{_\perp} ; \tau)$, then reduce to the same $11$ components as AGF.
Hence inserting Eqs. (C.4), (C.5) $\&$ (C.8) in AGF for the axisymmetric representations of $\Phi_{ij kl}$, $\Phi_{4j kl}$ and $\Phi_{4j 4l}$ respectively 
shows that the low frequency acoustic spectrum, (\ref{eq:IomWKB}), corresponding to the peak jet noise can be approximated by the following $3$ independent components of $\Phi_{\lambda j \mu l}$ :
\begin{equation}
\label{I_low}
\begin{split}
& 
I({\boldsymbol x}, {\boldsymbol y};\omega) 
\rightarrow
\left(\frac{\epsilon}{2 c{}_\infty^2 |\boldsymbol{x}|}\right)^2
\\
\times
&
\left[
4|\bar{G}_{12}|^2\Phi{}^*_{1212}
+ 2 Re\left\{ \bar{\Gamma}_{41} \bar{G}{}_{11}^*\Phi{}^*_{4111}\right\}
+
|\bar{\Gamma}_{41}|^2  \Phi{}^*_{4141}
\right]
\end{split}
\end{equation}
where the tensor $G_{ij}$ is the symmetric part of the propagator tensor (\ref{Prop_Exp}) when $\lambda = i$. 
The pre-factor in (\ref{I_low}) is what results after inserting (\ref{Prop_Exp}) into the equivalent propagator statement of the re-scaling in (\ref{eq:Scaled_G}) when $\tilde{\gamma}_{\lambda,j}(Y, U; T_0-T)$ appearing on the far right side and $\bar{\Gamma}_{\lambda,j}(Y, U)$ multiplied by appropriate pre-factor on the left in the second member of (\ref{eq:Scaled_G}).
Substituting this form of the scaled propagator into (\ref{eq:IomWKB}) results in formula (\ref{I_low}).
The scaled propagators in (\ref{I_low}) are also defined by the implicit function theorem statement of the form as in (\ref{G_implicit}),
\begin{equation}
\begin{split}
G_{12}
(y_1,r,\psi|{\boldsymbol x}; \omega)
=
&
\tilde{G}_{12}
(Y(y_1), U(y_1,r))
\\
= 
&
\frac{\partial \tilde{G}_1}{\partial r}
-
(\gamma-1)
\tilde{G}_4
\frac{\partial U}{\partial r},
\label{G_12}
\end{split}
\end{equation}
and, 
\begin{equation}
\begin{split}
G_{11}
(y_1,r,\psi|{\boldsymbol x}; \omega)
=
&
\tilde{G}_{11}
(Y(y_1), U(y_1,r))
\\
= 
&
\epsilon
\left(
\frac{\partial \tilde{G}_1}{\partial Y}
-
(\gamma-1)
\tilde{G}_4
\frac{\partial U}{\partial Y}
\right)
\label{G11_41}
\end{split}
\end{equation}
\begin{equation}
\Gamma_{41}
(y_1,r,\psi|{\boldsymbol x}; \omega)
=
\tilde{\Gamma}_{41}
(Y(y_1), U(y_1,r))
=  
\epsilon
\frac{\partial \tilde{G}_1}{\partial Y},
\label{G11_41b}
\end{equation}
when $(\tilde{G}_1,\tilde{G}_4)$, and therefore $\tilde{\nu}$, are inserted into (\ref{eq:Scaled_G}).
The second and third terms in square brackets in (\ref{I_low}) involving $\Phi{}_{4111}^*$ and $\Phi{}_{4141}^*$ components are referred to as the momentum/enthalpy flux coupling and the enthlapy flux terms respectively.
As we explained earlier, the absence of temperature fluctuations, $T^\prime \approx 0$, implies that the enthalpy fluctuation component of $v{}_\lambda^\prime$ is negligible, i.e. $v{}_4^\prime = o(1)$. 
Hence by (\ref{eq:Spec_Ten}), (\ref{eq:HFT}), (\ref{eq:Rijkl}) and the tensor relation defined below it, $\Phi{}_{4111}^*$ and $\Phi{}_{4141}^*$ are negligible in this case and the acoustic spectrum {\it for the peak jet noise} reduces to that involving only the momentum flux term $\Phi{}^*_{1212}$. 
This is consistent with Fig. 19a in Karabasov {\it et al}.\cite{Karab07} where the full numerical solution of the ALEE reveal that the
$G_{12}$ propagator component (which multiplies $\Phi{}^*_{1212}$ in $I({\boldsymbol x};\omega)$, Eq. \ref{I_om_last}) varies much more rapidly across the shear layer at the end of the potential core compared to any other component of the symmetric tensor, $G_{ij}$, at the peak frequency for the $30^\circ$ spectrum.

Karabasov's {\it et al}\cite{Karab07}.'s conclusions correspond (asymptotically) to the scaling derived in GSA; namely that, $\bar{G}_{12}=O(1)$ at fixed $\Omega =\omega/\epsilon = O(1)$ frequencies. 
The latter distinguished limit was derived by GSA using Karabasov \textit{et al}.'s\cite{KBH13} numerical solution to the ALEE as a guide to discern what spatial/temporal scaling ensures that the lowest order solution to these equations (and therefore the $\bar{G}_{12}$ propagator in \ref{G_12}) is everywhere different from that obtained by a locally parallel flow approximation.
But since Karabasov's\cite{KBH13} calculation was for an isothermal flow, it does raise the question whether this asymptotic scaling
continues to hold in heated flows?
While there is no evidence proving if the propagators, (\ref{G11_41}) $\&$ (\ref{G11_41b}), are numerically smaller than $|G_{12}|$ for a spatially spreading heated jet at $\omega \ll O(1)$, the analysis above (i.e. Eq. \ref{Prop_Exp} and Fig.\ref{SPL_n2n3vary} displayed later in the paper) indicates that it must be true. 

That is, since axisymmetric jet flows possess small spread rates\cite{Pope}, a slowly diverging mean flow approximation where the radial component, $V_r$, is asymptotically smaller than the streamwise, $U$, mean flow component (i.e. $U(Y,r)= O(1) +...$ and $V_r(Y,r)= O(\epsilon)$; see Eq. \ref{eq:Meanflow_exp}) shows that the only dominant balance that could allow $\bar{G}_{11}=O(1)$  is for $\bar{G}_1 = O(1/\epsilon)$ at its lowest order of expansion when ${g}{}_{\sigma 4}^{a}
({\boldsymbol y}, \tau| {\boldsymbol x}, t)$ evolves at $\tilde{T}=\epsilon \tau = O(1)$ times.
As we have explained, any other time-scaling would render the non-parallel flow effects as a higher-order correction to the locally parallel flow Green's function and only enters the leading order in the thin critical layer at frequencies larger than $\omega \ll O(1)$\cite{GanL}.
(Note that, $\bar{G}_{11}$ is used in this proof because this term is contained in both coupling and enthalpy flux propagators, \ref{G11_41} $\&$ \ref{G11_41b}).
But (\ref{eq:Scaled_G}) shows that this particular scaling for $\bar{G}_1$ results in an inconsistent asymptotic balance in (\ref{G_1}) since the combined Green's function, $\bar{\nu}$, and therefore $\bar{G}_4$, must expand like $O(1)$ at lowest order to match on to the outer wave equation solution. {We know this by taking the outer limit ($U\rightarrow 0$) of (\ref{Hyp3}) using matching conditions, (\ref{BC1}) $\&$ (\ref{BC2}).}
Hence, if we write (\ref{G_1}) in $(Y, U;\Omega)$ variables and insert (\ref{eq:Meanflow_exp}), 
\begin{equation}
\label{G_1disc}
\bar{G}_1
(Y, U)
=  
\frac{\partial \bar{\nu}} {\partial U}
-
\frac{\partial \widetilde{c^2}} {\partial U}
\bar{G}_4
+
\bar{S}_r(Y,U), 
\end{equation}
we can see that the only way  
$\bar{G}_1 = O(1/\epsilon)$ at lowest order is if $\bar{G}_r = O(1/\epsilon^2)$ in (\ref{Sr_exp}). 
This is because the first two terms in (\ref{G_1disc}) expand like $O(1)$, so $\bar{S}_r$ must go like $\sim 1/\epsilon$ to allow $\bar{G}_1 = O(1/\epsilon)$. 
But this result -- which would be equivalent to requiring that $\mathcal{F}(\tilde{\boldsymbol S})= O(1)$ after inserting $\bar{G}_r = O(1/\epsilon^2)$ into (\ref{Sr_exp}) $\&$ (\ref{Sterm_exp}) using (\ref{eq:Scaled_G}) -- is inconsistent with the $i=\phi$ component of (\ref{eq:GAA}a) and adjoint energy equation (\ref{eq:GAA}b) which, when taken together, show that 
$\bar{G}_{(r,\phi)} = 0$ at lowest order and, therefore, that $\bar{G}_1 = O(1)$  and ${\partial \bar{G}_1}/{\partial Y} = O(\epsilon)$ at this order.
Thus, the scaling $\bar{G}_{11} = O(\epsilon)$ at $\Omega = O(1)$ frequencies must hold. Our numerical calculations in Fig. \ref{SPL_n2n3vary} prove this to be the case. 

\subsection{Justification of reduced formula (\ref{I_low}) and further approximation to (\ref{I_low2})}

The spectral tensor $\Phi{}^*_{\lambda j \mu l}$ cannot be measured directly, however, there is extensive data for the physical space tensor, $R_{\lambda j \mu l}$, in isothermal flows at various subsonic $Ma$ (Karabasov {\it et al}. \cite{Karab2010}); isothermal water jets in (Pokora $\&$ McGuirk\cite{Pokora}); in subsonic heated co-axial jets (Gryazev {\it et al}.\cite{Gryazev2019}) and for supersonic mixing layers in isothermal and heated conditions (Sharma $\&$ Lele\cite{SharmLele}).
%
For example, Fig. 10 in Gryazev {\it et al}.\cite{Gryazev2019} shows the streamwise distribution of the LES extracted amplitudes of the temperature-associated components  of $R_{\lambda j \mu l}$ along the jet shear layer. 
They find that the ``{\it correlation amplitudes corresponding to the momentum/temperature terms $R_{4 j k l}$ are negligible in comparison with the temperature-temperature source terms, $R_{4 j 4 j}$}'' for a co-axial jet with subsonic core Mach number of $M_j = 0.877$, where $(j,k,l) =  (1,2,3)$ (see p.10 of their paper).
Gryazev {\it et al}.'s\cite{Gryazev2019} conclusions are more-or-less in line with the other available data sets. 
Fig. 16c in Sharma $\&$ Lele's\cite{SharmLele} LES study of a heated/isothermal mixing layer indicates that $R_{4242}$ will be almost one-half of $R_{4141}$ along the lip line of a splitter plate, when normalized by $R_{1111}$ (cf. Fig. 10b of Gryazev {\it et al}.) 
Moreover, $R_{4122}/R_{4111}$ is likely to be bounded by $R_{1122}/R_{1111}$ since `$41$' is at the location $\boldsymbol{y}$ in $R_{4122}$ $\&$ $R_{4111}$. Figs.10a $\&$ 10b in Karabasov {\it et al}.\cite{Karab2010} show that this ratio is $\approx 1/5$.
A similar conclusion can be found in Table $3$ of Sharma $\&$ Lele (2012) and Fig. 6.25 in Sharma\cite{Sharma2012}, 
all of which is basically consistent with Fig. 10 in Gryazev {\it et al}\cite{Gryazev2019}.
%

While Gryazev {\it et al}\cite{Gryazev2019} shows that all temperature-related correlations in the auto-covariance tensor $R_{\lambda j \mu l}$ are negligible for co-axial jets in their study, the PIV data shown in AGF (Figs. 1 $\&$ 2) from experiments at the NASA Glenn research center, indicates that $R_{4 111}$ remains important relative to $R_{4 141}$ for the single stream jet SP$49$ (where $Ma = 1.48$ and $TR=2.7$). That is, $R_{4 111}/c{}_\infty R_{1111} \sim R_{4 1 41}/c{}_\infty^2 R_{1111}$ at these jet speeds and temperatures. 
Hence we have neglected contribution of $ \Phi{}_{4212}^*$ (where $\mathcal{H}_{4221} ={H}_{4221}$),  $\Phi{}_{4122}^*$ and  $\Phi{}_{4242}^*$ in (\ref{I_low}) but retained $\Phi{}_{4111}^*$ and $\Phi{}_{4141}^*$ to allow (among other things) direct comparison between our results and AGF.
The asymptotic structure of the propagator and experimentally deduced scalings of the turbulence components in (\ref{I_low}) and Eqs.(26) $\&$ (27) of AGF are summarized in Table \ref{Table_asym}. 
Note that, $\mathcal{H}_{4122} =(2-\gamma)H_{4122}- (\gamma-1)H_{4111}/2 $ and $\mathcal{H}_{4111} = (3-\gamma)H_{4111}/2- (\gamma-1)H_{4122} \approx (3-\gamma)H_{4111}/2 $ after using (\ref{eq:HFT}) and the linear relations at end of $\S.$\ref{S:2}. 
There is some similarity in temporal de-correlation of $R_{4111}$ and $R_{4141}$ compared to $R_{1212}$ at fixed streamwise separation, $\eta_1$, in space-time structure shown in Fig. 19 in Sharma $\&$ Lele\cite{SharmLele} as well as in amplitude along lip line (Fig. 20 in Sharma $\&$ Lele) to give confidence in the possible universality of the normalized components of the generalized auto-covariance tensor. 
Such universality was argued by Semiletov $\&$ Karabasov\cite{Semil2016}. Their Figs. 2 $\&$ 3 illustrate a strong similarity in the normalized correlation curves for various components of $R_{ijkl}$. 
Hence, consistent with our model being a low-order representation of the acoustic spectrum, as a first approximation, we allow normalized components, $R_{4111} = n_2 R_{1212}$ and $R_{4141} =n_3 R_{1212}$ where $(n_2,n_3)$ are $O(1)$ constants. Therefore, we further approximate the acoustic spectrum formula in (\ref{I_low}) to the following formula:
\begin{widetext}
\begin{equation}
\label{I_low2}
I({\boldsymbol x}, {\boldsymbol y};\omega) 
\approx
\left(\frac{\epsilon}{2 c{}_\infty^2 |\boldsymbol{x}|}\right)^2
\left[
4|\bar{G}_{12}|^2
+ 
(3-\gamma)
n_2
Re\left\{ \bar{\Gamma}_{41} \bar{G}{}_{11}^* 
\right\}
+
n_3
|\bar{\Gamma}_{41}|^2 
\right]
 \Phi{}^*_{1212}.
\end{equation}
\end{widetext}

\begin{table*}
\caption{\label{Table_asym} Asymptotic structure of momentum/enthalpy flux coupling and enthalpy flux auto-covariance term in Eqs.(26) $\&$ (27) of AGF at $\Omega=O(1)$ frequencies.}
\begin{ruledtabular}
\begin{tabular}{ccc}
 Component  &  Propagator at lowest order& Scaling of $\mathcal{H}_{\nu j\mu l}$ \\ \hline
 $\mathcal{R}\{ \bar{\Gamma}_{41}(\bar{G}{}_{22}^* + \bar{G}{}_{33}^*) \Phi{}_{4122}^* \}$ 
 &
 $O(\epsilon^2)$  & $\mathcal{H}_{4122} \approx 0 $   \\
 $\mathcal{R}\{ \bar{\Gamma}_{41}\bar{G}{}_{11}^*\Phi{}_{4111}^* \}$ & 
 $O(\epsilon^2)$ 
 &
 $\mathcal{H}_{4111}  = O(1)$
 \\
 $\mathcal{R}\{ (\bar{\Gamma}_{42}\bar{G}{}_{12}^* + \bar{\Gamma}_{43}\bar{G}{}_{13}^*) \Phi{}_{4221}^* \}$ & $O(1)$ & $\mathcal{H}_{4221} \approx 0$ \\
 $ (|\bar{\Gamma}_{42}|^2 + |\bar{\Gamma}_{43}|^2) \Phi{}_{4242}^* $ &  $O(1)$ & $\mathcal{H}_{4242} \approx 0 $\\
 $ |\bar{\Gamma}_{41}|^2 \Phi{}_{4141}^* $  & $O(\epsilon^2)$  & $\mathcal{H}_{4141} = O(1)$  \\
\end{tabular}
\end{ruledtabular}
\end{table*}

\vspace{-0.5cm}

\section{Analysis of the acoustic spectrum of supersonic heated jets}
\label{S:5}
We investigate (\ref{I_low2}) numerically in this section using a formula for the component of the turbulence spectrum, $\Phi{}^*_{1212}$, derived in App.(\ref{App:B}) that is based on a relatively simple turbulence model for real-space function $R_{1212}$. 
The latter model is validated in App.(\ref{App:C}) against an LES database for similar jets (Bres {\it et al}.\cite{Bres17}).
%

\subsection{Fluent RANS simulations of mean flow for jet in Table (\ref{tab:table1}) }

\begin{figure}[h]
    \centering
    \begin{subfigure}[b]{0.48\textwidth}
        \centering
        \includegraphics[width=\textwidth]
{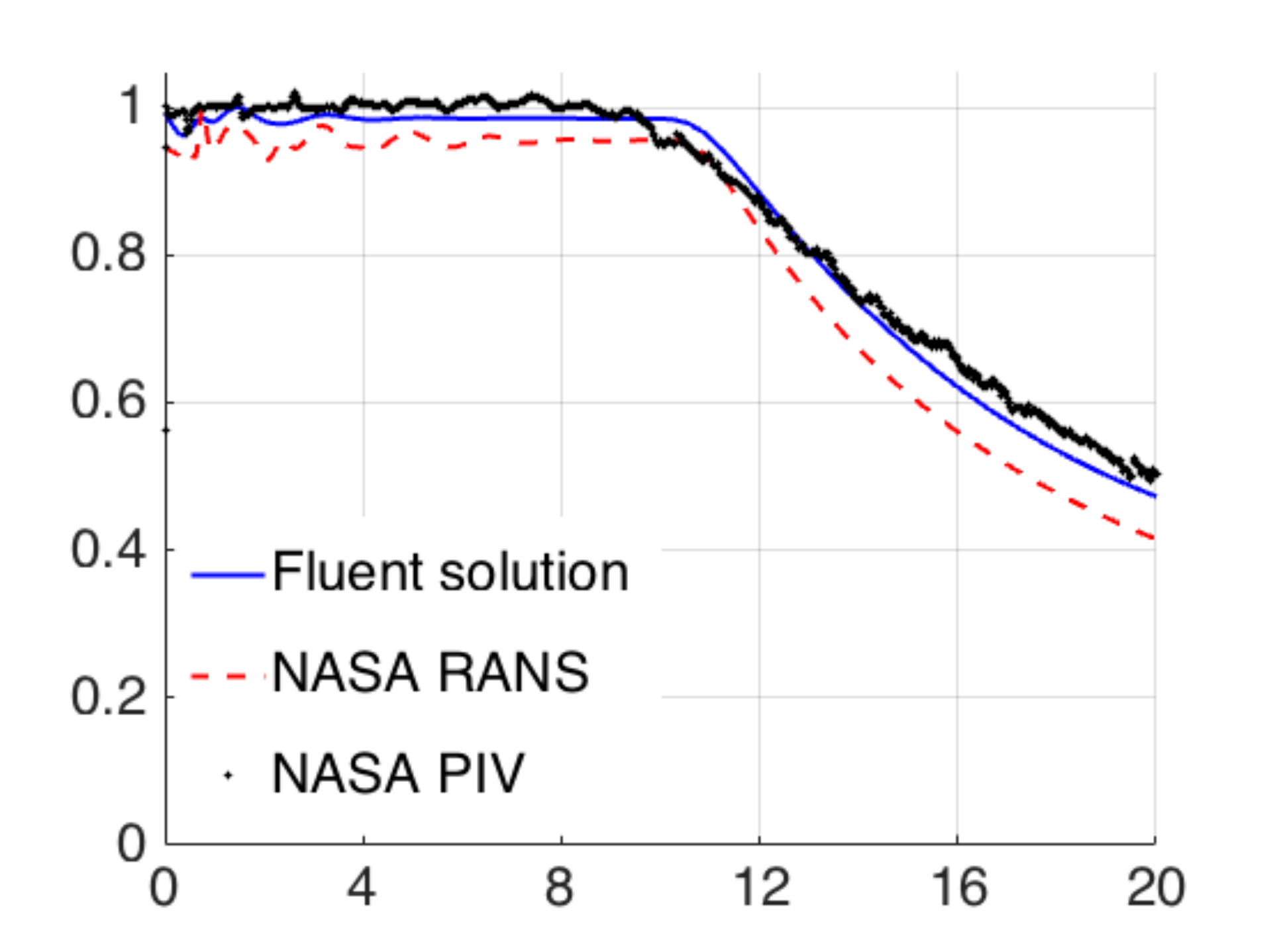}       
         \caption{}
        \label{fig5_1a}
    \end{subfigure}
    \begin{subfigure}[b]{0.45\textwidth}
        \centering
     \includegraphics[width=\textwidth]{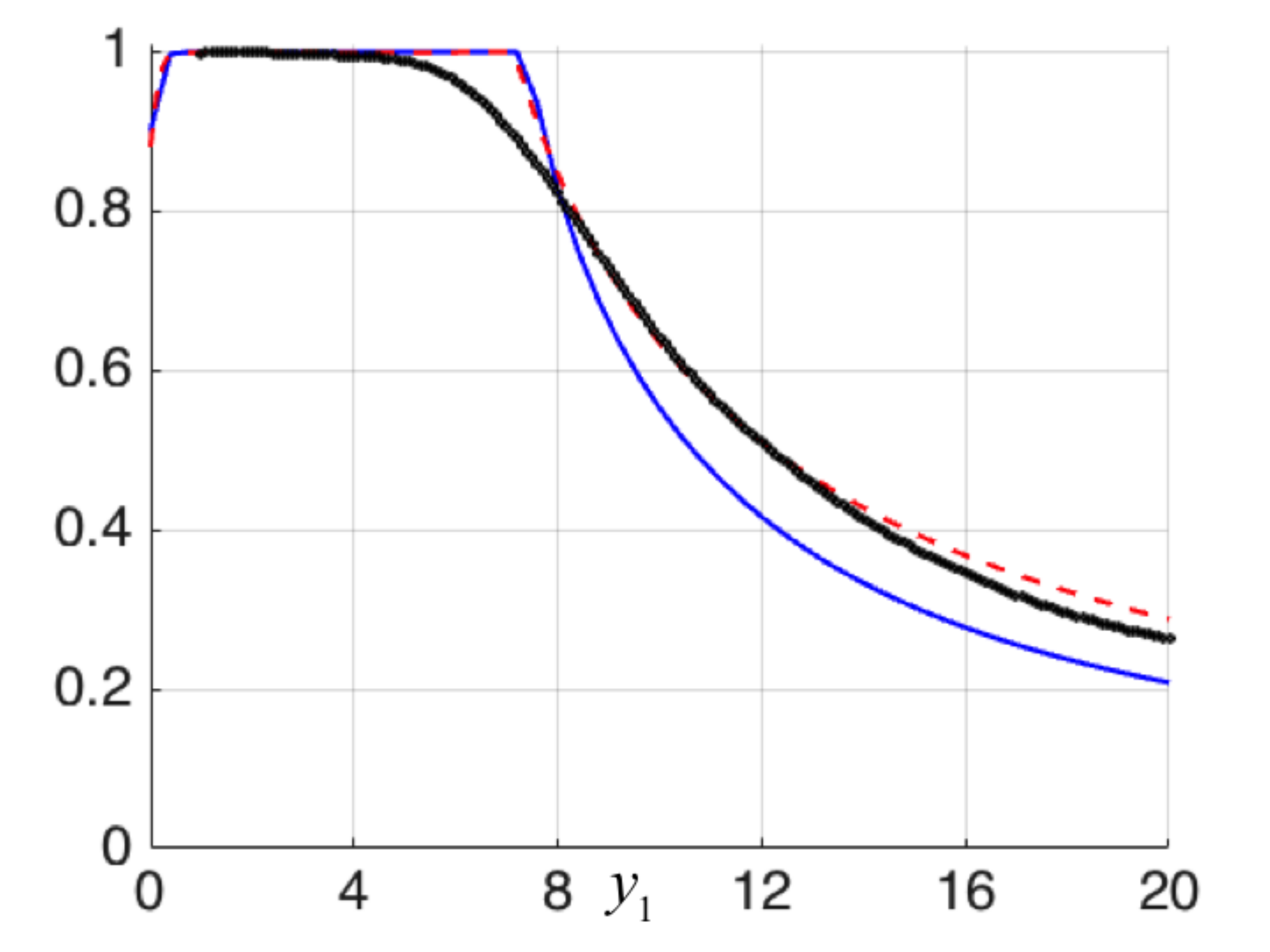}
        \caption{}
        \label{fig5_1b}
    \end{subfigure} \\
    \begin{subfigure}[b]{0.45\textwidth}
        \centering  
                \includegraphics[width=\textwidth]{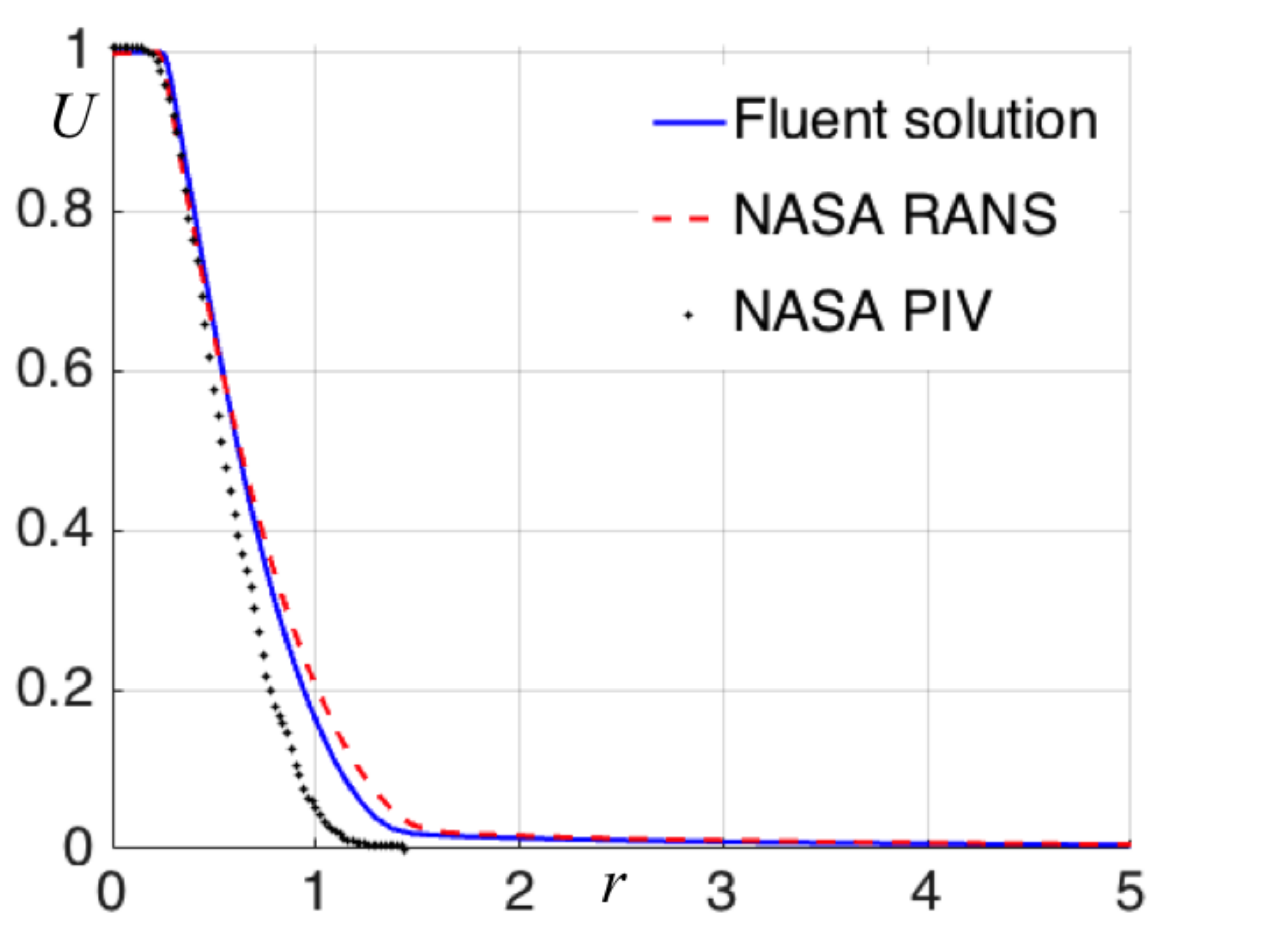}
        \caption{}
        %
        %
        \label{fig5_1d}
    \end{subfigure}
    %
    \begin{subfigure}[b]{0.45\textwidth}
        \centering
        \includegraphics[width=\textwidth]{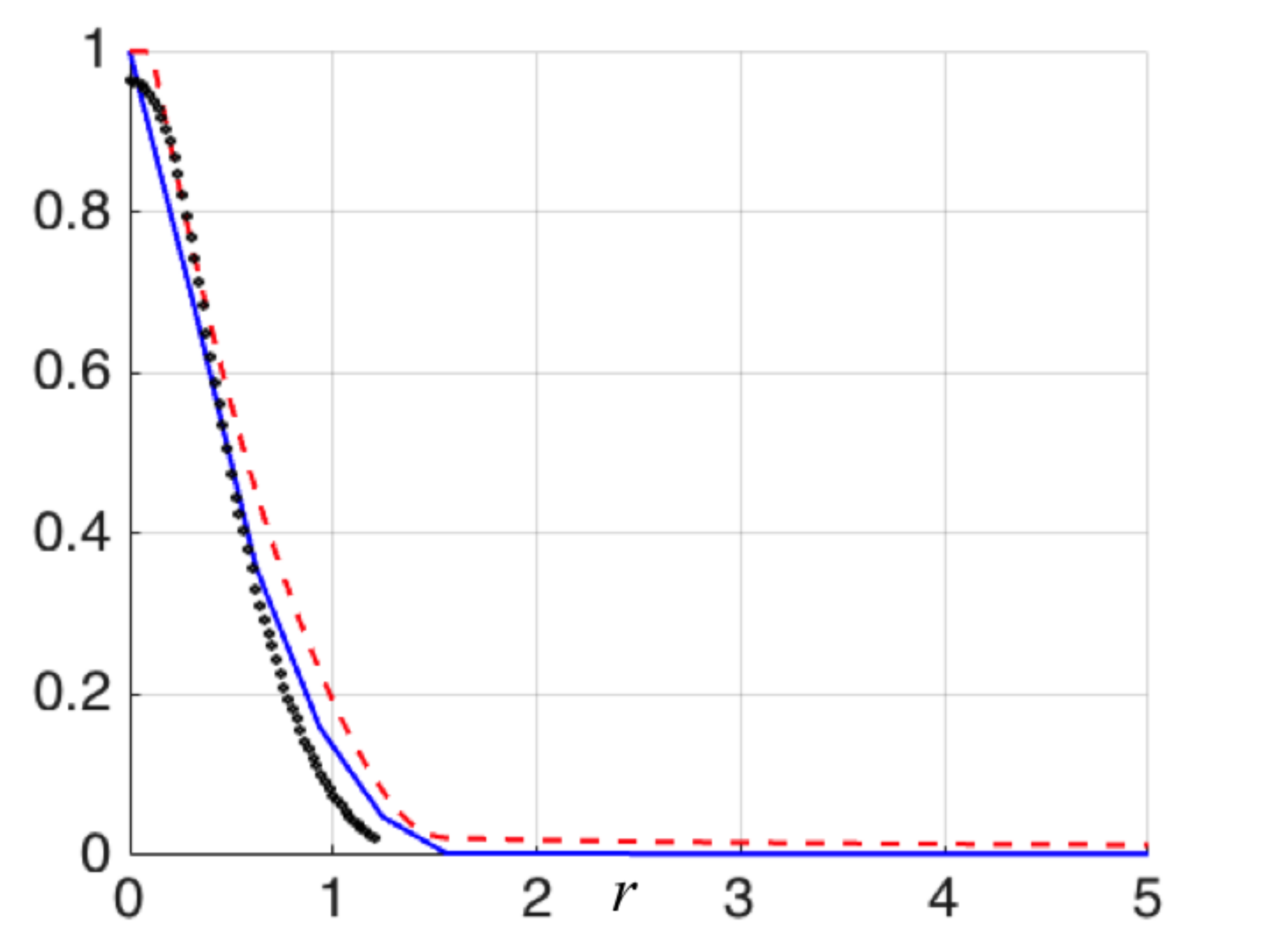}
        \caption{}
        \label{fig5_1e}
    \end{subfigure}
 \caption{Streamwise and radial variation of $U(y_1,r)$ at fixed $r$ and $y_1$ respectively for SP$90$ ($Ma=1.5$ $\&$ $TR=1$) and SP$49$ ($Ma=1.5$ $\&$ $TR=2.7$). (a). SP90: $r = 0$; (b). SP49: $r = 0$; (c). SP90: $y_1 = 6$; (d). SP49: $y_1 = 6$.}
    \label{fig5_1}
\end{figure}

The mean flow field for the Green's function calculation is found from a steady RANS calculation using {\sc Fluent}. In Fig. \ref{fig5_1} we compare the streamwise and radial profiles of the mean flow component, $U(y_1,r)$, 
against PIV data from the NASA Glenn Research Center together with RANS solutions obtained using the {\sc Wind} code (Nelson $\&$ Power\cite{Nelson2001}).
We use the same computational grids (based on a structured mesh with rectangular cells) for the {\sc Fluent} calculation that were used in the {\sc Wind} code solutions but with a more optimized turbulence model.
The SP$49$ domain consists of approximately $297,536$ cells (converging nozzle) while SP$90$ domain has $59,600$ cells (convergent-divergent nozzle). Lower mesh density was required for SP$90$ to achieve a converged solution. 
%
%
CFD calculations are then implemented using the usual pressure-based (ambient) far-field conditions on the left, top, bottom and right boundaries. Total pressure conditions are specified at the nozzle inlet to obtain the required $TR$ and $Ma$. No-slip boundary conditions are applied on the nozzle walls with symmetry boundary conditions at the jet axis (consistent with an axisymmetric mean flow field).
%
%
Both domains were simulated in {\sc Fluent} using the density-based, steady-state solver using Menter's Shear Stress Transport (SST) turbulence model. No appreciable differences found when using the $(k-\epsilon)$ model in {\sc Fluent}. 

The streamwise and radial variation of $U$ is shown in Fig. \ref{fig5_1}. 
We compare profiles of the normalized streamwise velocity obtained from RANS ({\sc Fluent} $\&$ {\sc Wind} code) with PIV data measured at the NASA Glenn Research Center (Bridges $\&$  Wernet 2017). 
There will always be some differences between RANS and PIV measurements of jet flow. 
%
Having said this, the centerline $r=0$, streamwise velocity (Figs. \ref{fig5_1a} $\&$ \ref{fig5_1b}) and radial distribution at $y_1 = 6$, (\ref{fig5_1d} $\&$ \ref{fig5_1e}) do compare favorably with NASA PIV data. 
In general, Fig. \ref{fig5_1} shows that both RANS solutions are basically the same, with {\sc Fluent} showing slightly closer agreement to PIV. 

The centerline profile of $U$ (Fig. 5.1a) shows a small degree of oscillations at $0\leq y_1\leq 4$ due to slightly imperfectly expanded conditions (i.e. consecutive expansions and compressions that were not present in the SP49 case). {\sc Fluent} gives a much smoother solution for SP$90$ with amplitude of the oscillations being approximately  $0.5\%$ compared to PIV data (the WIND solution is at less than $2\%$) in region, $y_1 < 2$. 

\begin{figure}
    \centering
    \begin{subfigure}[b]{0.42\textwidth}
        \centering
        \includegraphics[width=\textwidth]{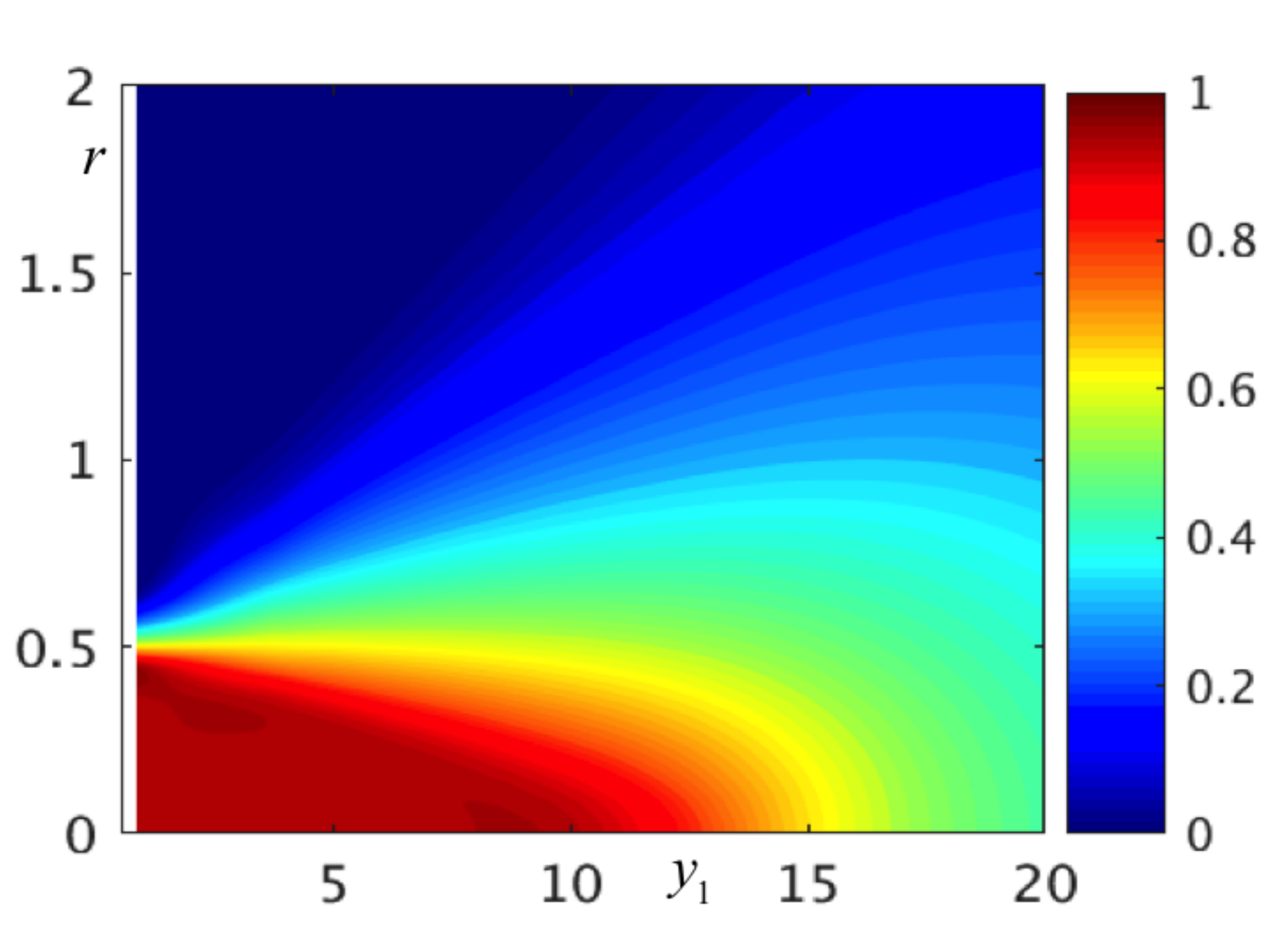}
        \caption{}
        \label{fig5_3a}
    \end{subfigure}
    \begin{subfigure}[b]{0.42\textwidth}
        \centering
        \includegraphics[width=\textwidth]{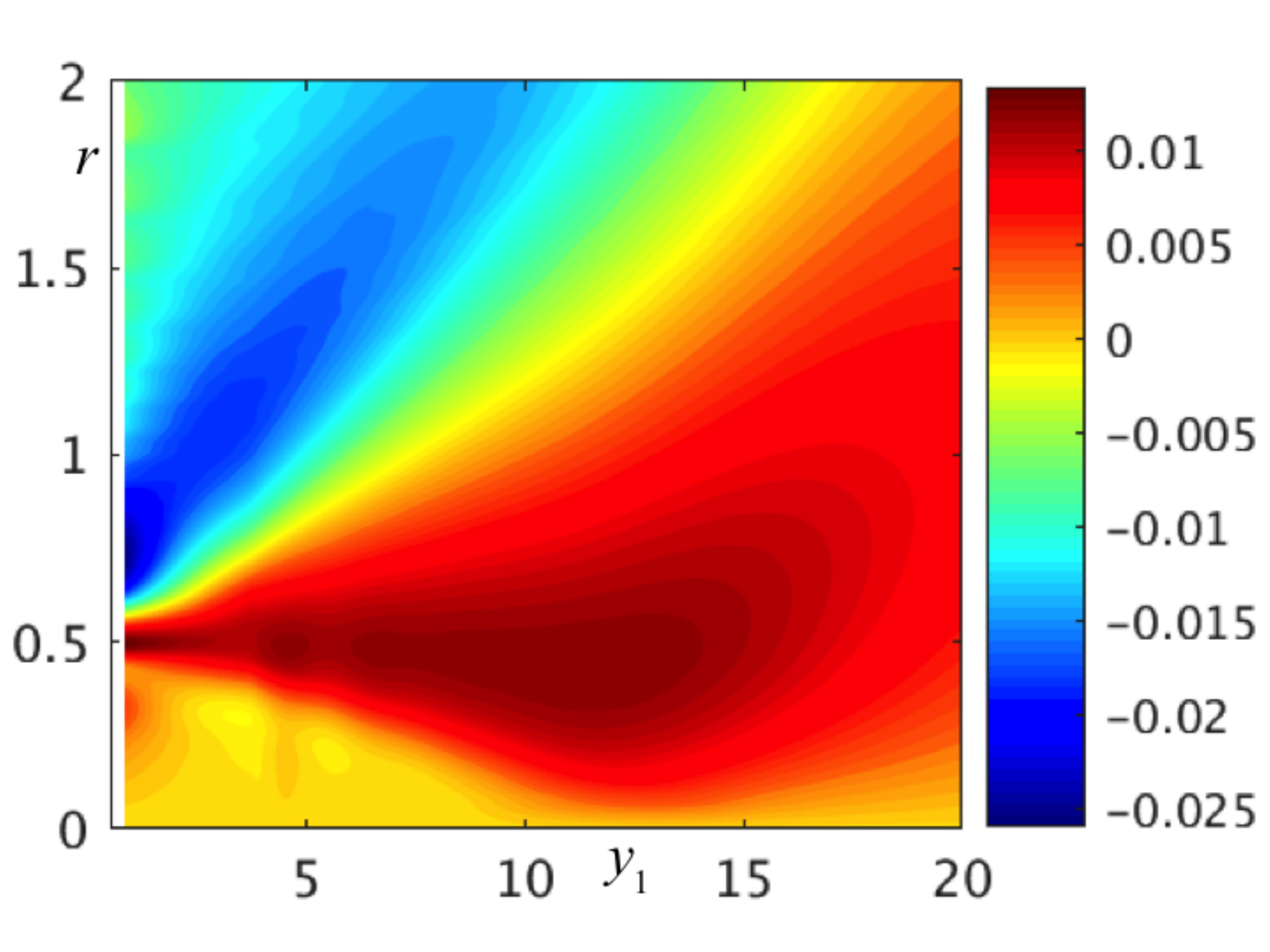}
        \caption{}
        \label{fig5_3b}
    \end{subfigure} \\
    \begin{subfigure}[b]{0.42\textwidth}
        \centering
        \includegraphics[width=\textwidth]{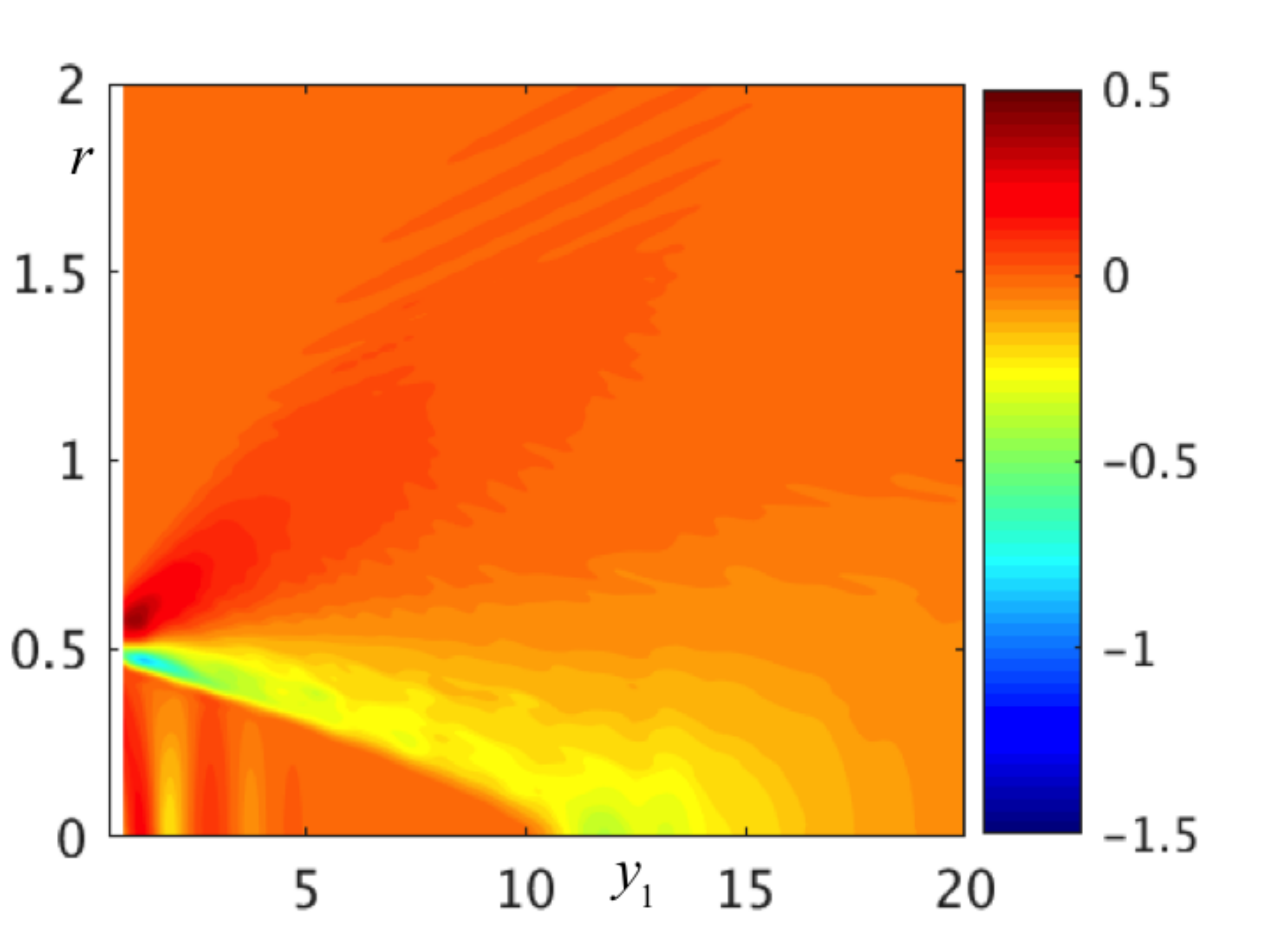}
        \caption{}
        \label{fig5_3c}
    \end{subfigure}
        \centering
    \begin{subfigure}[b]{0.42\textwidth}
        \centering
        \includegraphics[width=\textwidth]{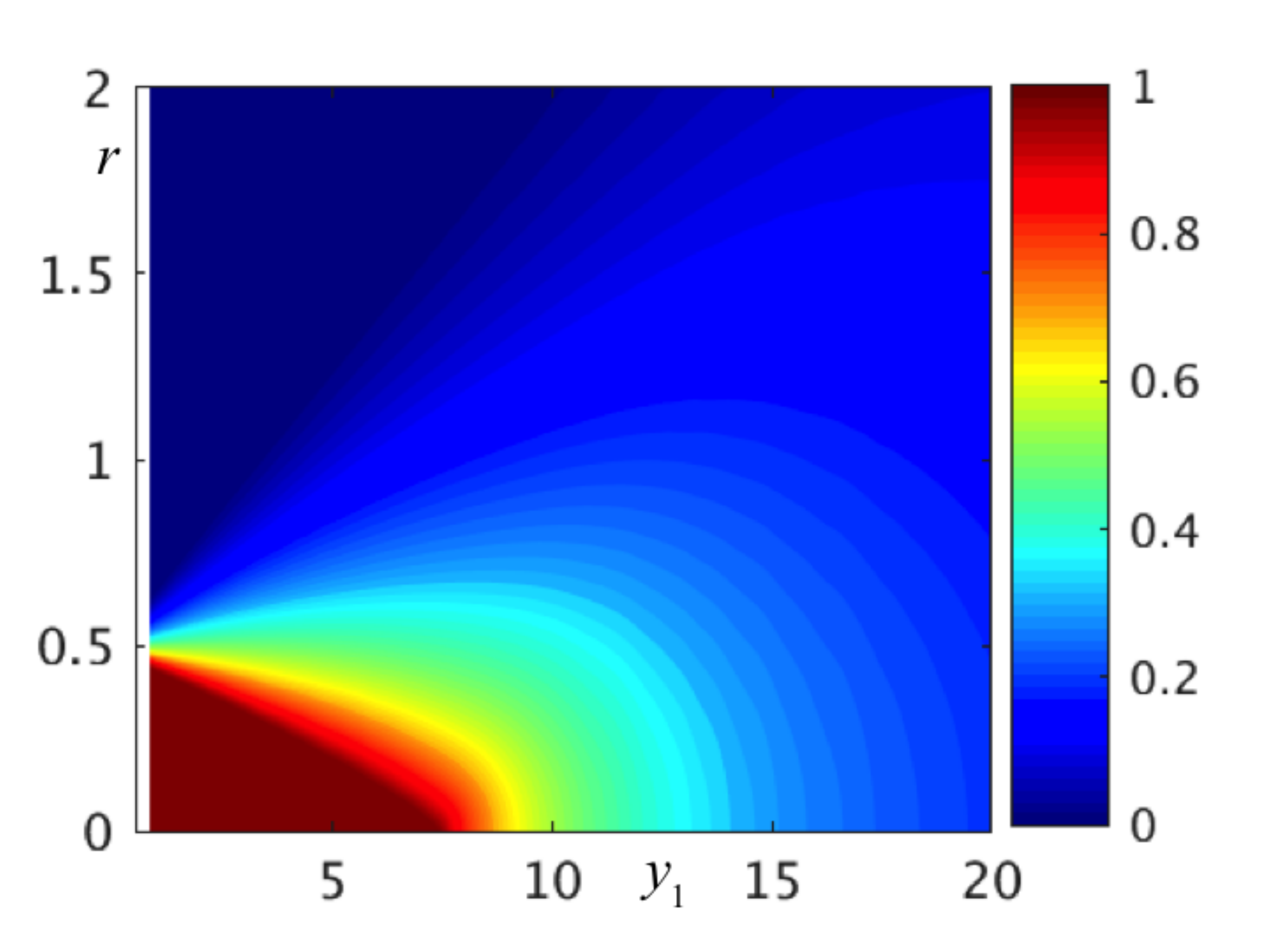}
        \caption{}
        \label{fig5_3d}
    \end{subfigure} \\
    \begin{subfigure}[b]{0.42\textwidth}
        \centering
        \includegraphics[width=\textwidth]{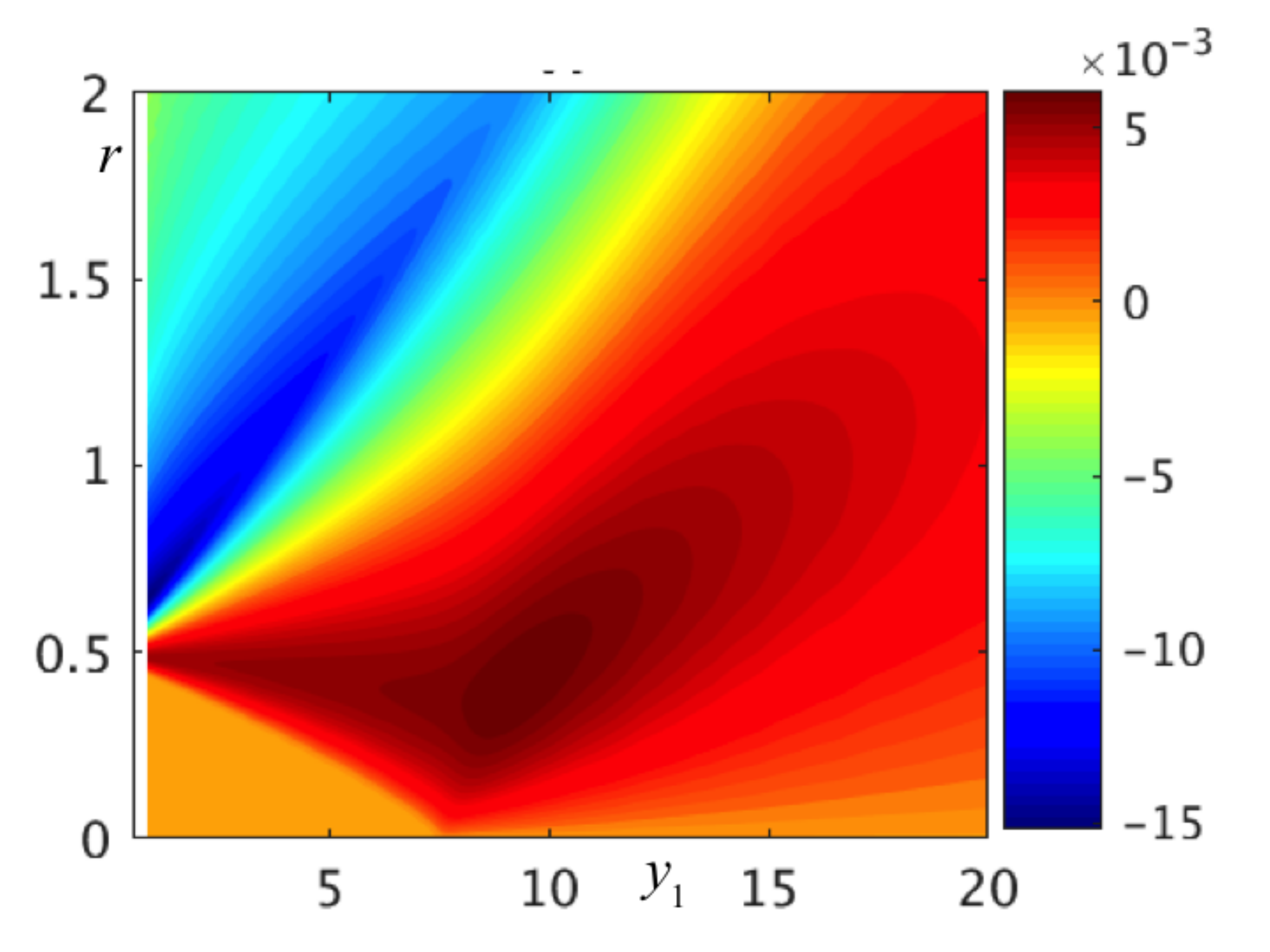}
        \caption{}
        \label{fig5_3e}
    \end{subfigure}
    \begin{subfigure}[b]{0.42\textwidth}
        \centering
        \includegraphics[width=\textwidth]{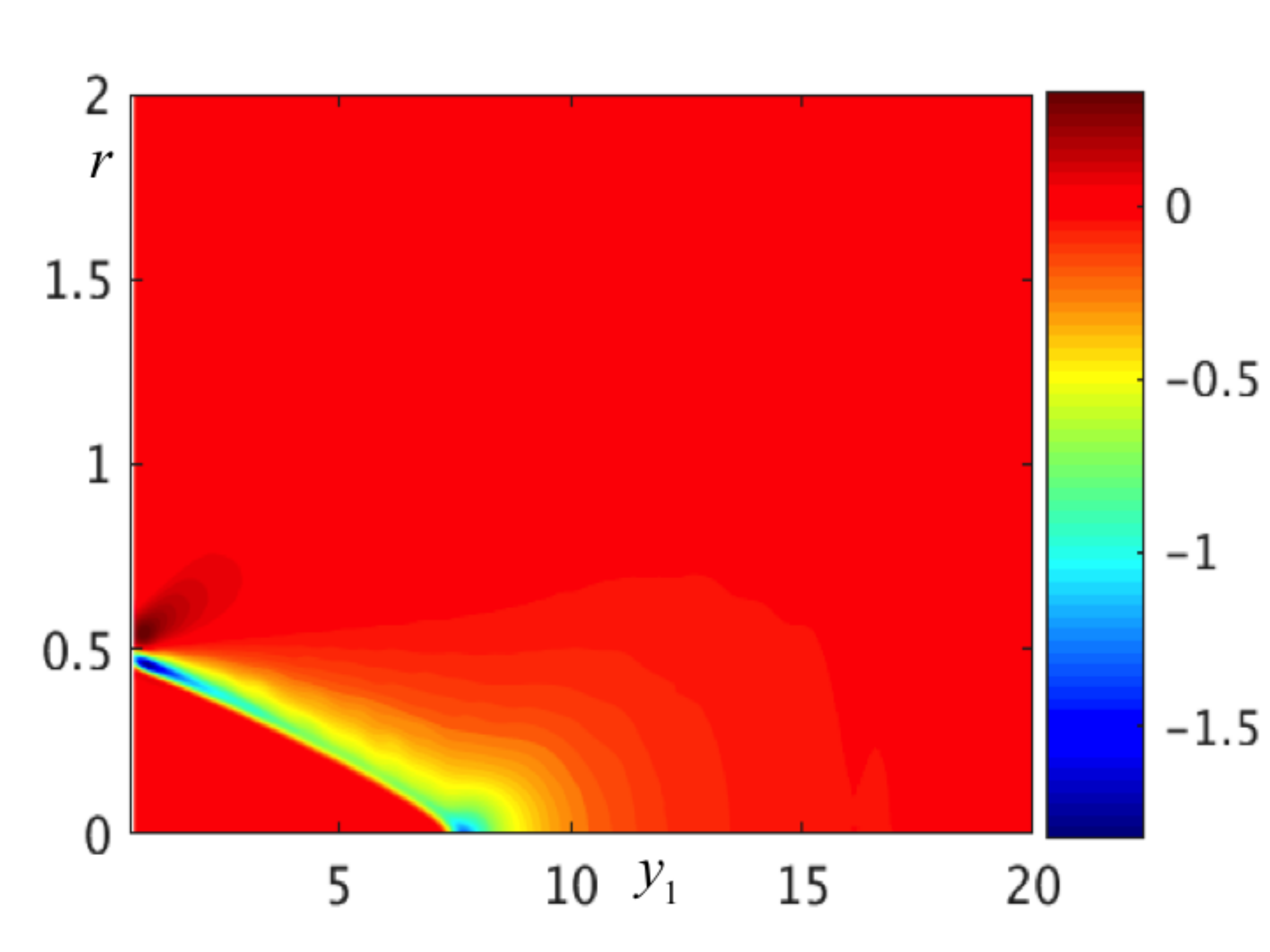}
        \caption{}
        \label{fig5_3f}
    \end{subfigure}
    \caption{Spatial distribution of mean flow components: $\tilde{v}_i=(U, V_r)$ and streamwise mean flow advection, $\bar{X}_1$ for SP$90$ ($Ma=1.5$ $\&$ $TR=1$) and SP$49$ ($Ma=1.5$ $\&$ $TR=2.7$). (a). SP90: $U(y_1,r)$; (b). SP90: $V_r(y_1,r)$; (c). SP90: $X_1(y_1,r)$; (d). SP49: $U(y_1,r)$; (e). SP49: $V_r(y_1,r)$; (f). SP49: $X_1(y_1,r)$. }
    \label{fig5_3}
\end{figure}

In Fig. \ref{fig5_3} we show the $(y_1,r)$ spatial distribution of the mean flow $\tilde{v}_i = \{U, V_r\}$ obtained directly by {\sc Fluent} simulations. $\bar{X}_1$ in Figs. \ref{fig5_3c} $\&$ \ref{fig5_3f} is determined by (\ref{subX1}) using the RANS mean flow and central differencing in $r$ to determine mean flow gradient, $\partial U/\partial r$.  
The contours of SP$90$ in Fig. \ref{fig5_3} show slight oscillations because $M_J>1$ for this jet. 
Note that Figs. \ref{fig5_1a} $\&$ \ref{fig5_1b} shows that the oscillations are present in the PIV data as well. But this does not introduce a significant impact to the subsequent acoustic predictions. We investigated this by filtering out the small oscillations (using the scheme of Vasilyev {\it et al.}\cite{Vasilyev}) in the same manner as Ref. \onlinecite{Afs2009}.
The effect on the predictions in peak noise direction of $\theta = 30^\circ$ was found to be less than $0.5$dB when the filtered mean flow was used for the propagator calculations in the acoustic spectrum formula (\ref{I_low2}) instead of the non-filtered flow (used in Figs. \ref{fig5_3a}  $\&$ \ref{fig5_3c}) whilst keeping the turbulence model (\ref{eq:SpecPhi1212_A4}) fixed. 
We treated this effect as negligible since the acoustic data itself has an error of the order of about $1$ dB\cite{Bridges06,BridgesWern}.

In general, the RANS simulations recover the general features of heated turbulent jets.
That is, the length of the potential core in Figs. \ref{fig5_3a} and \ref{fig5_3d} for SP$90$ relative to SP$49$ is reduced by approximately $30\%$ at the most intense region (which lies at $y_1\sim 7$ for SP$49$ compared to $y_1 \sim 10$ for SP$90$). Similarly, at its maximum, the potential core spreads out faster for SP$49$ to $y_1\approx 9$ compared with $y_1\sim 14$ for SP$90$.
While the radial mean velocity component, $V_r$, is significantly smaller than the streamwise component, $U$, the former does affect the magnitude and structure of coefficient, $\bar{X}_1$, defined below (\ref{eq:X_exp}). 
The peak value of $\bar{X}_1$ remains focused at the nozzle lip line with negative values along the interface between the potential core and the mixing region above the shear layer. For SP$49$, on the other hand, $\bar{X}_1$ is more localized and negative compared to SP$90$ (cf. Figs.\ref{fig5_3c} and \ref{fig5_3f} respectively).

\begin{figure}[h]
        \begin{subfigure}[b]{0.42\textwidth}
                \includegraphics[width=1.2\linewidth]{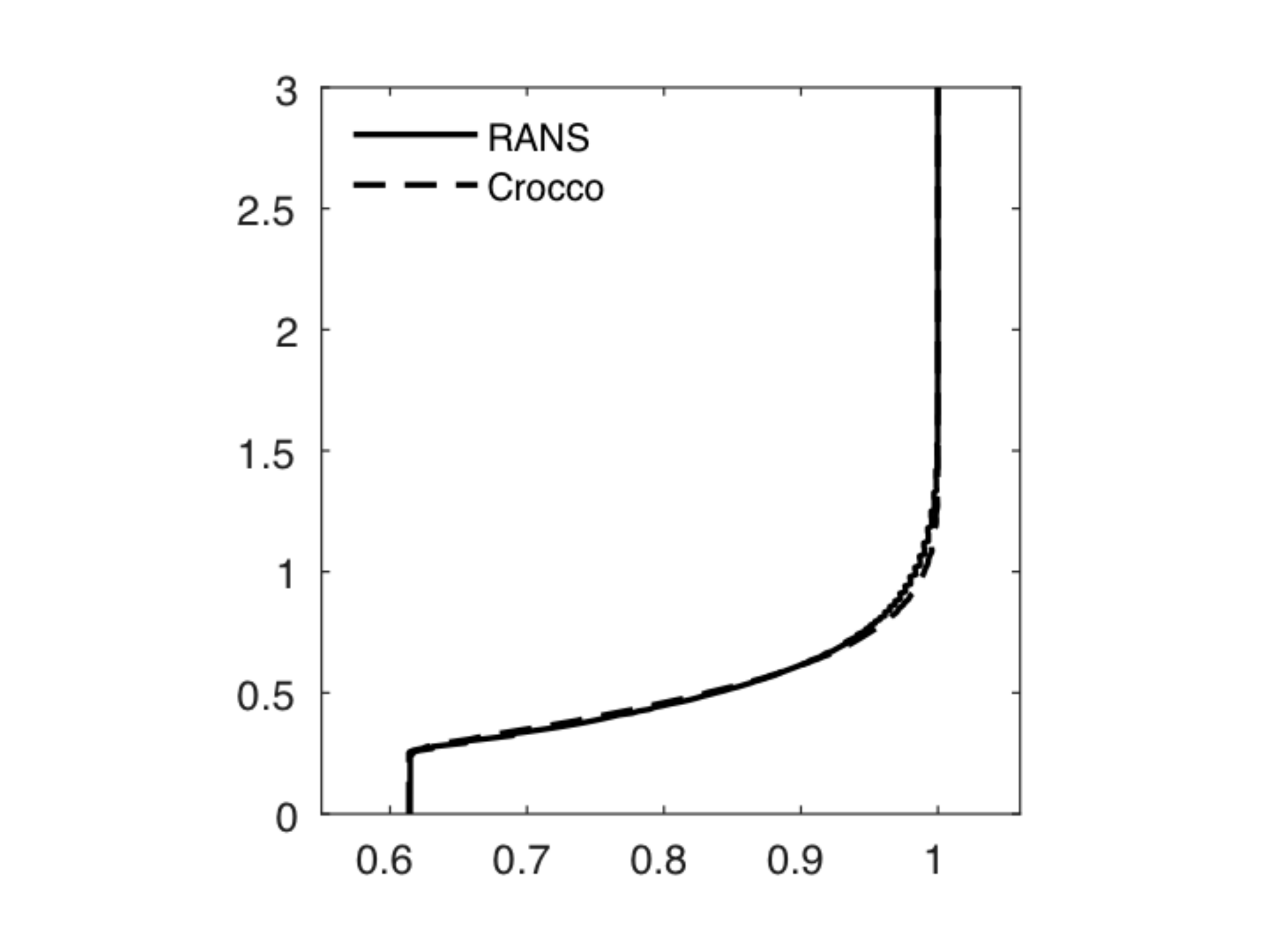}
        \caption{}
                \label{fig:5.5b}
        \end{subfigure}%
        \begin{subfigure}[b]{0.42\textwidth}
                \includegraphics[width=1.2\linewidth]{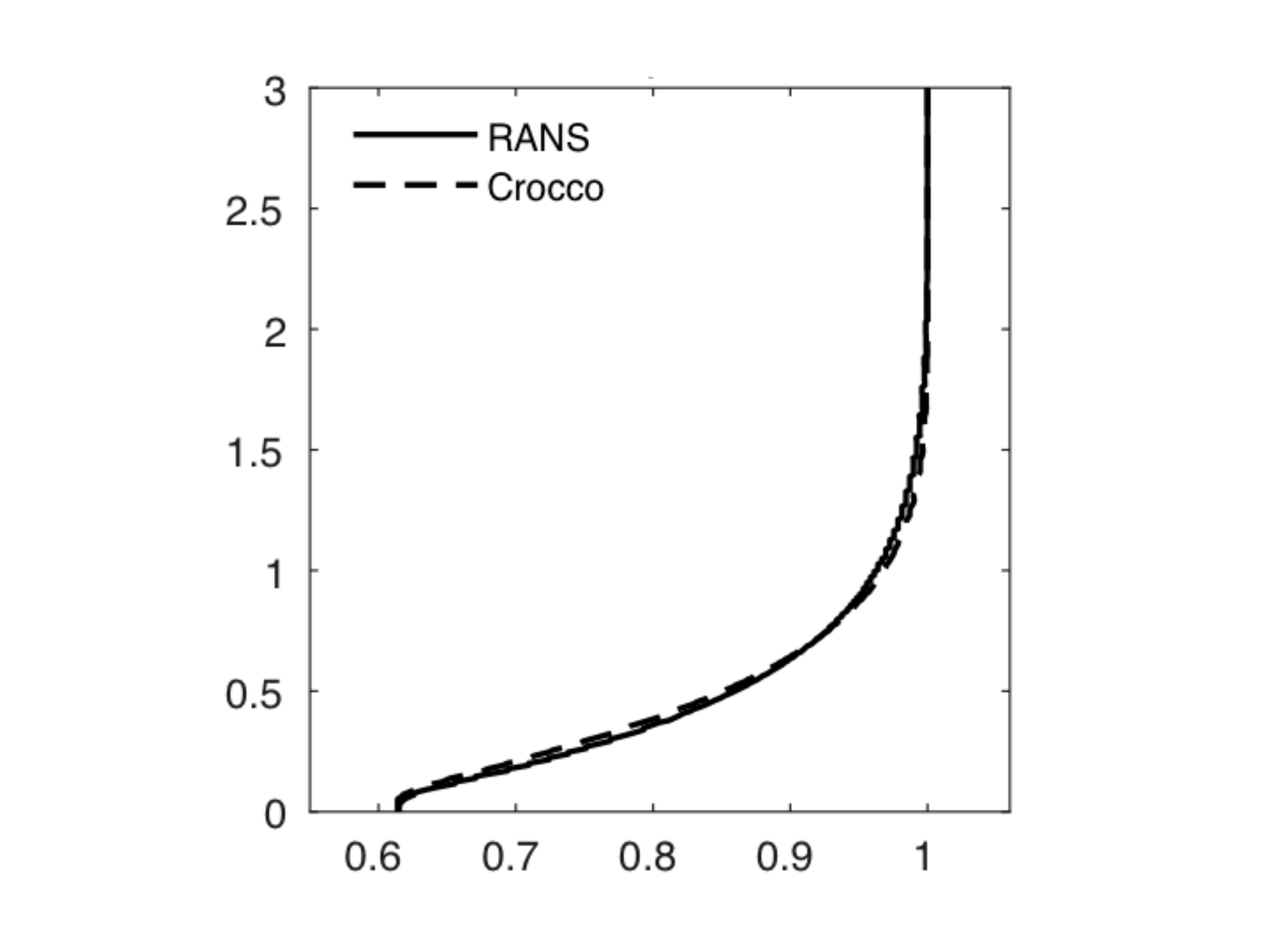}
        \caption{}
                \label{fig:5.5c}
        \end{subfigure} \\
        \begin{subfigure}[b]{0.42\textwidth}
                \includegraphics[width=1.2\linewidth]{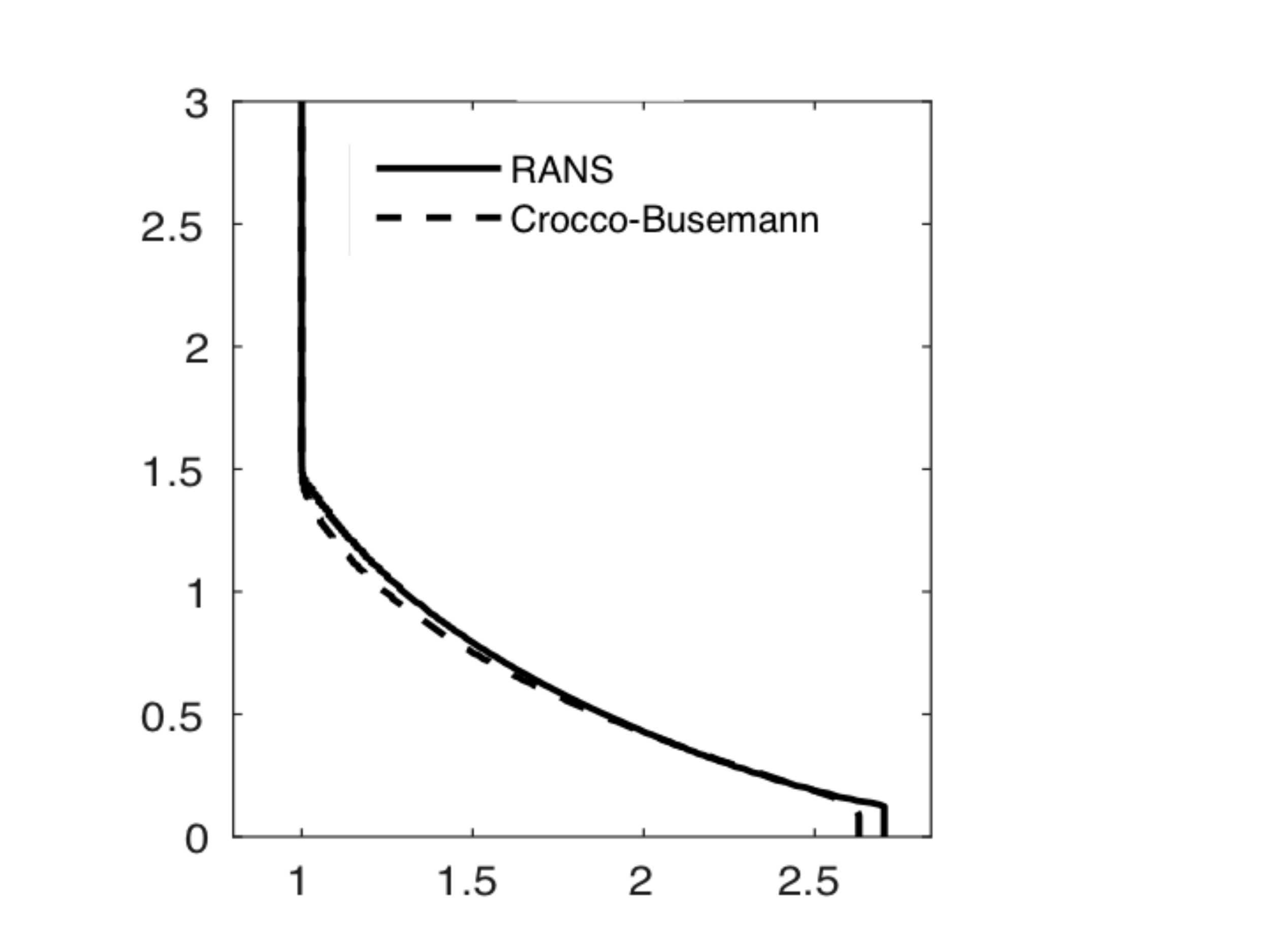}
        \caption{}
                \label{fig:5.5f}
        \end{subfigure}%
        \begin{subfigure}[b]{0.42\textwidth}
                \includegraphics[width=1.2\linewidth]{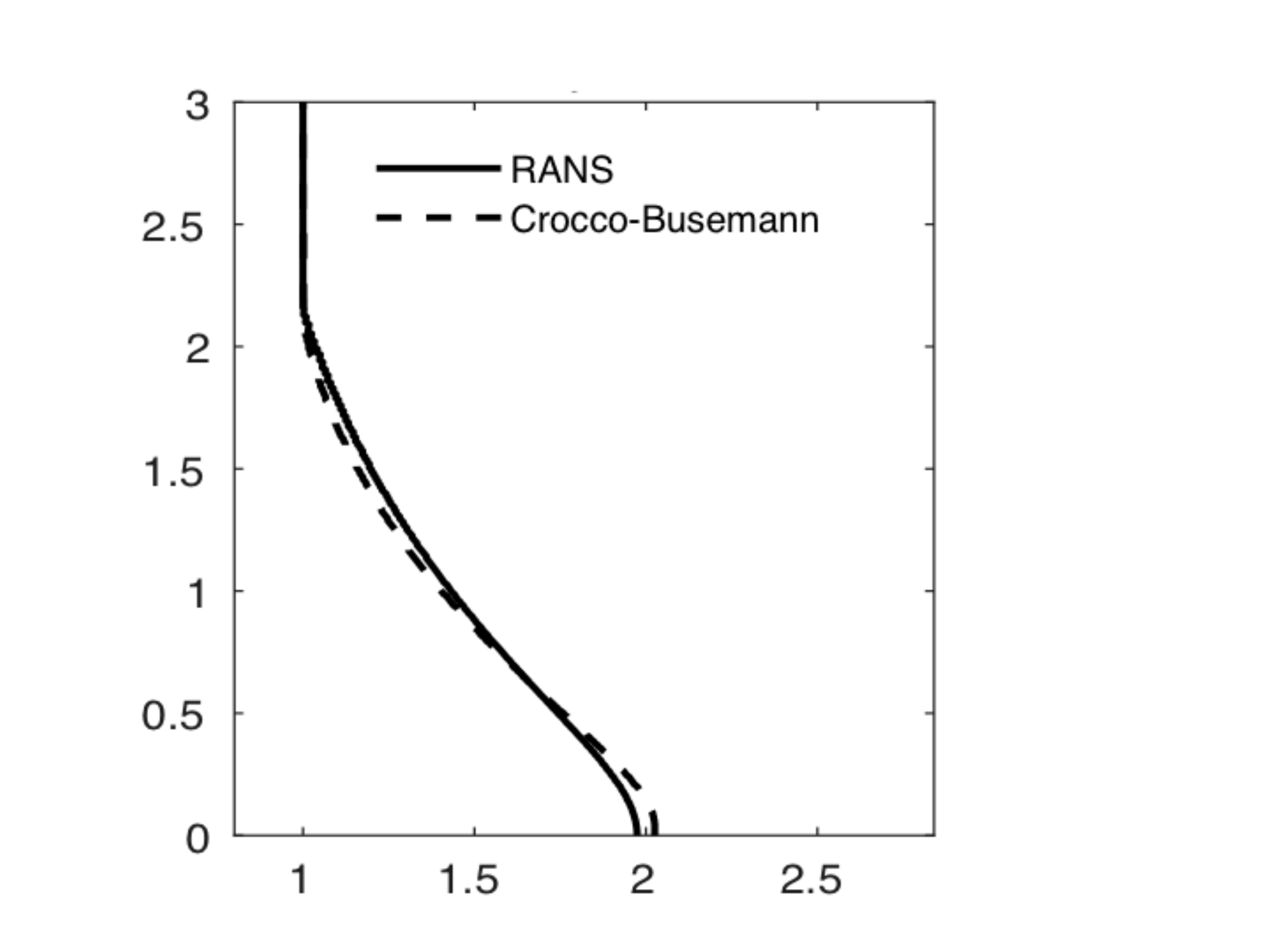}
        \caption{}
                \label{fig:5.5g}
        \end{subfigure}%
        \caption{Verification of the Crocco relation (Eq. 5.33 in GSA) and Crocco-Busemann  (\ref{CB_reln}) and against RANS mean flow for SP$90$ ($Ma=1.5$ $\&$ $TR=1$) and SP$49$ ($Ma=1.5$ $\&$ $TR=2.7$) respectively at various points in the jet. (a).SP90: $y_1=6$; (b). SP90: $y_1=10$; (c). SP49: $y_1=6$; (d). SP49: $y_1=10$.}
        \label{fig:Crocco+CB}
\end{figure}

In Fig. \ref{fig:Crocco+CB} we compare the Crocco (Eq. $5.33$ in GSA) and Crocco-Busemann relation (\ref{CB_reln}) to the RANS-based $\widetilde{c^2}$ for SP$90$ and SP$49$ respectively. It is clear that both of these approximations are accurate enough for aero-acoustic calculations with the Crocco relation (for SP$90$) having a maximum error of 2$\%$ (consistent with Dahl's\cite{Dahl} results). But this is only at large streamwise distance at $y_1 = 14$ from the nozzle exit, which is far downstream from the region of maximum turbulence. 
The Crocco-Busemann relation is equally accurate at most locations within the jet and there is only a small discrepancy near the core region $r<0.1$ at $y_1<6$ (possibly due to non-unity Prandtl number in the RANS calculations) giving an error of about 4$\%$ localized here.

\begin{figure}
    \centering
    \begin{subfigure}[b]{0.42\textwidth}
        \centering
        \includegraphics[width=\textwidth]{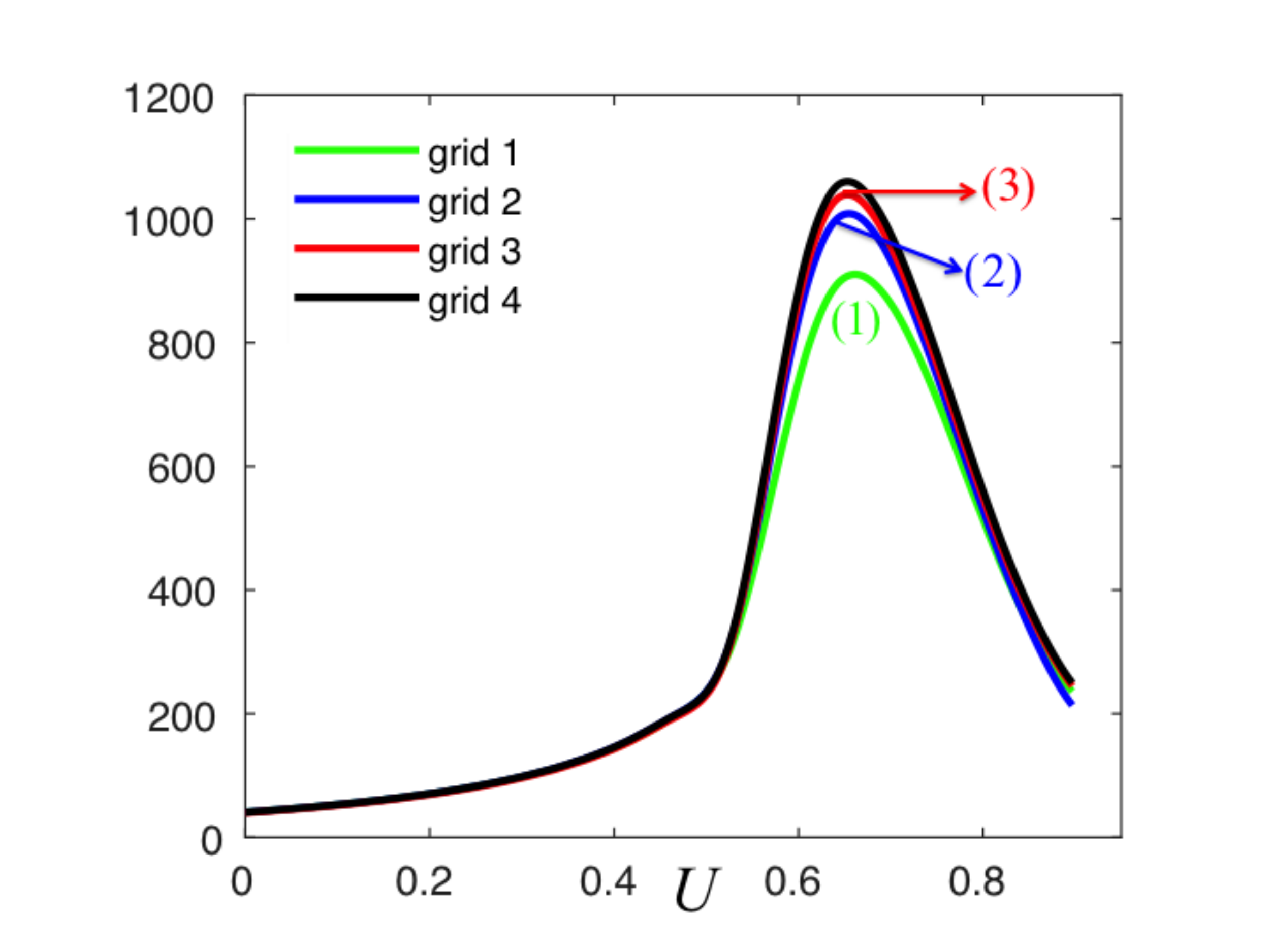}
        \caption{  }
        \label{fig:nu_sp90}
    \end{subfigure}
    \begin{subfigure}[b]{0.42\textwidth}
        \centering
        \includegraphics[width=\textwidth]{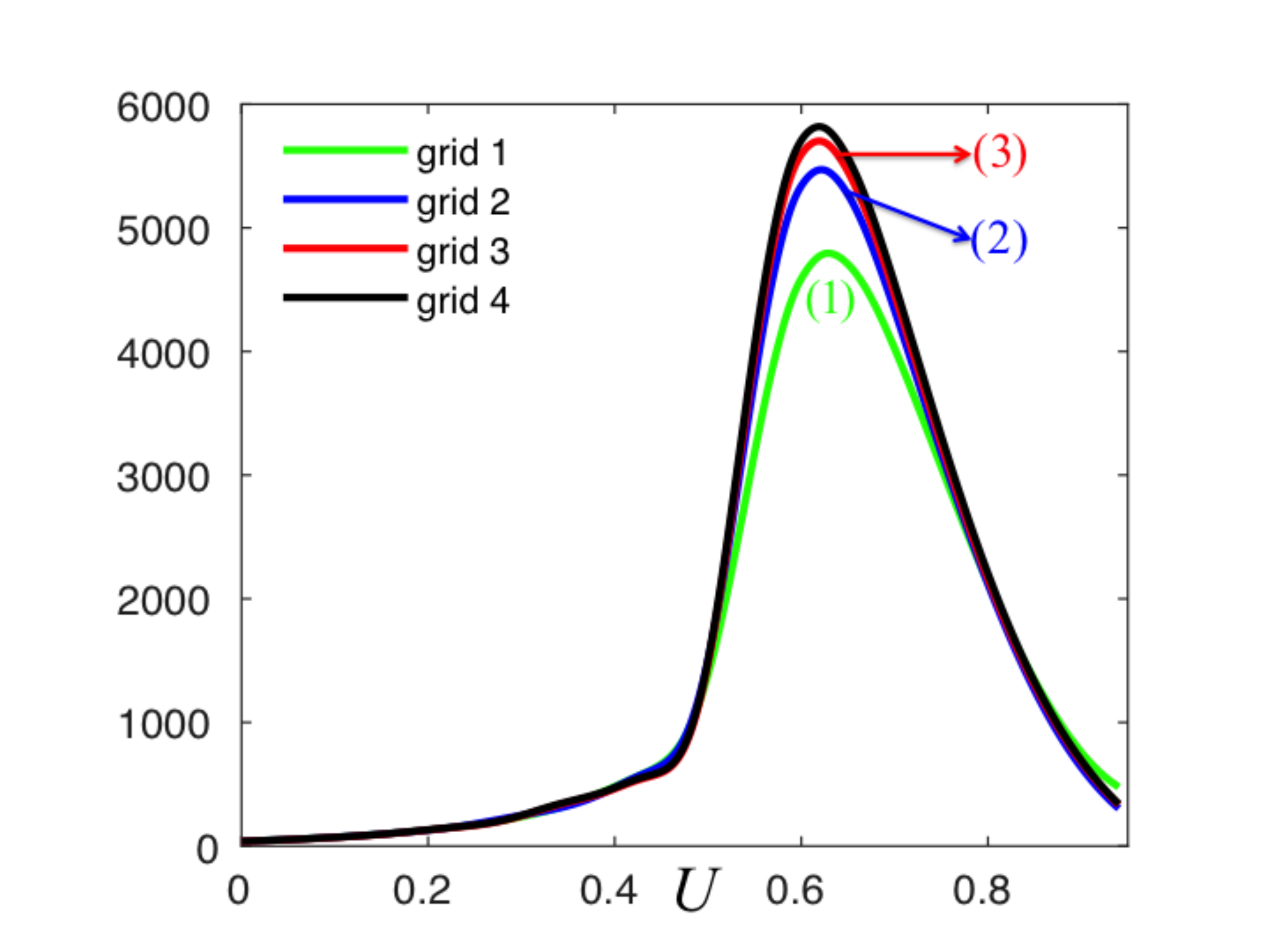}
        \caption{ }
        \label{fig:nu_sp49}
    \end{subfigure} \\
            \centering
    \begin{subfigure}[b]{0.42\textwidth}
        \centering
        \includegraphics[width=\textwidth]{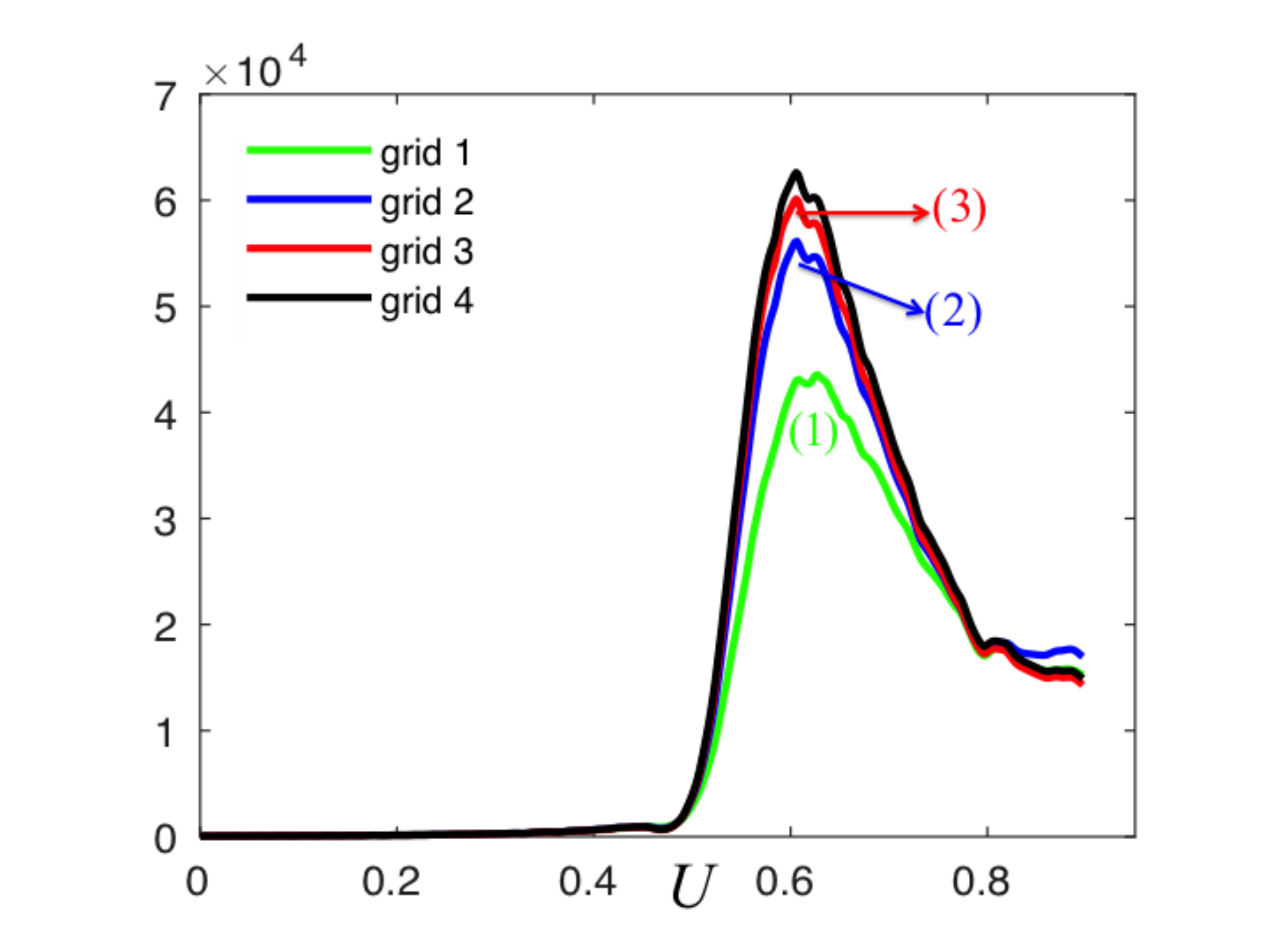}
        \caption{}
        \label{fig:dnu_sp90}
    \end{subfigure}
    \centering
    \begin{subfigure}[b]{0.42\textwidth}
        \centering
        \includegraphics[width=\textwidth]{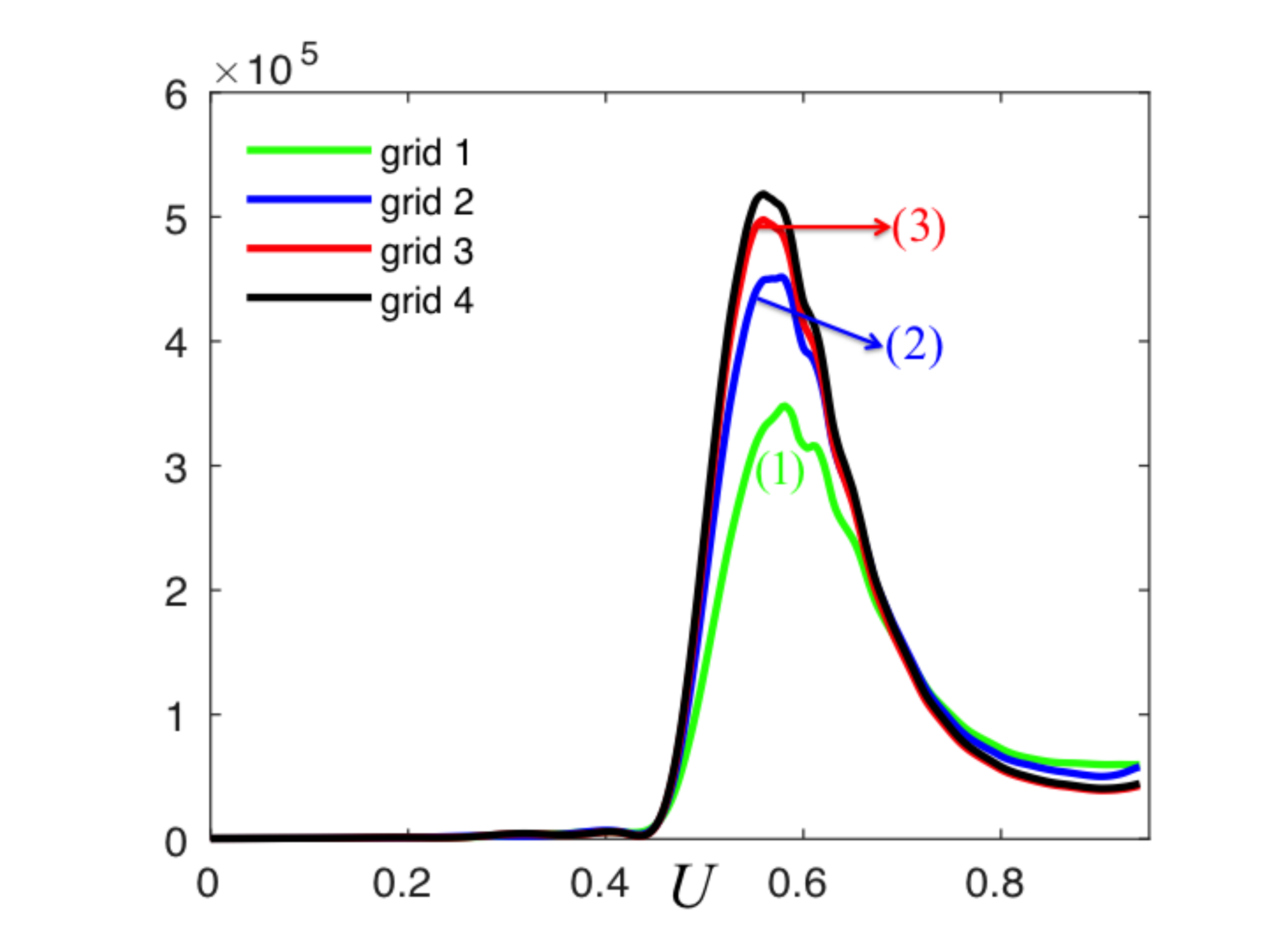}
        \caption{}
        \label{fig:dnu_sp49}
    \end{subfigure}
    \hfill
           \caption{Convergence of $\bar{\nu}(Y,U)$ and $\partial\bar{\nu} (Y,U)/\partial U$ at Y = 1.0 and $(\Omega,\theta)=(0.2, 30^\circ)$ for SP$90$ and SP$49$ (table \ref{tab:table1}). Grid dimensions are: grid 1 - $400\times250$ $(100000)$; grid 2 - $450\times300$ $(135000)$; grid 3 - $500\times350$ $(175000)$ and grid 4 - $550\times400$ $(220000)$. (a). SP90: $\bar{\nu}$; (b). SP49: $\bar{\nu}$; (c). SP90: ${\partial \bar{\nu} }/{\partial U}$; (d). SP49: ${\partial \bar{\nu} }/{\partial U}$. }
    \label{fig:converge}
\end{figure}

In Fig. \ref{fig:converge} we investigate the grid independence of the solution $\bar{\nu}(Y,U)$ determined by (\ref{Hyp3})--(\ref{BC2}).
The $\bar{\nu}$ solution and its derivative, $\partial\bar{\nu}/\partial U$, are quite well converged using grid 2 ($450\times300$: $144,000$ points, see Fig.\ref{fig:converge} caption) with only a very slight deviation near the inner boundary, $U\rightarrow 1$ of less than 2$\%$. 
An increase in grid points does remedy the difference (cf. grid $3$ and $4$ in Fig. \ref{fig:converge}) in this region of the jet however, given that our experiments reveal a commensurate rise in computation time of $(1-2)$ hours when increasing the grid resolution up to $(550\times400)$ (i.e., grid $4$: $220,000$ points) performed on an AMD Opteron $6274$ processor, the remaining calculations in the paper are therefore based on the $(500\times350)$ grid $3$, which is still virtually identical to the numerical solution to (\ref{Hyp3})--(\ref{BC2}) obtained using highest dimension grid $4$.
The dimensions of the grids used in the refinement study are given in the caption of Fig. \ref{fig:converge}.

\subsection{Spatial structure of propagator terms, (\ref{G_12}), (\ref{G11_41}) $\&$ (\ref{G11_41b})}

\begin{figure}[h]
    \centering
    \begin{subfigure}[t]{0.42\textwidth}
        \centering
        \includegraphics[width=\textwidth]
        {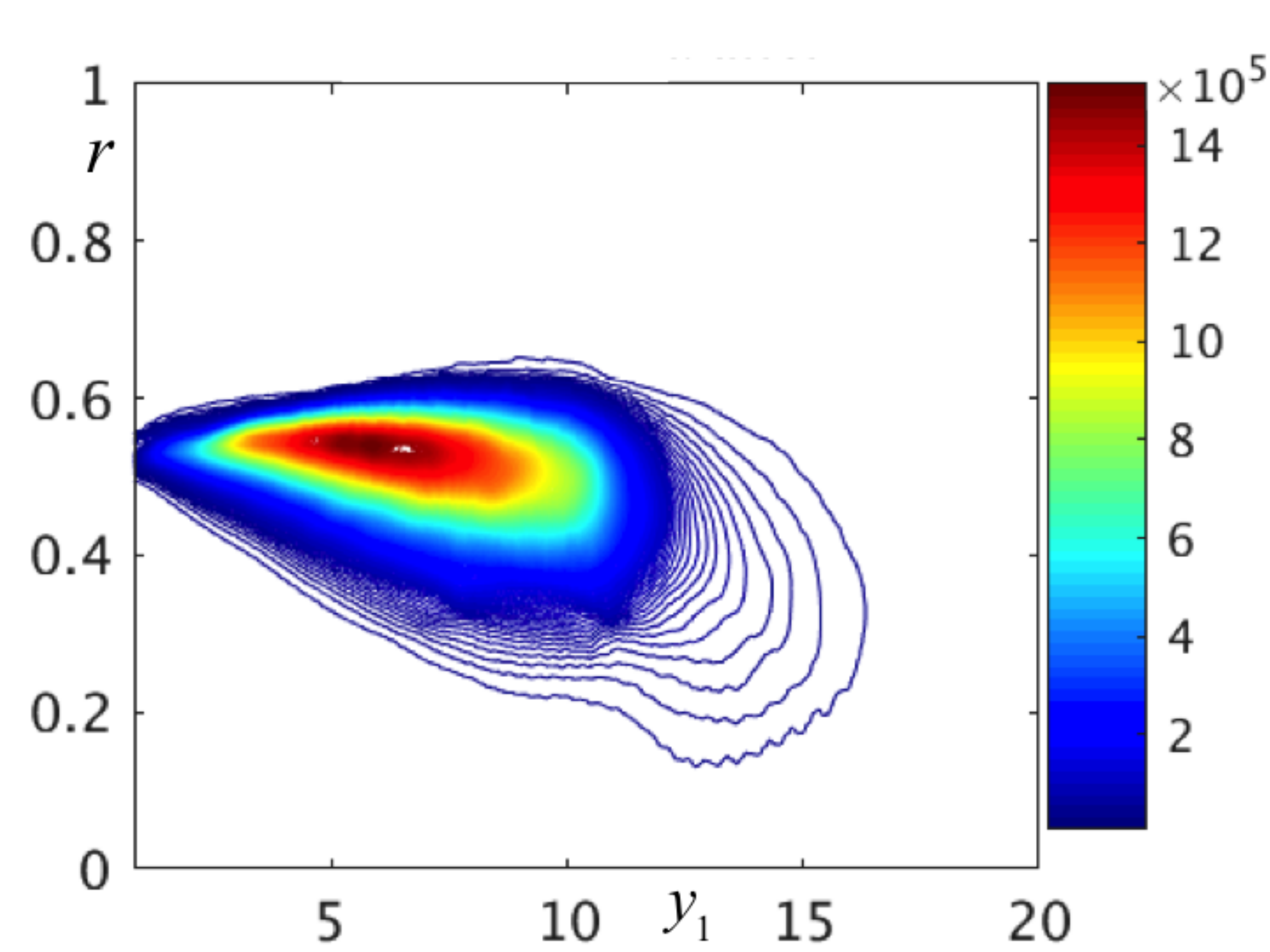}
        \caption{}
        \label{fig:G12_SP90_NP}
    \end{subfigure}
    \begin{subfigure}[t]{0.42\textwidth}
        \centering
        \includegraphics[width=\textwidth]{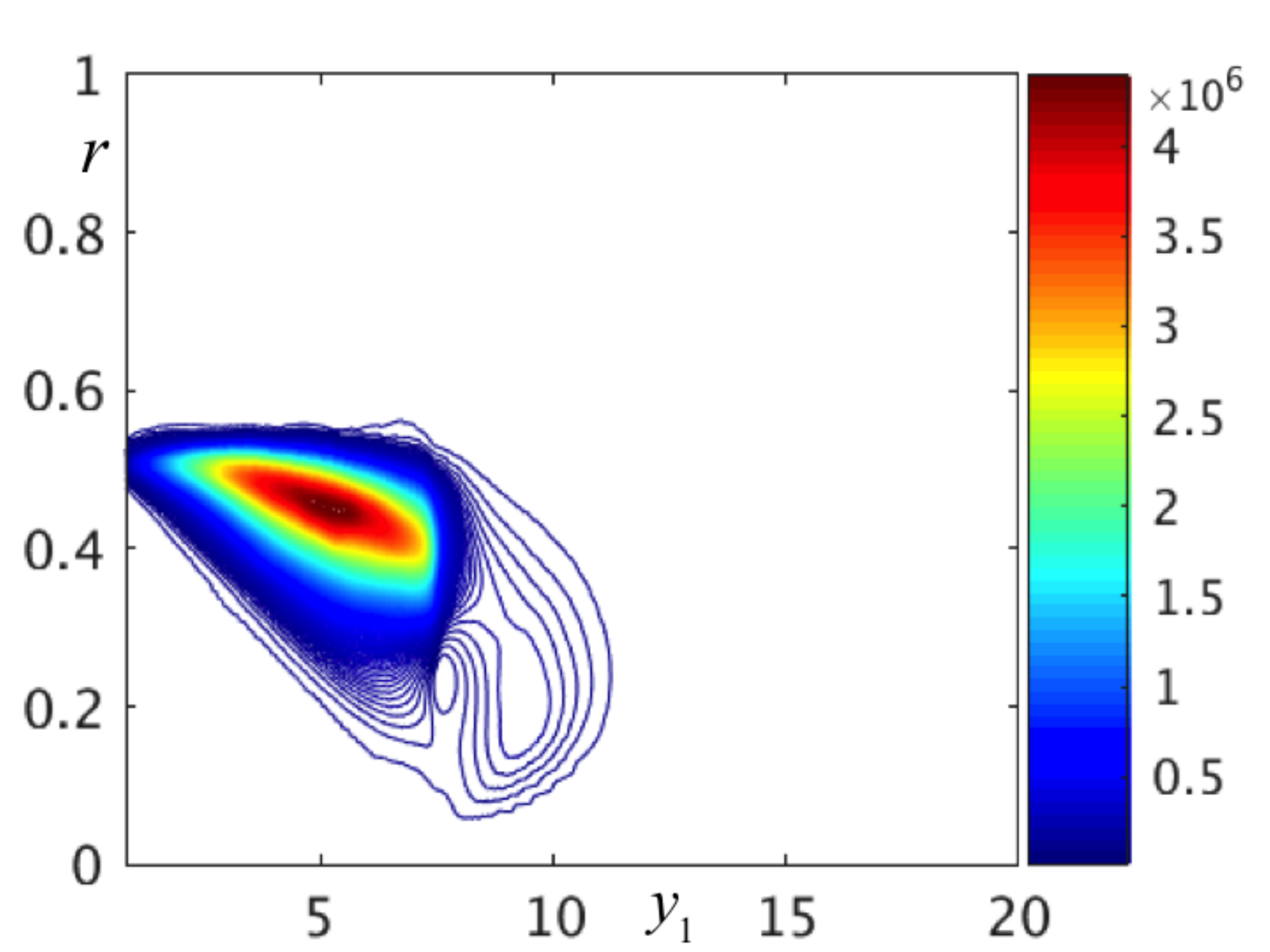} 
        \caption{}
          \label{fig:G12_SP49_NP}
    \end{subfigure} \\
            \centering
    \begin{subfigure}[t]{0.42\textwidth}
        \centering
        \includegraphics[width=\textwidth]
{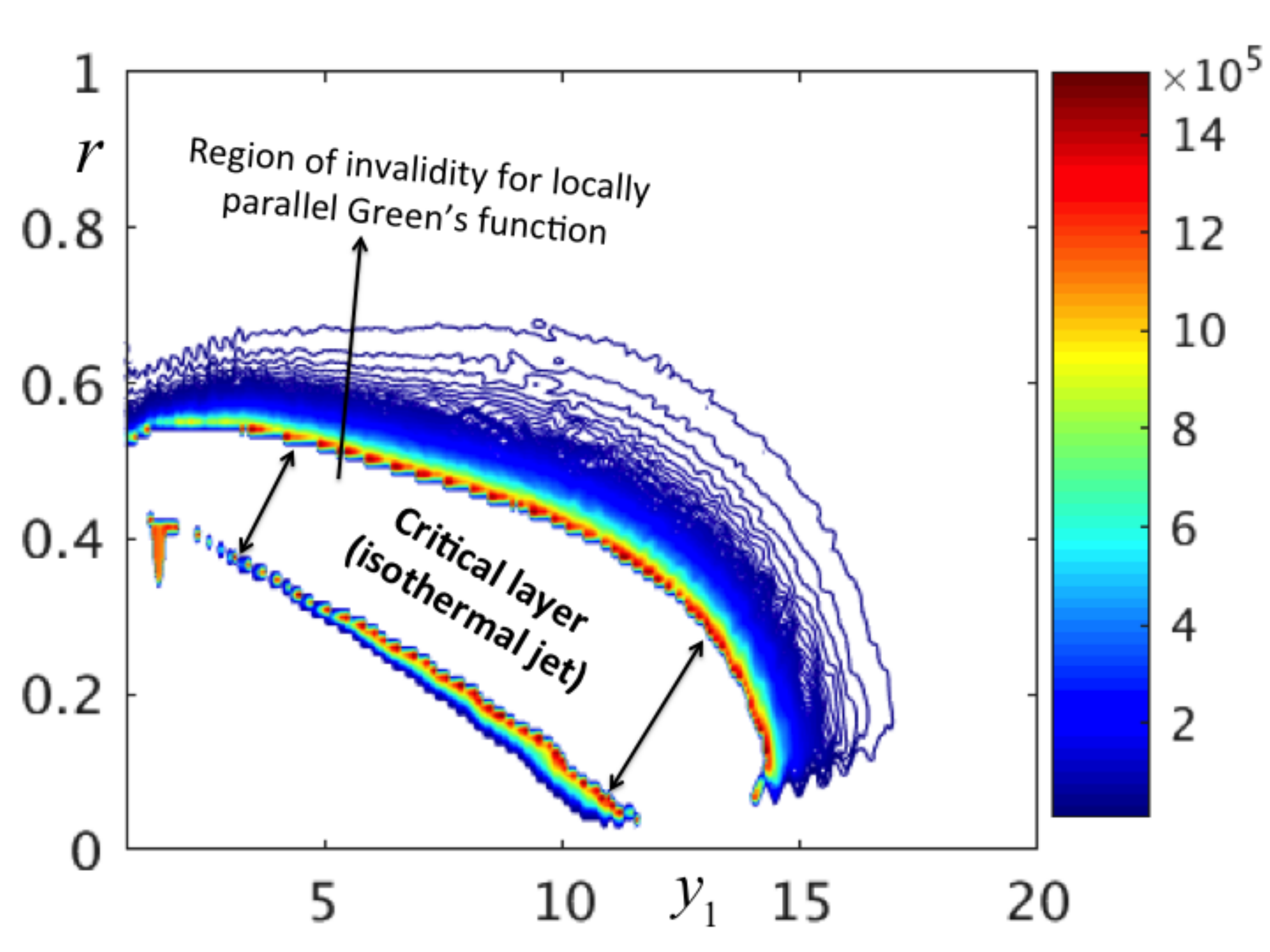}        
        \caption{}
          \label{fig:G12_SP90_P}
    \end{subfigure}
    \centering
    \begin{subfigure}[t]{0.42\textwidth}
        \centering
        \includegraphics[width=\textwidth]        {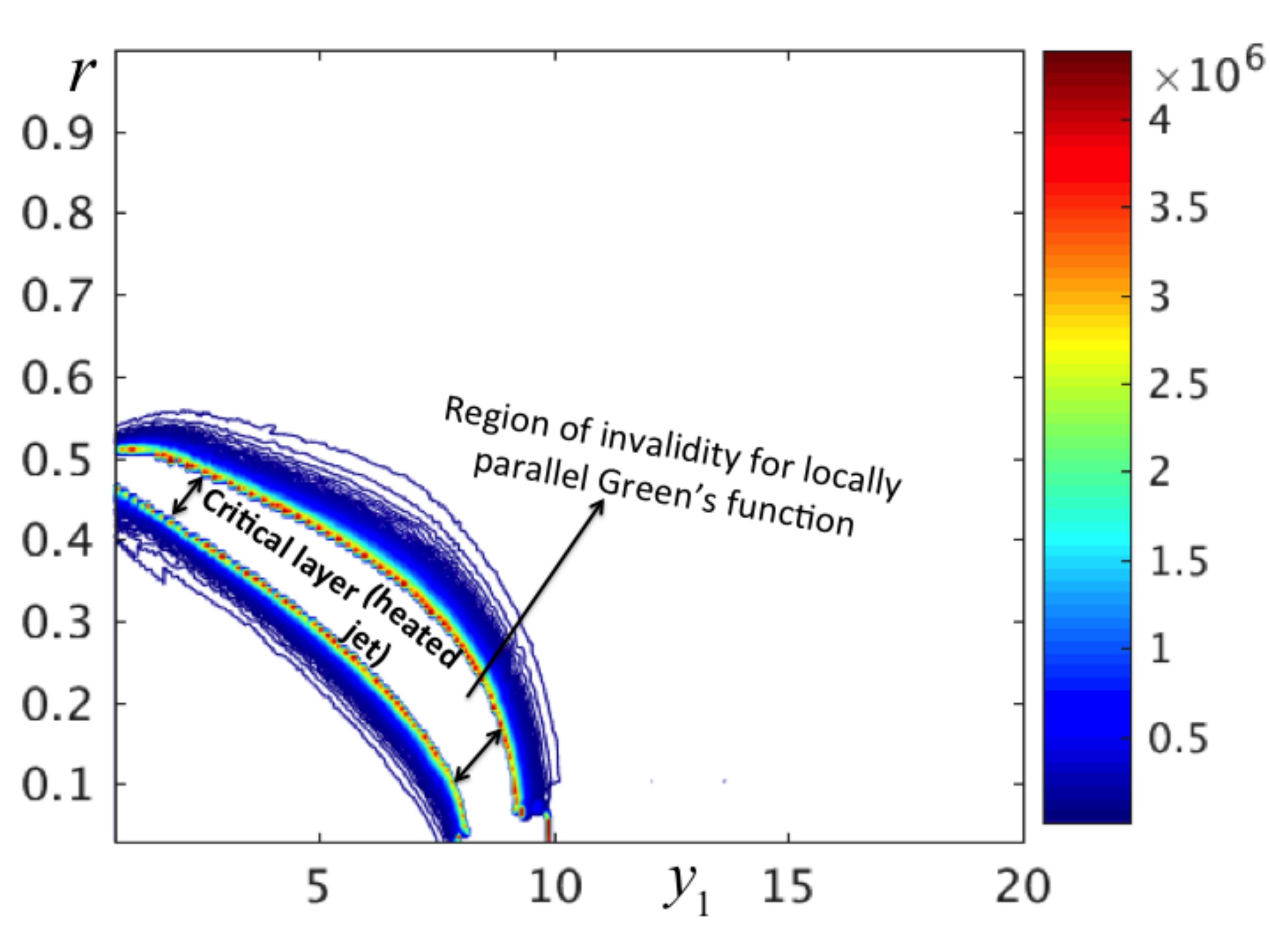}
        \caption{}
    \label{fig:G12_SP49_P}
    \end{subfigure}
    \hfill
           \caption{Spatial structure of momentum-flux propagator $|G_{12}|^2$ in (\ref{G_12}) using  Non-parallel (N-P) and locally-parallel (P) solution to $\bar{\nu}(Y,U)$ via (\ref{Hyp3}) at $(St,\theta) = (0.2, 30^\circ)$. (a) $\&$ (c) are N-P $\&$ P for SP$90$; (b) $\&$ (c) N-P $\&$ P for SP$49$.
           }
    \label{fig:prop_G12}
\end{figure}

The spatial structure of momentum flux propagator, $|\bar{G}_{12}|^2$, in (\ref{I_low}) is shown in Fig. \ref{fig:prop_G12} at the peak frequency and observation angle of $(St,\theta)= (0.2, 30^\circ)$ for SP$90$ and SP$49$.
The scaled frequency $\Omega$ is now $\Omega=2\pi St/\epsilon$ where the Strouhal number $St$ is based on jet exit diameter and velocity.
The mean flow in the calculation is non-parallel inasmuch as $\bar{X}_1 \neq 0$ in the solution to (\ref{Hyp3}). 
The slope of the upper most level curve in Fig. \ref{fig5_3a} and Fig. \ref{fig5_3d} gives the spread rates: $\epsilon \approx 0.09$ for SP$90$ and $0.12$ for SP$49$; this allows us to transform between slow variable, $Y$, and the physical variable, $y_1$ that will be needed in the computation of (\ref{eq:Iom2}).

The non-parallel flow structure of $|\bar{G}_{12}|$ in Figs. \ref{fig:G12_SP90_NP} $\&$ \ref{fig:G12_SP49_NP} possesses a single (almost) streamwise-aligned lobe which peaks between $2<y_1 <8$ for SP$90$ (Fig. \ref{fig:G12_SP90_NP}) continuing to more-or-less the end of the potential core at $y_1 \sim 12$.
This is different to Karabasov {\it et al}'s (\cite{Karab2010}, Fig. 13a) calculations and GSA's qualitative analysis both of which showed that a double-peak structure of $|\bar{G}_{12}|$ is recovered in subsonic flows. 
%
%
We can explain this by considering the asymptotic structure of the propagator for a locally parallel mean flow. Here the asymptotic solution of $|\bar{G}_{12}|^2$ is independent of $y_1$ and proportional to $(\partial U/\partial r)^2/(1-M(y_1,r)\cos\theta)^6$  (Goldstein 1975; Goldstein $\&$ Leib\cite{GanL}; Afsar\cite{Afs2010}). Therefore, the initial shear layers (of large $\partial U/\partial r$) dominate the spatial structure of $|\bar{G}_{12}|$  at subsonic $Ma$.
In supersonic conditions, however, the locally parallel-based $|\bar{G}_{12}|^2$ is singular at the critical layer location $(y_1, r)$ at given observation angle $\theta=\theta_c$ at the peak frequency.
This is exemplified by a white region between the edge of the potential core and shear layer in Figs. \ref{fig:G12_SP90_P} $\&$ \ref{fig:G12_SP49_P} where $|\bar{G}_{12}|^2$ is infinite and therefore invalid; the edge of this region has a (peak) red contour line according to the figure legend.
On the other hand, the non-parallel flow solution to (\ref{Hyp3}) is large in magnitude because it prevents the locally parallel solution to $\bar{\nu}$ (i.e. when $\bar{X}_1 =0$ in \ref{Hyp3}), and $|\bar{G}_{12}|$ from being singular everywhere in the jet (cf. Figs.17 and 21 in GSA).

The magnitude of $|\bar{G}_{12}|$ increases  for SP$49$, which is more concentrated over a shorter (streamwise/radial) region for both non-parallel and locally parallel (Figs. \ref{fig:G12_SP90_NP} $\&$ \ref{fig:G12_SP90_P} cf. \ref{fig:G12_SP49_NP} $\&$ \ref{fig:G12_SP49_P} respectively) solutions. 
%
For SP$90$, $Ma = M_J$, hence a critical layer first comes into play in the locally parallel flow solution to $\bar{\nu}(Y,U)$ at $\theta_c = 48^\circ$ where $M(y_1,r)=M_J = Ma= 1.5$. But there will also be critical layer for all $\theta \leq 48^\circ$ in this solution owing to the reduction in $M(y_1,r)$ with increasing $r$ at fixed streamwise location, $y_1$.
SP$49$ does, of course, have a subsonic jet Mach number in the core region because $TR>1$, nevertheless a critical layer in $\bar{\nu}(Y,U)$ continues to persist since the locally parallel flow solution to (\ref{Hyp3}) possesses the pre-factor $(1 - U\cos\theta/c_\infty)^{-1}$ (Eq.7.1 in GSA).
Hence a critical layer exists in both $\bar{\nu}(Y,U)$ and $|\bar{G}_{12}|$ (the latter having $3$ inverse Doppler factors) for SP$49$ when the non-parallel flow term in (\ref{Hyp3}) is set to zero.
Our calculations show that the critical far-field location, $\theta_c$, for SP$49$ of $\theta_c \lessapprox 51^\circ$, starts at a slightly higher value than SP$90$ but decreases much more rapidly owing to the faster decay of the mean flow with radial location, $r$, at fixed $y_1$ consistent with the greater spread rate for the heated jet.

\begin{figure}
    \centering
    \begin{subfigure}[b]{0.42\textwidth}
        \centering
        \includegraphics[width=\textwidth]
        {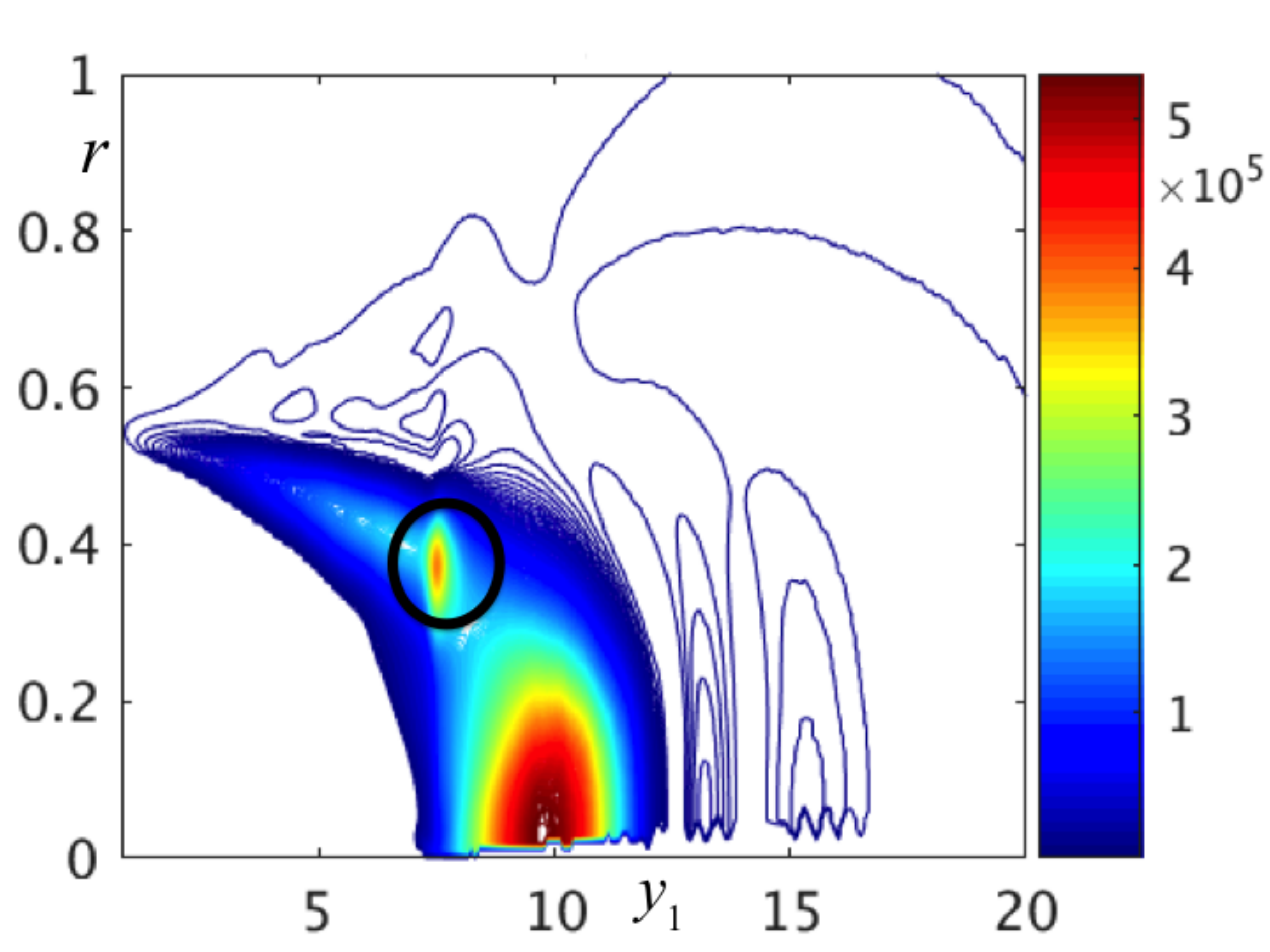}
        \caption{}
        %
       \label{fig:I2_SP49_NP}
    \end{subfigure}
    \begin{subfigure}[b]{0.42\textwidth}
        \centering
        \includegraphics[width=\textwidth]
        {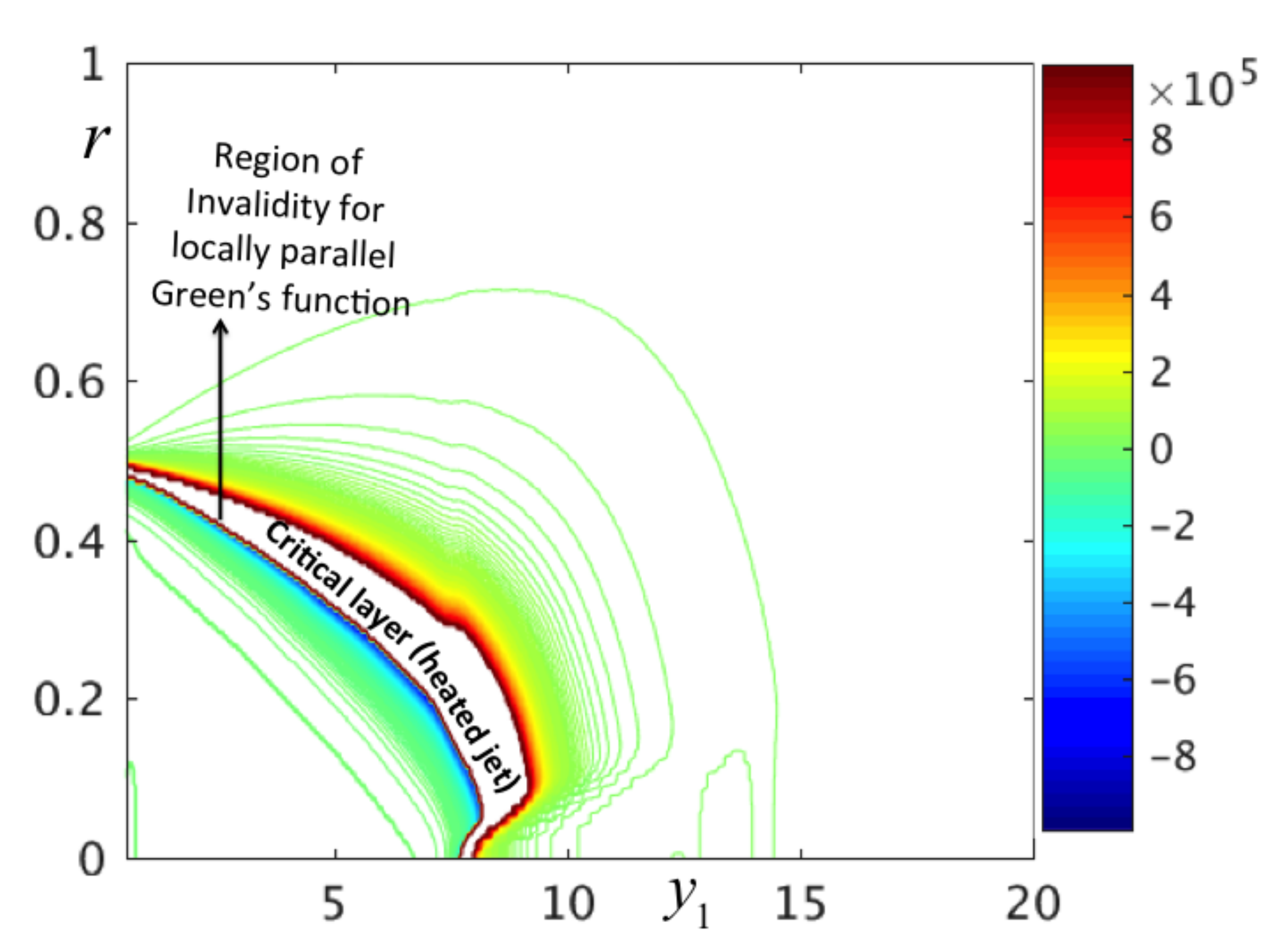}
        \caption{}
       \label{fig:I2_SP49_P}
    \end{subfigure} \\
            \centering
    \begin{subfigure}[b]{0.42\textwidth}
        \centering
        \includegraphics[width=\textwidth]
        {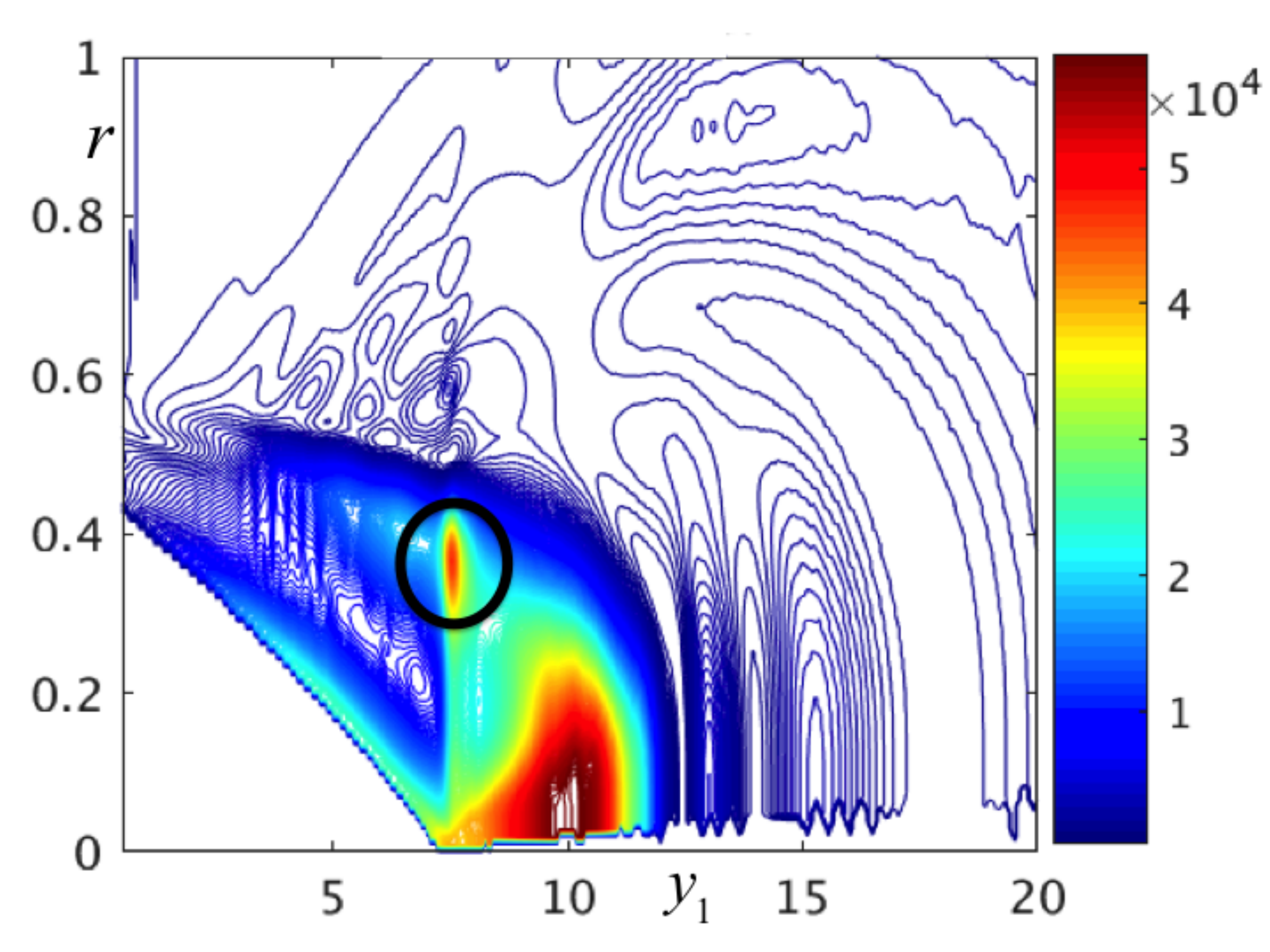}
        \caption{}
       \label{fig:I3_SP49_NP}
    \end{subfigure}
    \centering
    \begin{subfigure}[b]{0.42\textwidth}
        \centering
        \includegraphics[width=\textwidth]
        {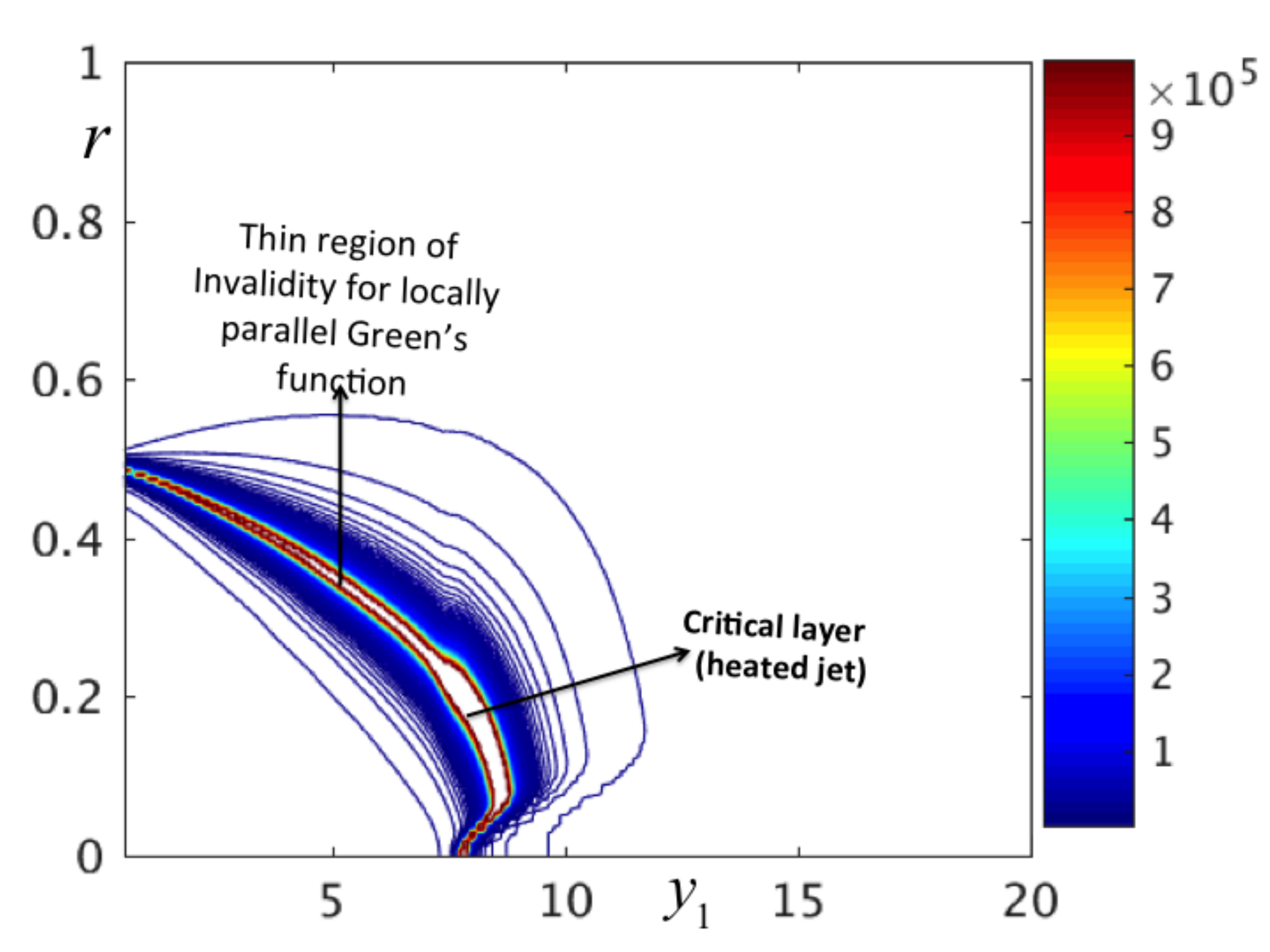}
        \caption{}
       \label{fig:I3_SP49_P}
    \end{subfigure}
    \hfill
                     \caption{Spatial structure of propagator terms in (\ref{G11_41}) $\&$ (\ref{G11_41b}) for SP$49$ at $(St,\theta) = (0.2, 30^\circ)$.
                     (a). 2$ Re\left\{ \bar{\Gamma}_{41} \bar{G}{}_{11}^*\right\}$: N-P;
                      (b). 2$ Re\left\{ \bar{\Gamma}_{41} \bar{G}{}_{11}^*\right\}$: P;
                       (c). $|\bar{\Gamma}_{41}|^2$: N-P;
                        (d). $|\bar{\Gamma}_{41}|^2$: P. See caption of Fig. \ref{fig:prop_G12}.
                     %
                     }
    \label{fig:Prop_I2_I3}
\end{figure}

The temperature-associated propagator terms in Fig. \ref{fig:Prop_I2_I3} have the most dramatic change in spatial structure at the peak frequency and observation angle of $(St,\theta) = (0.2, 30^\circ)$. In the case of locally parallel flow  both $ Re \bar{\Gamma}_{41} \bar{G}{}_{11}^*$ and $|\bar{\Gamma}_{41}|^2$ are consistent with parallel flow estimates of AGF.
That is, in the sense that the coupling term propagator, $ Re \bar{\Gamma}_{41} \bar{G}{}_{11}^*$, in Fig. \ref{fig:I2_SP49_P} has a negative region along the critical layer edge between the shear layer and potential core. This was predicted in AGF using the scaled adjoint Lilley Green's function at $\omega=O(1)$ frequencies (determined by a solution to the adjoint Rayleigh equation) given by Eqs. (4.20), (5.20), (5.22), (C.2) $\&$ (C.4) in Ref. \onlinecite{GanL}. 
When the latter two formulae in Ref. \onlinecite{GanL} that define the propagator components are inserted into (\ref{I_low2}) using (\ref{G11_41}) and (\ref{G11_41b}) in a locally parallel flow, we find that  $ Re \bar{\Gamma}_{41} \bar{G}{}_{11}^*$ is proportional to $(1-M(r) \cos\theta)^{-3}$ and therefore changes sign as $\theta \rightarrow 0$ when ($Ma, TR) >1$ (AGF, pp.2526--2528). 
Note also that AGF's asymptotic estimates did not take into account the correction factors that are required to render the parallel flow Green's function uniformly valid at supersonic speeds across the critical layer. 
But Eq. (4.52), the last term in Eqs. (6.14), (C.16) $\&$ (C.21) all in Ref. \onlinecite{GanL} show that the peak noise component of the acoustic spectrum involving $\Phi{}^*_{1212}$ possesses a correction that, while altering the magnitude of the propagator, remains positive-definite for an isothermal flow as $\theta \rightarrow 0$ for $\omega = O(1)$ and at $Ma = O(1)$. 
Although these correction factors have not been worked out in a heated jet, it seems reasonable to estimate their effect as producing a similar result to the isothermal case; i.e. that the correction changes only the magnitude of the propagator term and not its sign (since jet heating only affects the propagator by a non-uniform variation in $\widetilde{c^2}$). 

In non-parallel flow, the propagator structure is quite different however. That is, our calculation in Fig. \ref{fig:I2_SP49_NP} reveals that $Re \bar{\Gamma}_{41} \bar{G}{}_{11}^*$ {\it remains entirely positive-definite for the supersonic heated jet we have considered here}. Moreover, the peak in the spatial structure of this term is shifted from the shear layer (as in the locally parallel case of Figs. \ref{fig:I2_SP49_P} $\&$ \ref{fig:I3_SP49_P}) to much further downstream. That is at $y_1 \sim 10$ and $r<0.2$ for $Re \bar{\Gamma}_{41} \bar{G}{}_{11}^*$ in Fig. \ref{fig:I2_SP49_NP} and 
at a similar location for $|\bar{\Gamma}_{41}|^2$ in  Fig. \ref{fig:I3_SP49_NP}, where the coupling term propagator has the greater magnitude compared to the enthalpy flux propagator.
As we show later, the auto-covariance component, $R_{1212}$, turns out to be weak in this downstream region.
Thus, contrary to what AGF found, our results indicate that there cannot be any cancellation in the acoustic spectrum formula (\ref{I_low2}) at low frequencies and small observation angles (i.e. for the peak noise) due to the momentum flux/enthalpy flux coupling term because its propagator will \textit{always} be positive when non-parallel flow effects are taken into account. Our numerical calculations show that the positive-definiteness of the coupling term remains true at higher frequencies and for even larger observation angles as well, but the non-parallel flow asymptotic theory developed in $\S.$\ref{S:3} has less direct validity at these locations. 

\vspace{-0.5cm}
\subsection{Spectral tensor component, $\Phi{}^*_{1212}
({\boldsymbol y}, {\boldsymbol k} ; \omega)$}

Since $\mathcal{H}_{1212} \equiv H_{1212}$, the spectral tensor component,  $\Phi{}^*_{1212}(\boldsymbol{y},k_1 ,k{}_T^2 ,\omega)$, is explicitly related to $R_{1212}$ using (\ref{eq:HFT}), (\ref{eq:Rijkl}), the linear transformation below (\ref{eq:Rijkl}) and the space-time Fourier transform, (\ref{eq:Spec_Ten}), as follows:
\begin{equation}
\label{eq:SpecPhi1212}
\Phi{}^*_{1212}
({\boldsymbol y}, {\boldsymbol k} ; \omega)
=
\frac{1}{2\pi}
\int
\limits_{V_\infty({\boldsymbol \eta})}
\int\limits_{-\infty}^{\infty}
e^{i(\boldsymbol{k}.\boldsymbol{\eta} - \omega\tau)}
{R}_{1212}
({\boldsymbol y},\eta_1, \eta_T, \tau)
\,d\tau
\,d{\boldsymbol \eta},
\end{equation}
where $\eta_T = |\boldsymbol{\eta}_T|$. 
We let ${R}_{1212}
({\boldsymbol y},\eta_1, \eta_T, \tau)$ be represented by the following functional form 
\begin{equation}
\label{eq:R1212_model}
  \begin{split}
\frac{{R}_{1212}
({\boldsymbol y},\eta_1, \eta_T, \tau)}
{{R}_{1212}
({\boldsymbol y},{\boldsymbol 0}, 0)}
&
=
\\
&
\left[
a_0
+ 
a_1
\tau
\frac{\partial}{\partial \tau}
+
a_2
\eta_1
\frac{\partial}{\partial \eta_1}
+
...
\right]
e^{\alpha-X(\eta_1, \eta_T, \tau)}
 \end{split}
\end{equation}
We do not include explicit convective streamwise variable $\eta_1 - U_c\tau$ (where $U_c$ is the convection velocity) in (\ref{eq:R1212_model}) or mixed higher-order derivatives as Eqs. (47) $\&$ (48) in Ref. \onlinecite{LG11} possess.
The numerical analysis in Appendix \ref{App:C} justifies this by showing that model (\ref{eq:R1212_model}) gives a slightly more realistic estimation of $c_\perp$ relative to Eqs.(47) $\&$ (48) in Ref. \onlinecite{LG11} when comparing models of ${R}_{1212}
({\boldsymbol y},{\boldsymbol 0}, 0)$ against LES-extracted correlation function data of a similar jet flow. 

The leading term ($a_0$) in square brackets in (\ref{eq:R1212_model}) gives a  cusp for the auto-correlation of ${R}_{1212}
({\boldsymbol y},{\boldsymbol 0}, \tau)$ as $\tau \rightarrow 0$ and the derivative terms, bounded by pre-factors $a_1,a_2$, allow for anti (i.e. negative)-correlations with increasing $\tau$ and streamwise separation, $\eta_1$, respectively. Inspired by Leib $\&$ Goldstein\cite{LG11} we use the separation function, \newline $X(\eta_1, \eta_T, \tau) = \sqrt{\alpha^2 + {\eta{}_1^2}/{l{}_1^2} + {(\eta_1 - U_c\tau)^2}/{l{}_0^2} + f(\eta_T) }$ where an algebraic form of the transverse decay function, $f(\eta_T)\sim \eta{}_T^m$  (with integer values of $m$), is chosen to allow (\ref{eq:R1212_model}) to decay fast enough in $\eta{}_T$. 
A model of this type was found to agree with the structure of the high-order correlation functions in the jet measured by, among others, Harper-Bourne\cite{HB07,HB09,HB10}. 
The length scales $(l_0,l_1)$ are therefore turbulence correlation lengths to appropriately normalize $X(\eta_1, \eta_T, \tau)$ in (\ref{eq:R1212_model}). 
They are taken to be proportional to the local RANS length scales in (\ref{Eq:lengthscales}) with pre-factors $(c_0,c_1)$  that we discuss shortly. 
$\alpha$ is an $O(1)$ parameter in (\ref{eq:R1212_model}), it is introduced to give a more rounded ($\alpha> 0$) cusp of the auto-correlation of ${R}_{1212}
({\boldsymbol y},\eta_1, \boldsymbol{\eta}_T, \tau)$ (see Ref. \onlinecite{GanL} and Afsar {\it et al}. \cite{Afsetal2017}) at
$\boldsymbol{\eta} = 0$ and $\tau = 0$. 
The faster decay in $(\eta_1,\tau)$ away from the cusp of the auto-correlation of ${R}_{1212}
({\boldsymbol y},\eta_1, \eta_T, \tau)$ is a feature of higher-order turbulence
correlations in more homogeneous settings (see Fig. 36d in Yaglom $\&$ Monin\cite{M&Y} on p.249 $\&$f. taken from Frenkiel $\&$ Klebanoff\cite{Kleb}) but it has also been found in (non-homogeneous) jet flows\cite{HB10,M&Z}; see Fig. 19 in Pokora $\&$ McGuirk\cite{Pokora}.
%

\begin{figure}
  \centering
    \begin{subfigure}[h]{0.38\textwidth}
        \centering
        \includegraphics[width=\textwidth]{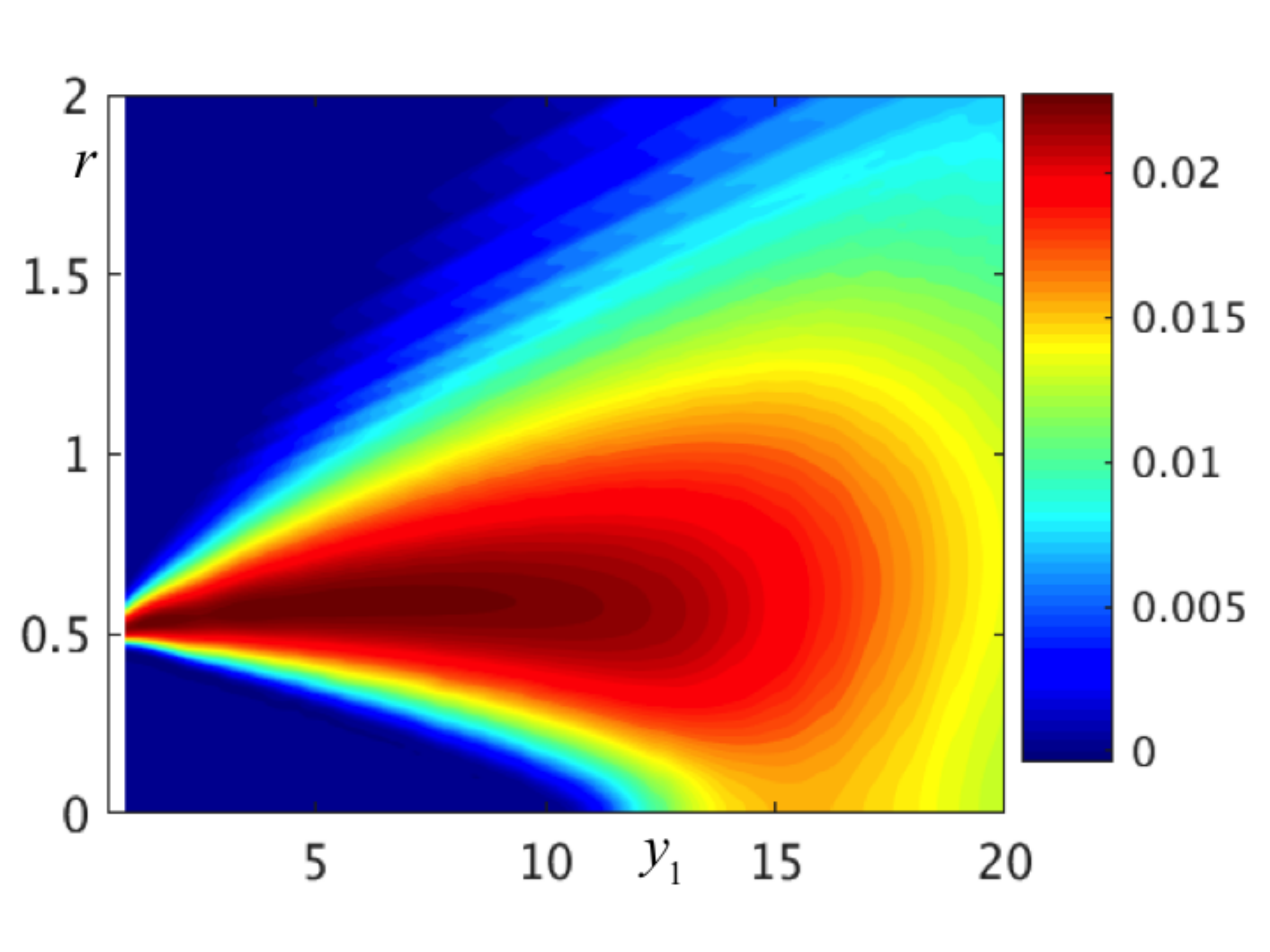}
        \caption{}
        \label{fig5_10a_TKESP90}
    \end{subfigure}
      \centering
    \begin{subfigure}[h]{0.38\textwidth}
        \centering
        \includegraphics[width=\textwidth]{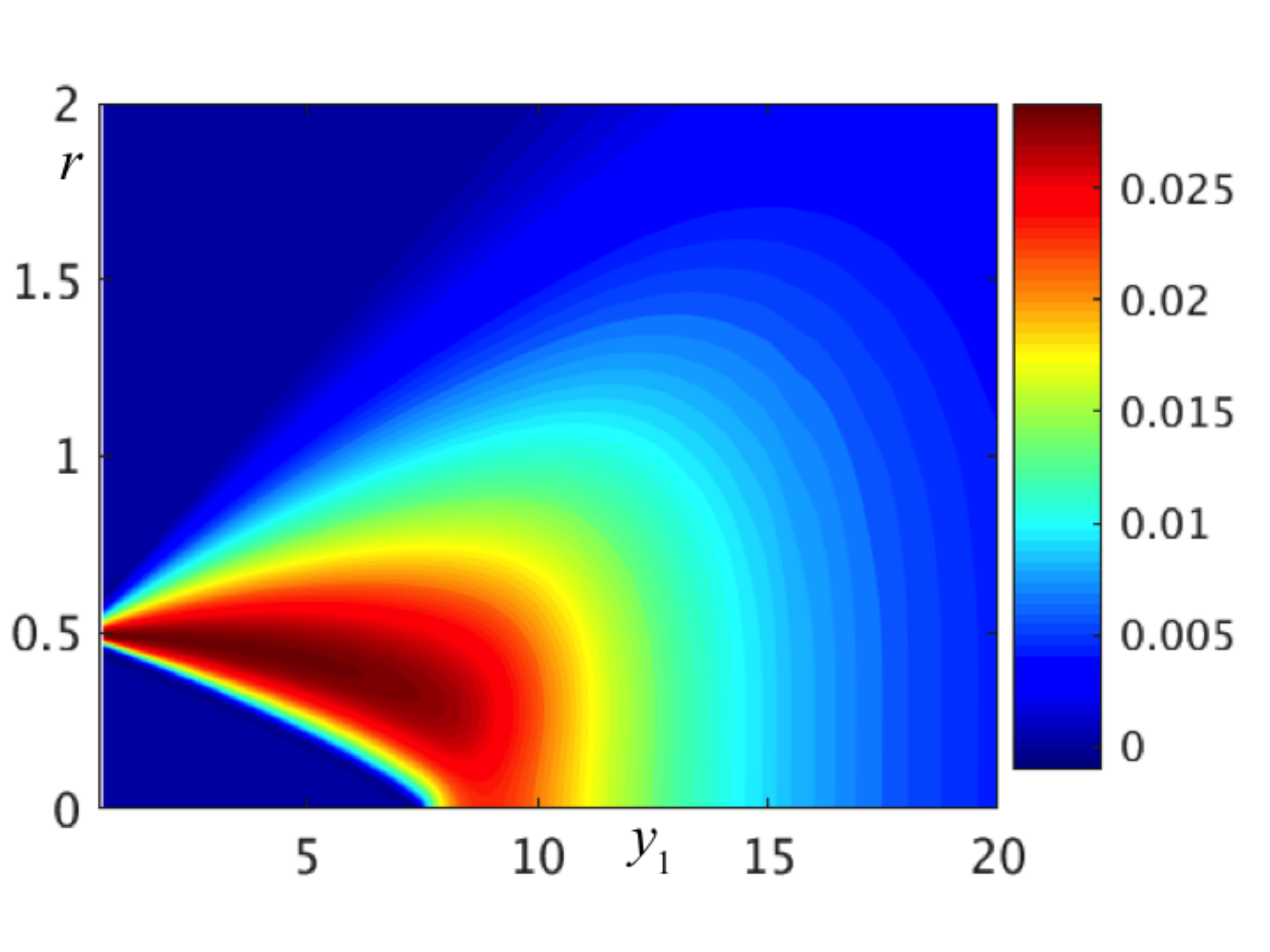}
        \caption{}
        \label{fig5_10b_TKE_SP49}
    \end{subfigure} \\
    \centering
    \begin{subfigure}[h]{0.38\textwidth}
        \centering
        \includegraphics[width=\textwidth]{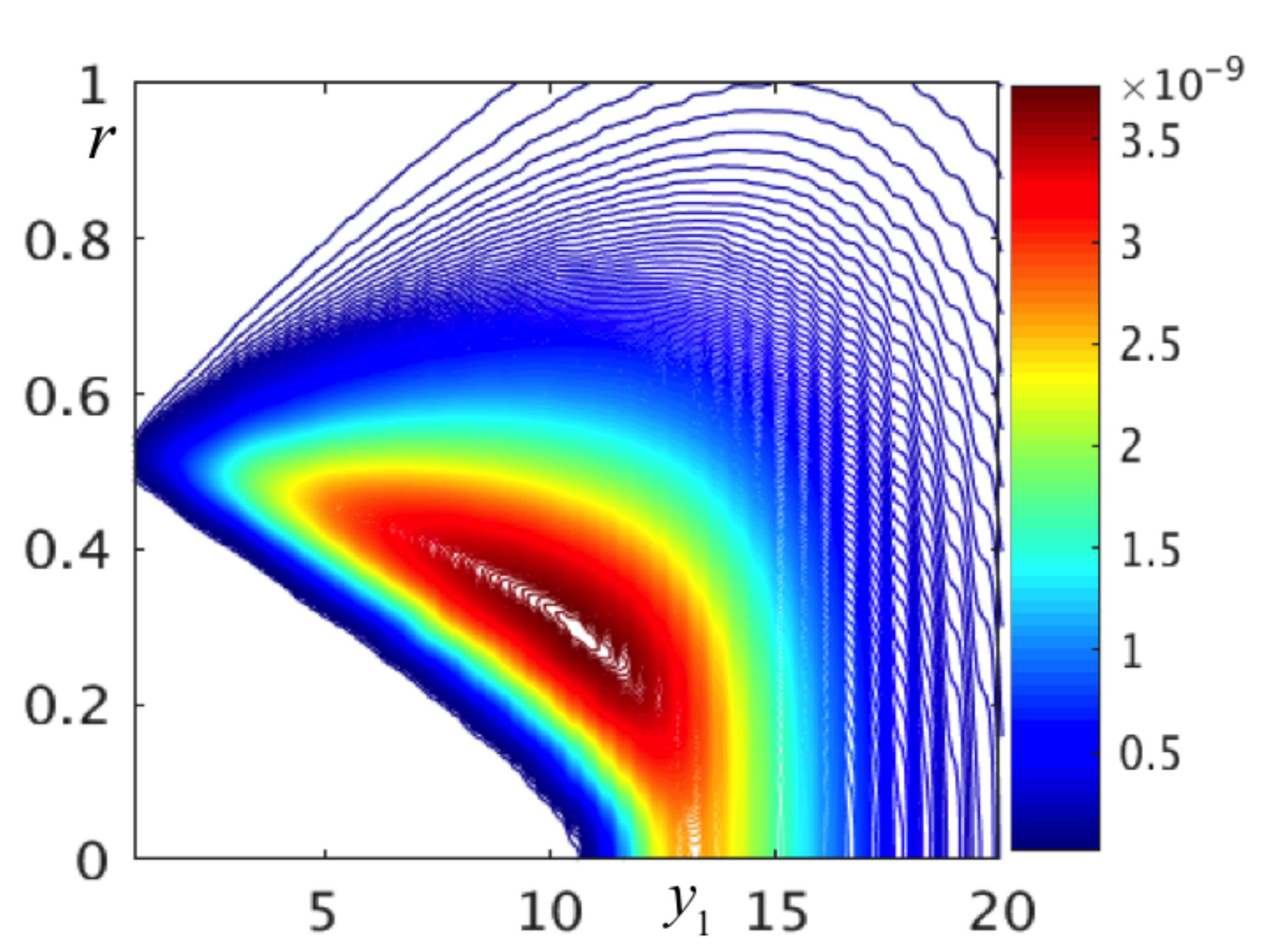}
        \caption{}
        \label{fig5_10c_Phi1212_SP90}
    \end{subfigure}
    \begin{subfigure}[h]{0.38\textwidth}
        \centering
        \includegraphics[width=\textwidth]{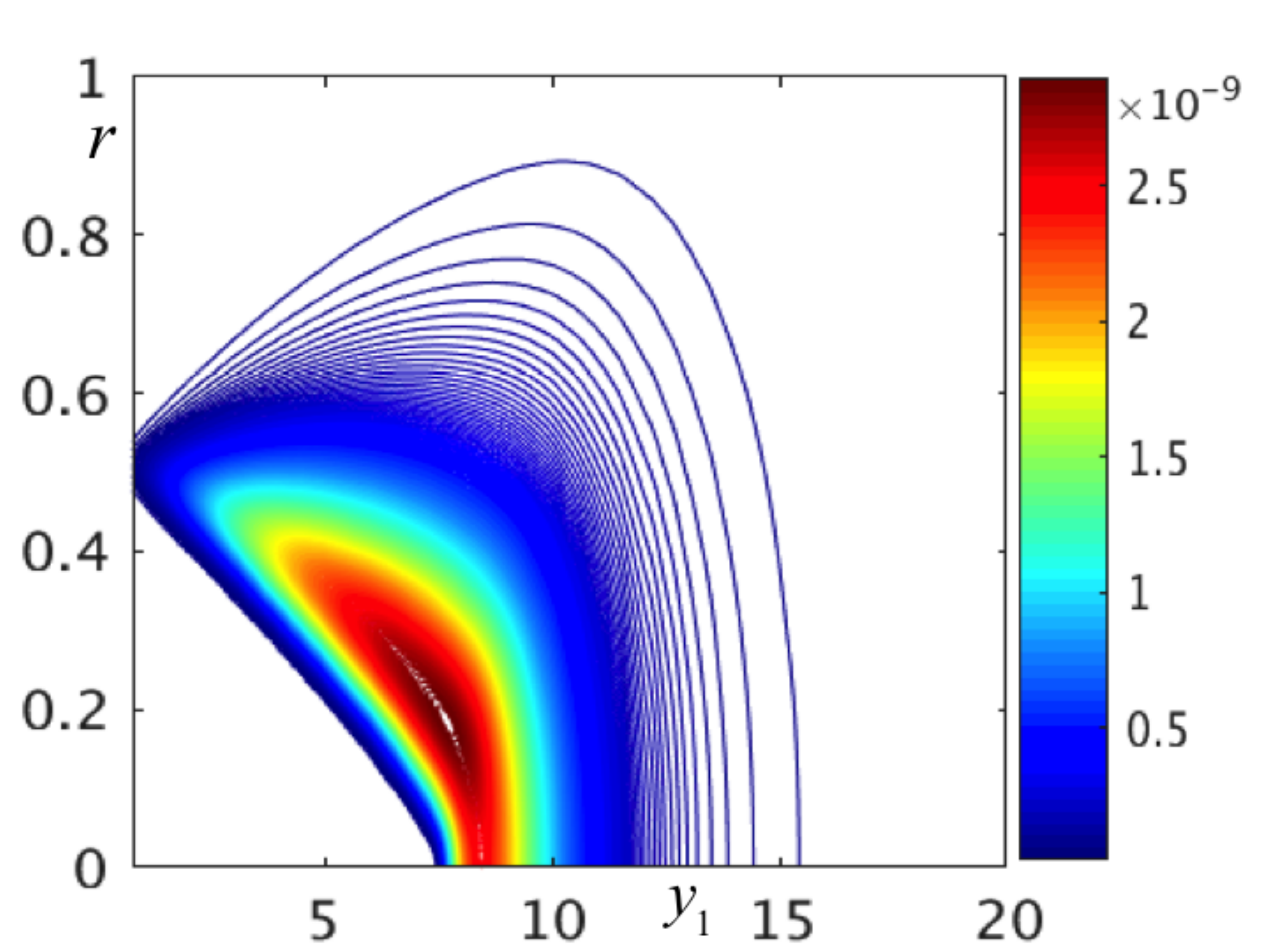}
        \caption{}
        \label{fig5_10d_Phi1212_SP49}
    \end{subfigure}
        \hfill
           \caption{Spatial distribution of TKE, $k(y_1,r)$ and spectral tensor component, $\Phi{}^*_{1212}$, at $(St,\theta) = (0.2, 30^\circ)$; see Table \ref{Table_params} and caption to Fig. \ref{fig5_11_SPLpreds} for turbulence parameters used in (\ref{eq:SpecPhi1212_A4}.
(a). $\&$ (c). show $(k, \Phi{}^*_{1212})$ for SP$90$; 
(b). $\&$ (d). show $(k, \Phi{}^*_{1212})$ for SP$49$.
           }
    \label{fig5_10_TKE_PHI1212}
\end{figure}

%
For an axisymmetric jet the acoustic spectrum is
\begin{equation}
\label{I_om_last}
I({\boldsymbol x};\omega) 
=
2\pi
\int_{r}
\int_{y_1}
I({\boldsymbol x}, {\boldsymbol y};\omega) 
r
\,d {y}_1
\,d r
\end{equation}
where $I({\boldsymbol x}, {\boldsymbol y};\omega)$, given by (\ref{I_low2}), is equal to the product of propagators defined by (\ref{G_12}), (\ref{G11_41}) $\&$ (\ref{G11_41b}) and spectral tensor component, $\Phi{}^*_{1212}
({\boldsymbol y}, k_1, k{}_T^2 ; \omega)$, is worked out explicitly in Appendix \ref{App:C}. 
Our numerical tests showed that there was very little effect on the acoustic spectrum when taking non-zero values for $(\alpha,\tilde{k}_{_T})$  in (\ref{eq:R1212_model})
across Strouhal numbers $0.01\leq St \leq 1.0$.
Therefore, we focus on the algebraically compact formula (\ref{eq:SpecPhi1212_A4}) where $(\alpha,\tilde{k}_{_T}) = 0$ but $(a_1,a_2)$ are slightly non-zero. 
The contour plots of the turbulent kinetic energy $k(y_1,r)$ and $\Phi{}^*_{1212}
$ at the peak noise location $(St,\theta) \approx (0.2, 30^\circ)$ show that jet heating causes greater concentration of contour lines as well as a reduction in magnitude.
This reduction is about $0.7$ between the maxima of $\Phi{}^*_{1212}$
for SP$49$ relative to SP$90$.
However we expect that the localization of contour lines in Fig. \ref{fig5_10_TKE_PHI1212} with heating shall produce a bigger impact to the acoustic spectrum after multiplication by the propagator terms since Figs. \ref{fig:prop_G12} $\&$ \ref{fig:Prop_I2_I3}   show that the latter also display a degree of localization and/or redistribution of peak contour lines  with jet heating that do not entirely coincide with $\Phi{}^*_{1212}$ in Fig. \ref{fig5_10_TKE_PHI1212}.
Note that, any oscillation in Fig. \ref{fig5_10c_Phi1212_SP90} will have a largely negligible impact on the predictions because the propagator is zero in the region where it occurs (cf. Figs. \ref{fig:G12_SP90_NP} $\&$  \ref{fig5_10c_Phi1212_SP90}).

\subsection{Analysis of acoustic predictions}
Low-order/fast computations using RANS-based jet noise models (Leib $\&$ Goldstein\cite{LG11}, Karabasov {\it et al}.\cite{Karab2010}, Afsar\cite{Afs2010} etc.) take $R_{\lambda j \mu l} ({\boldsymbol y}, {\boldsymbol 0}; 0) = a_{\lambda j \mu l} \bar{\rho}^2 (\boldsymbol y) k^2(\boldsymbol y)$
where the density ($\bar{\rho}$) and turbulence kinetic energy (TKE, $k$) fields are obtained from a local RANS solution in which $a_{\lambda j \mu l}$ could also be a function of $\boldsymbol y$ but is usually approximated by a single value on the shear layer location $r = 0.5$ at the end of the potential core (Fig. 4 in Semiletov $\&$ Karabasov\cite{Semil2016}). This value of $a_{\lambda j \mu l}$ can be found by examining the spatial distribution of  
$R_{\lambda j \mu l}  ({\boldsymbol y}, {\boldsymbol 0}; \omega)$ measured in either experiment or via appropriate LES calculation\cite{Karab2010}. 

The prediction model (\ref{I_low2}) requires $8$ independent parameters: $6$ are required to quantify the turbulence structure in the model of $\Phi{}^*_{1212}
({\boldsymbol y}, k_1, k{}_T^2 ; \omega)$, (\ref{eq:SpecPhi1212_A4}): i.e. turbulence length scales $(l_1,l_0,l_\perp)$ and anti-correlation parameters $(a_1, a_2)$ as well as the amplitude constant $a_{1212}$.
Finally, parameters $(n_2, n_3)$ bound the coupling and enthalpy flux terms in the acoustic spectrum formula, (\ref{I_low2}).
But the functional form of $R_{1212}$, (\ref{eq:R1212_model}), depends on $(l_1, l_0)$ through the ratio $l_1/l_0$ (see \ref{eq:R1212_exp}). Indeed as shown in \ref{App:C}, $(l_1/l_0, a_1, a_2 )$ and $a_{1212}$ can easily be determined by appropriate comparison to turbulence data.
Also, the peak radiated sound of SP$49$ turns out to be insensitive to parameters $(n_2, n_3)$ (see Fig. \ref{SPL_n2n3vary}).
Essentially, then, there are two free parameters: $l_0$ or $l_1$ (determined once the ratio, $l_1/l_0$, is fixed after comparison to correlation function data for $R_{1212}$) and $l_\perp$.
By (\ref{Eq:lengthscales}), this requires determining coefficients $c_0$ or $c_1$ and $c_\perp$ when $c_1/c_0$ is fixed.

Since no turbulence data exists for the SP$90$ and SP$49$ jets, we have compared our model (\ref{eq:R1212_model}) to space-time data of $R_{1212}$ for two round jets at a fixed $M_J=1.5$ and varying $TR$ (with one being isothermal; see Br\'es {\it et al}.\cite{Bres17}). These jets were analyzed in Ref. \onlinecite{Afsetal2017} and, at least for the isothermal case, are expected to exhibit a consistent turbulence structure as SP$90$.
But even if there is a difference in Reynolds number between SP$90$/SP$49$ and the Br\'es {\it et al}.\cite{Bres17} jets, Fig. 6b in Karabasov {\it et al}.\cite{Karab2010} shows that the Reynolds number effect does not introduce an appreciable impact on the streamwise spatial and/or temporal decay of (at least) $R_{1111}({\boldsymbol y},\eta_1, \boldsymbol{\eta}_T, \tau)$. This conclusion can be extended to $R_{1212}$ also because the normalized space-time variation of this component of auto-covariance tensor is similar to $R_{1111}$ (see Fig.1 in Semiletov $\&$ Karabasov\cite{Semil2015}).

\begin{figure}
    \centering
    \begin{subfigure}[h]{0.4\textwidth}
        \centering
        \includegraphics[width=\textwidth]
{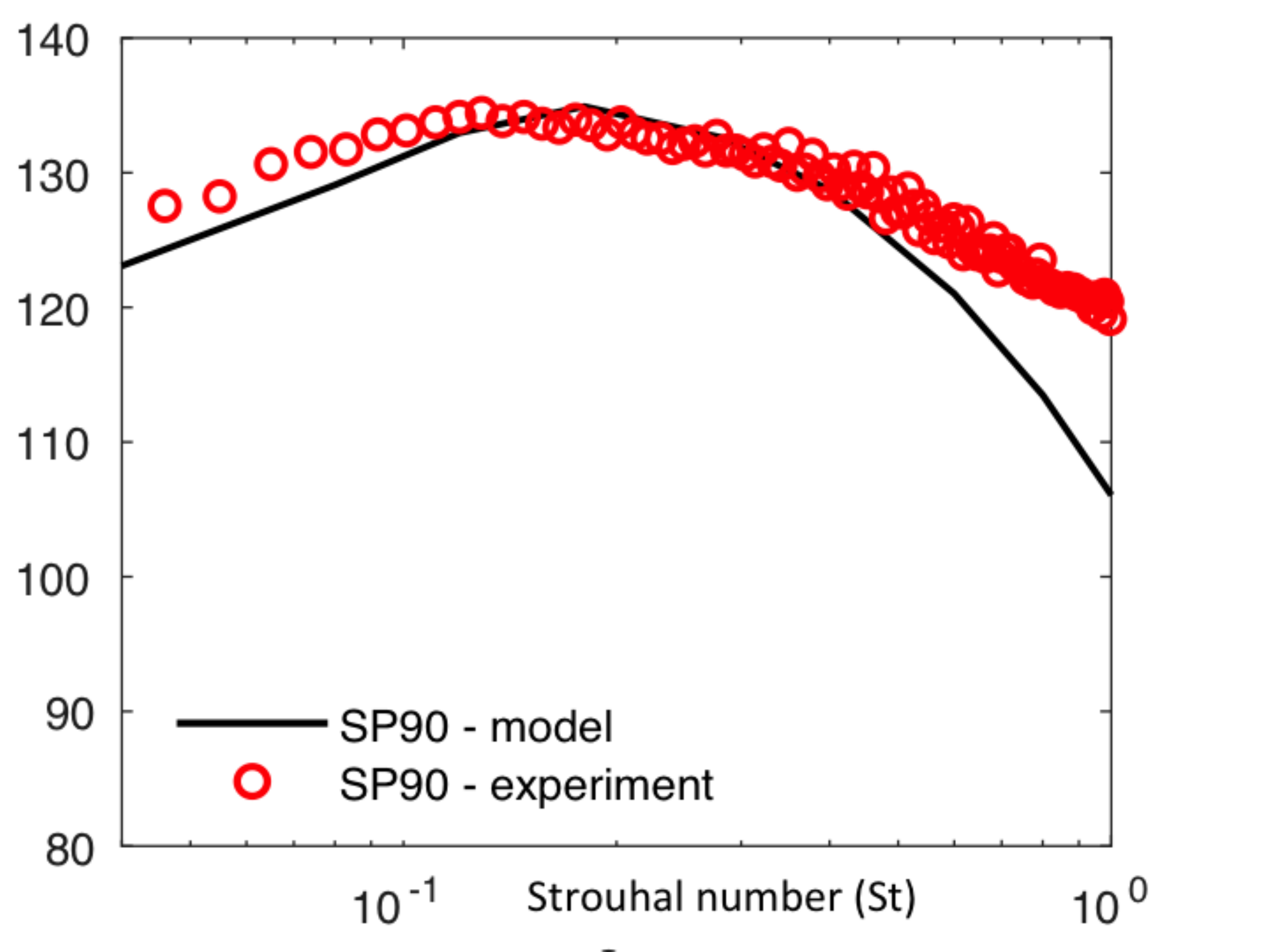}       
        \caption{}
        \label{fig5_11a_SP90theta=23}
    \end{subfigure}
    \begin{subfigure}[h]{0.4\textwidth}
        \centering
        \includegraphics[width=\textwidth]{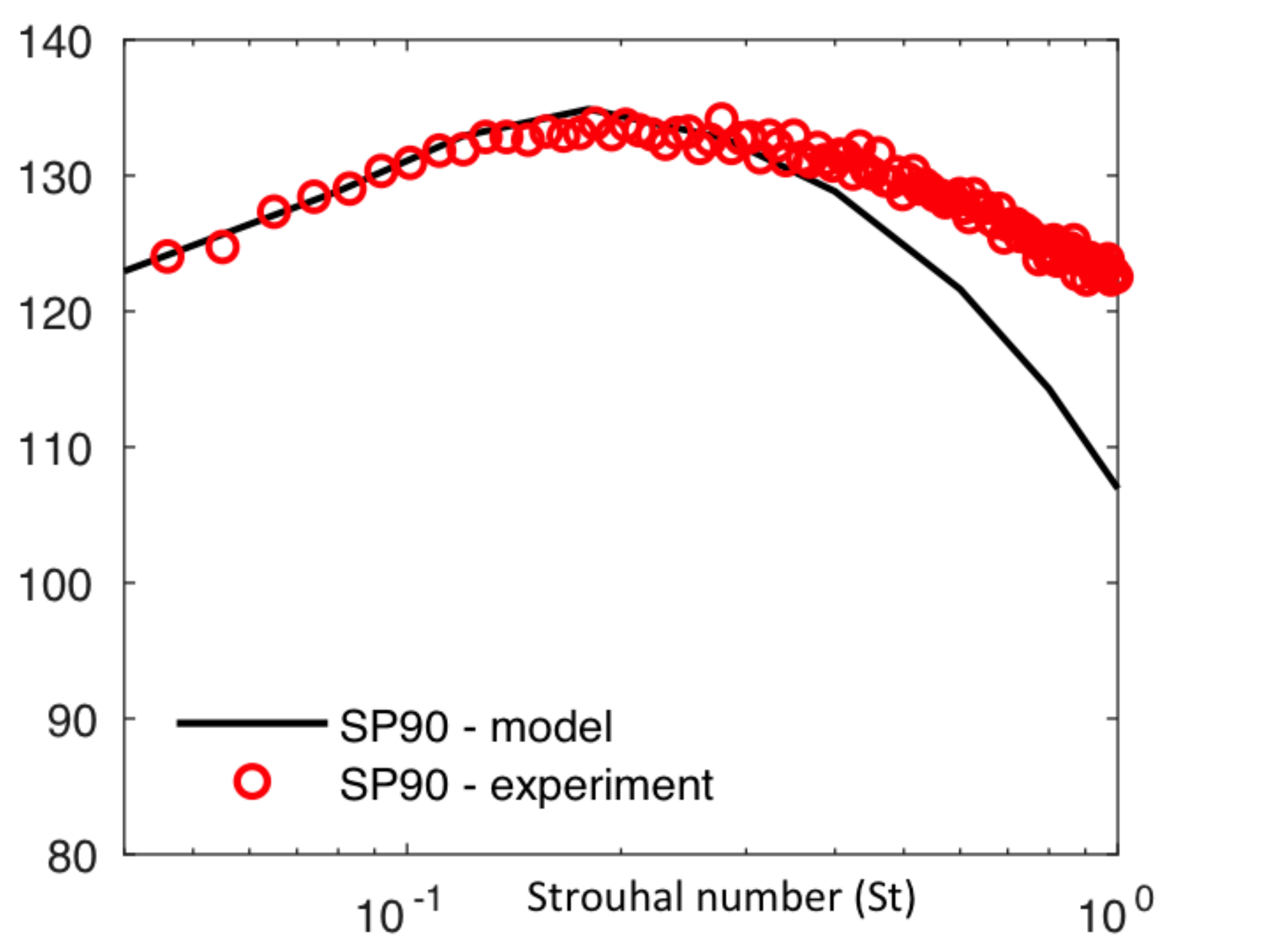}       
        \caption{}
        \label{fig5_11b_SP90theta=28}
    \end{subfigure} \\
    \begin{subfigure}[h]{0.4\textwidth}
        \centering
        \includegraphics[width=\textwidth]{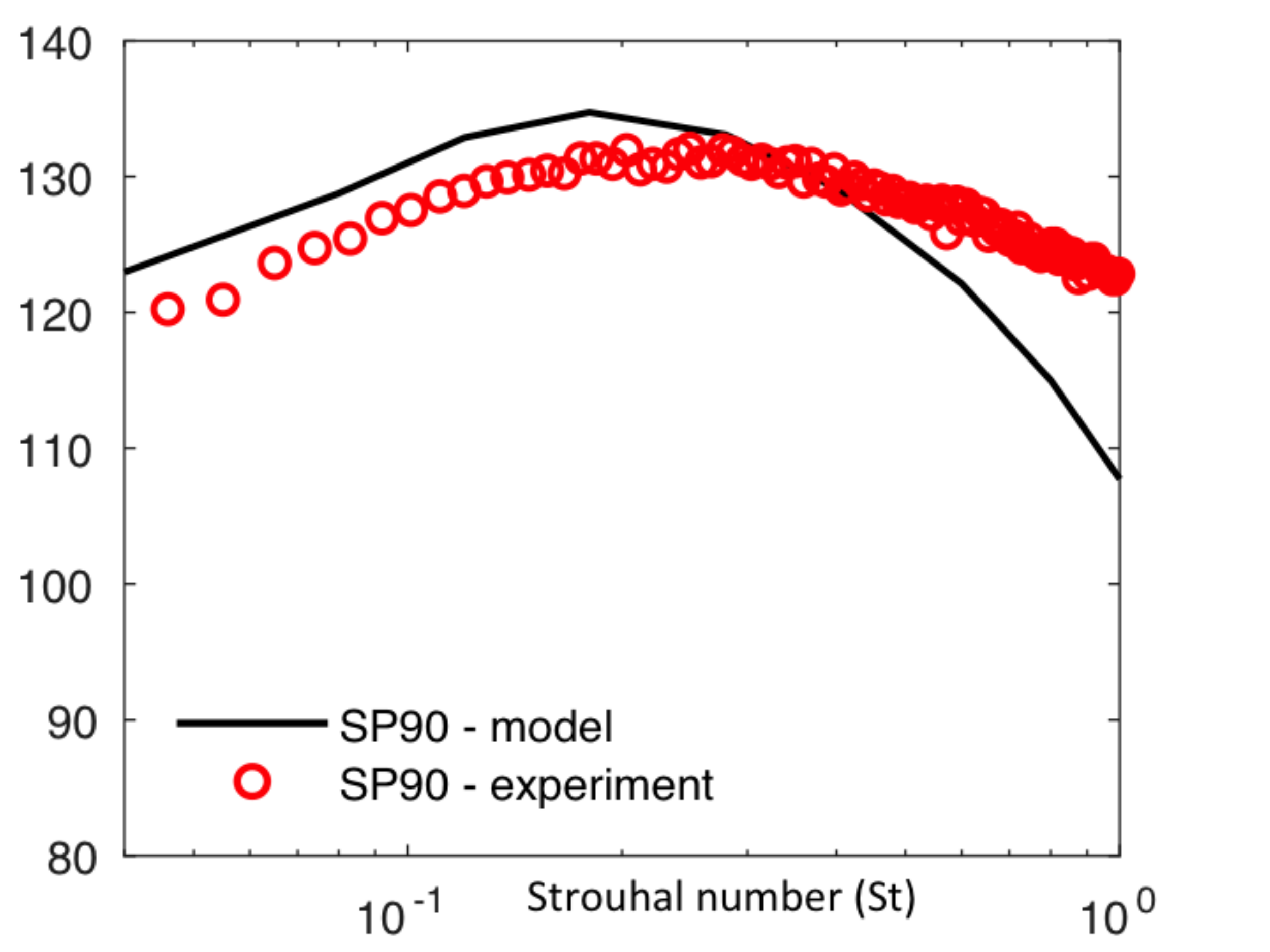}       
        \caption{}
        \label{fig5_11c_SP90theta=33}
    \end{subfigure}
    %
    \begin{subfigure}[h]{0.4\textwidth}
        \centering
        \includegraphics[width=\textwidth]{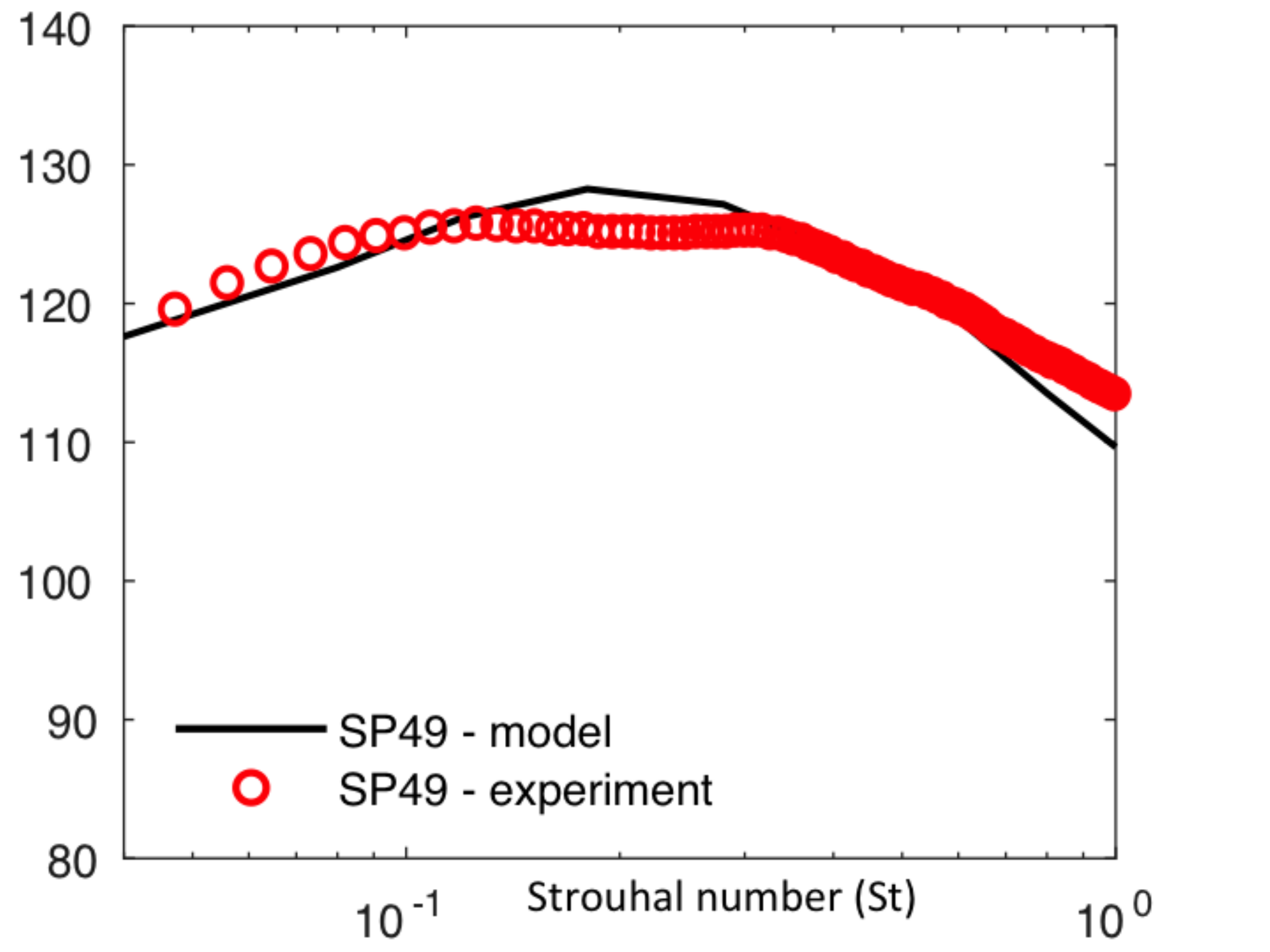}       
        \caption{}
        \label{fig5_11d_SP49theta=25}
    \end{subfigure} \\
    %
    \begin{subfigure}[h]{0.4\textwidth}
        \centering
        \includegraphics[width=\textwidth]{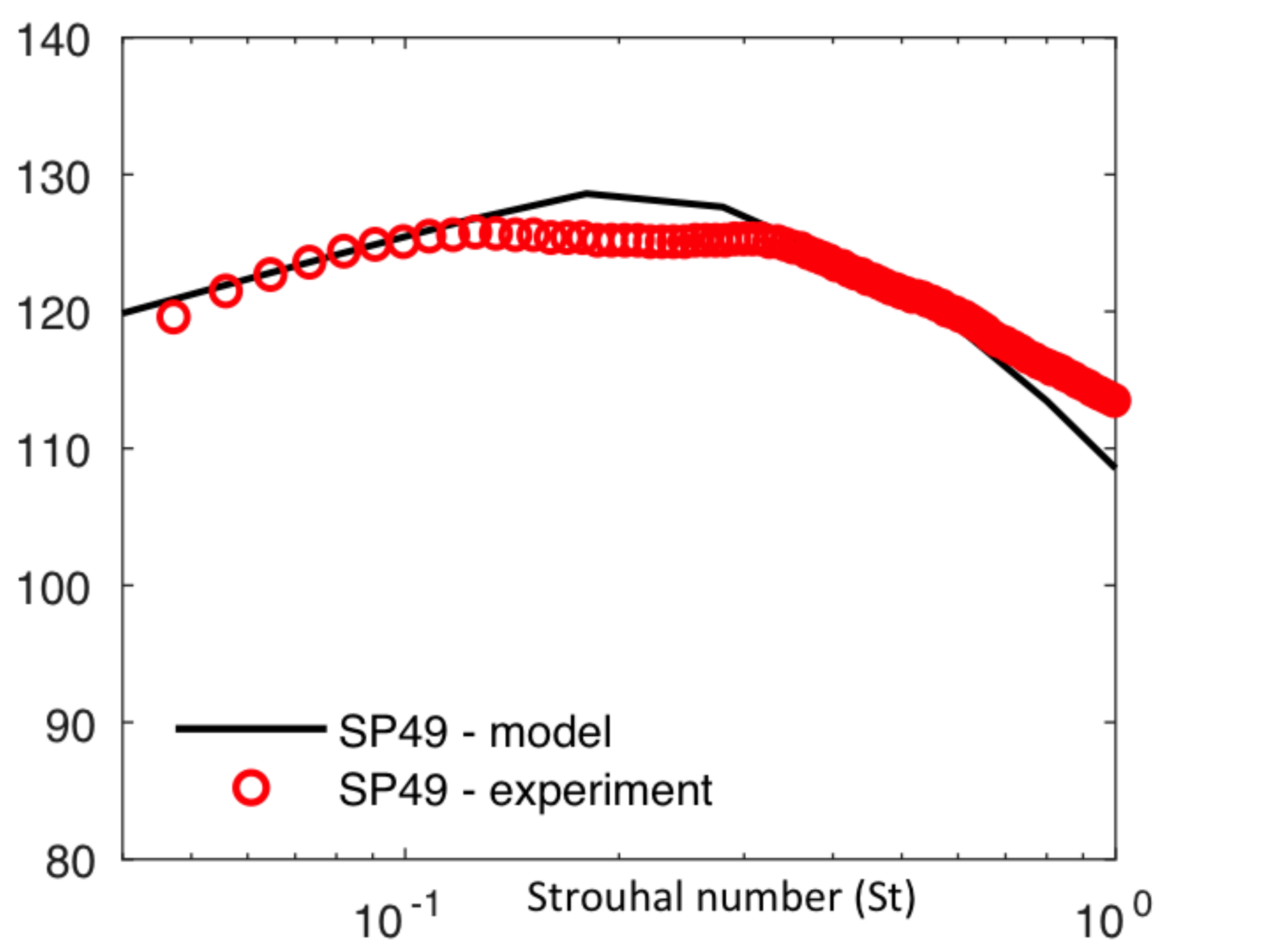}       
        \caption{}
        \label{fig5_11e_SP49theta=30}
    \end{subfigure}
    %
    \begin{subfigure}[h]{0.4\textwidth}
        \centering
        \includegraphics[width=\textwidth]{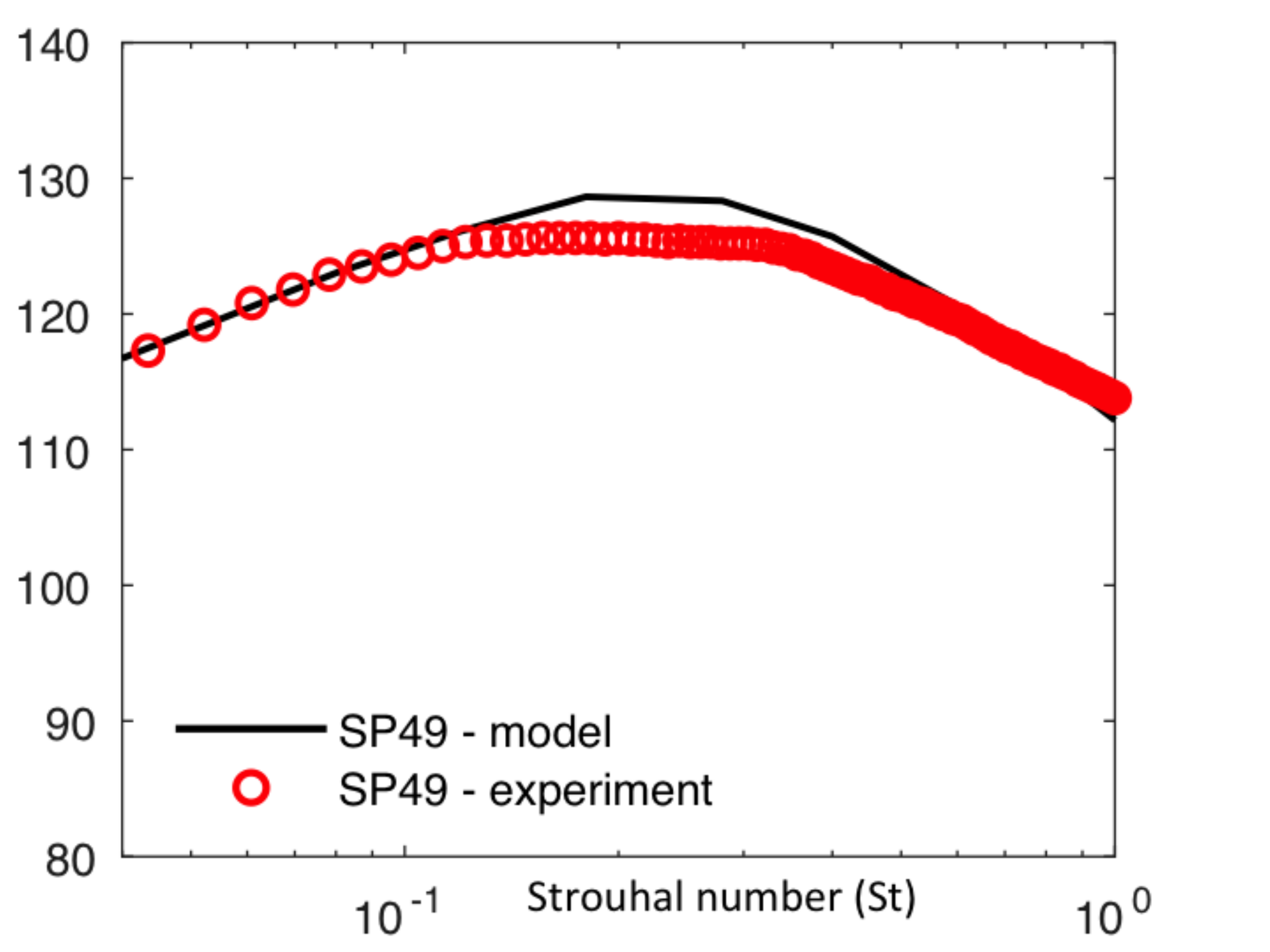}       
        \caption{}
        \label{fig5_11f_SP49theta=35}
    \end{subfigure}
 \caption{Comparison of prediction with NASA experiments using acoustic spectrum formula (\ref{I_om_last}) with spectral tensor model determined by (\ref{eq:SpecPhi1212_A4}) using parameters in Table (\ref{Table_params}). Propagators in (\ref{I_om_last}) are determined by (\ref{Hyp3}), (\ref{BC1}), (\ref{BC2}), (\ref{G_12}), (\ref{G11_41}) and (\ref{G11_41b}). The coefficients $(n_2, n_3)$ in (\ref{I_low2}) are chosen to be $(2.5, 4.0)$ respectively. 
 (a). SP$90$: $\theta = 23.3^\circ$; (b). SP$90$: $\theta = 28.6^\circ$; (c). SP$90$: $\theta = 33.9^\circ$; (d). SP$49$: $\theta = 25^\circ$; (e). SP$49$: $\theta = 30^\circ$; (e). SP$49$: $\theta = 35^\circ$.
 %
 }
    \label{fig5_11_SPLpreds}
\end{figure}

\begin{figure}
  \centering
    \begin{subfigure}[h]{0.4\textwidth}
        \centering
        \includegraphics[width=\textwidth]{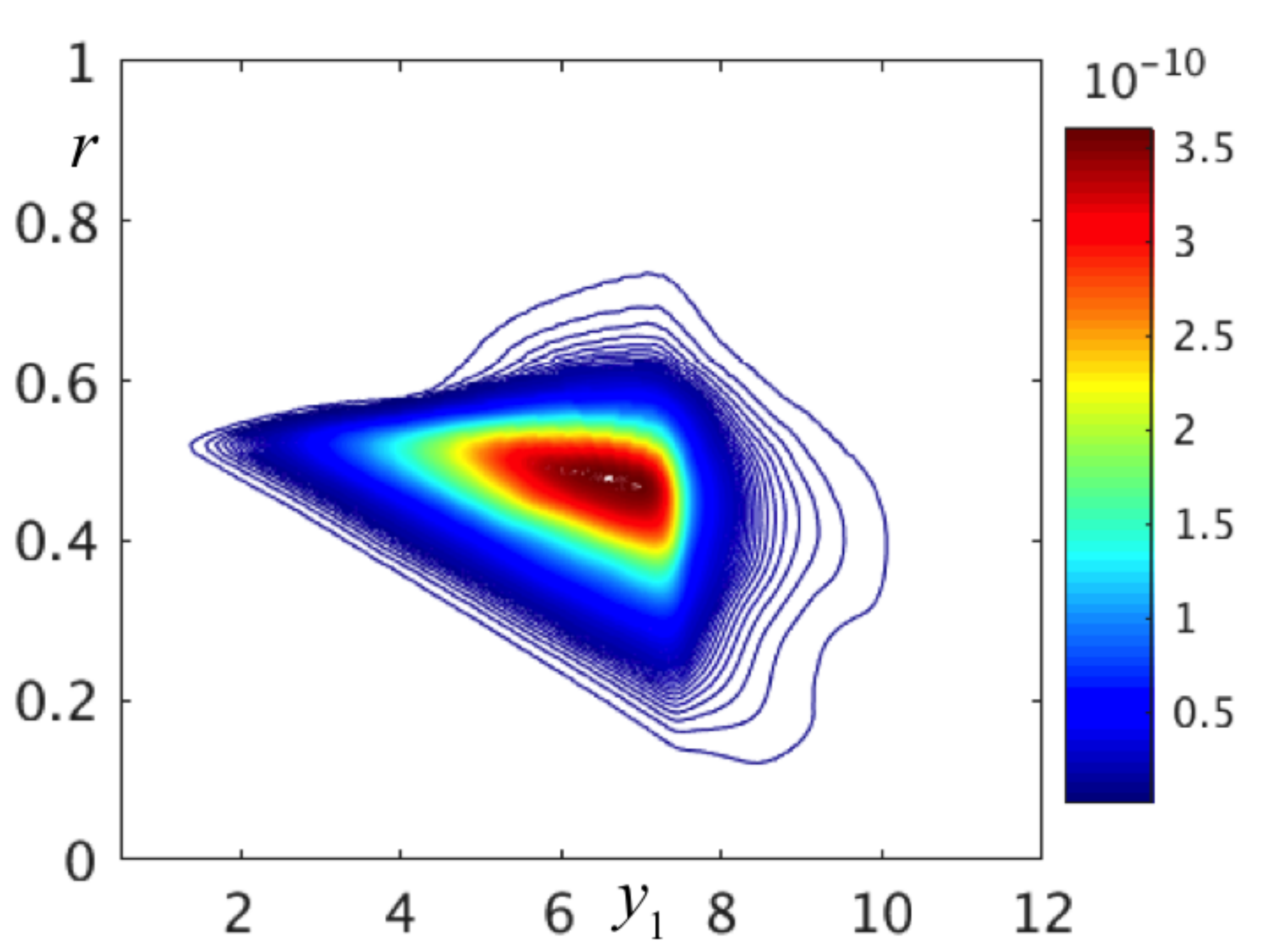}
         \caption{}
        \label{rILow_G12sqrdonly}
    \end{subfigure}
      \centering
    \begin{subfigure}[h]{0.4\textwidth}
        \centering
        \includegraphics[width=\textwidth]
        {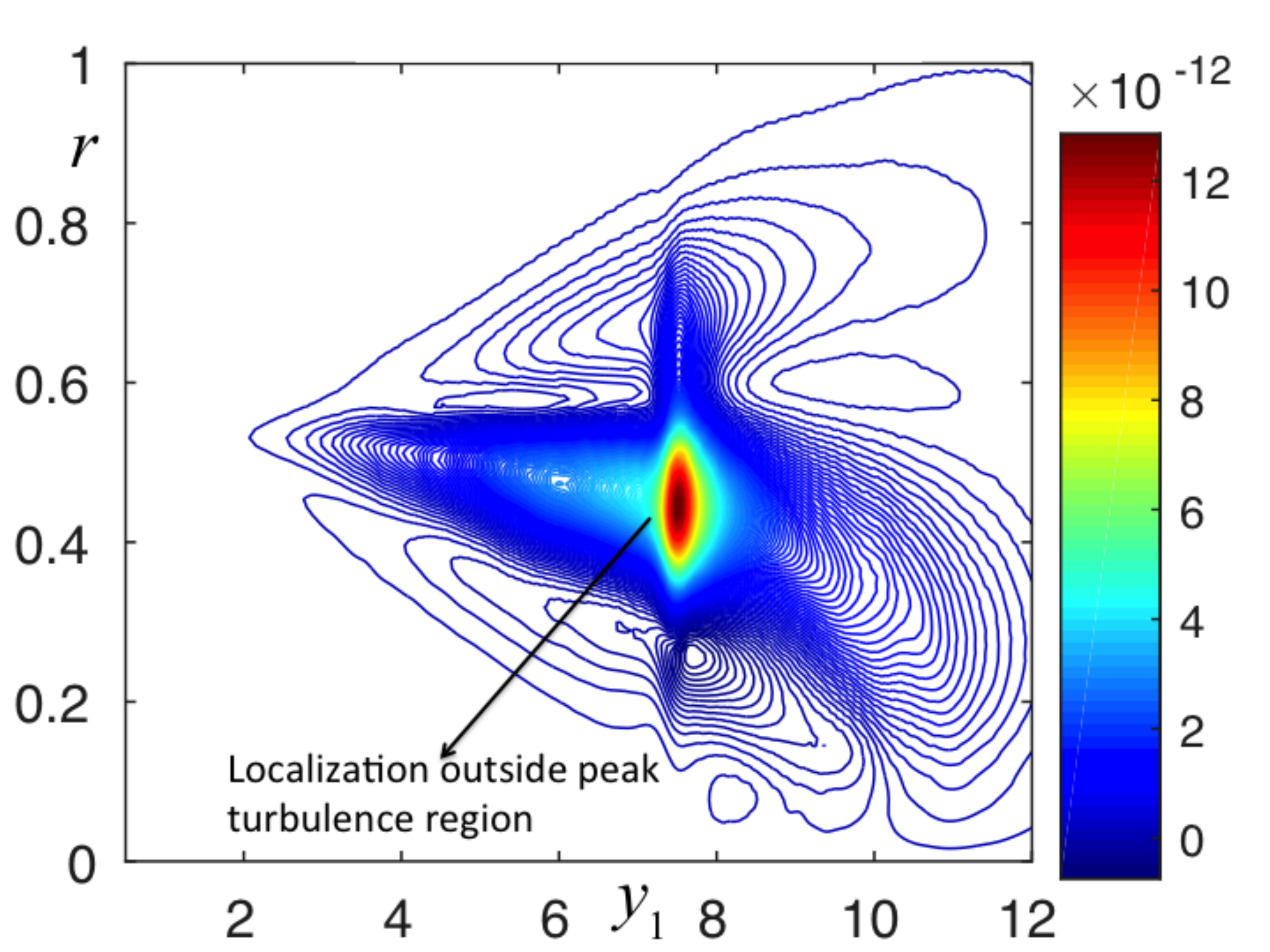}
        \caption{}
%
        \label{rILow_I2only}
    \end{subfigure} \\
    \centering
    \begin{subfigure}[h]{0.4\textwidth}
        \centering
        \includegraphics[width=\textwidth]{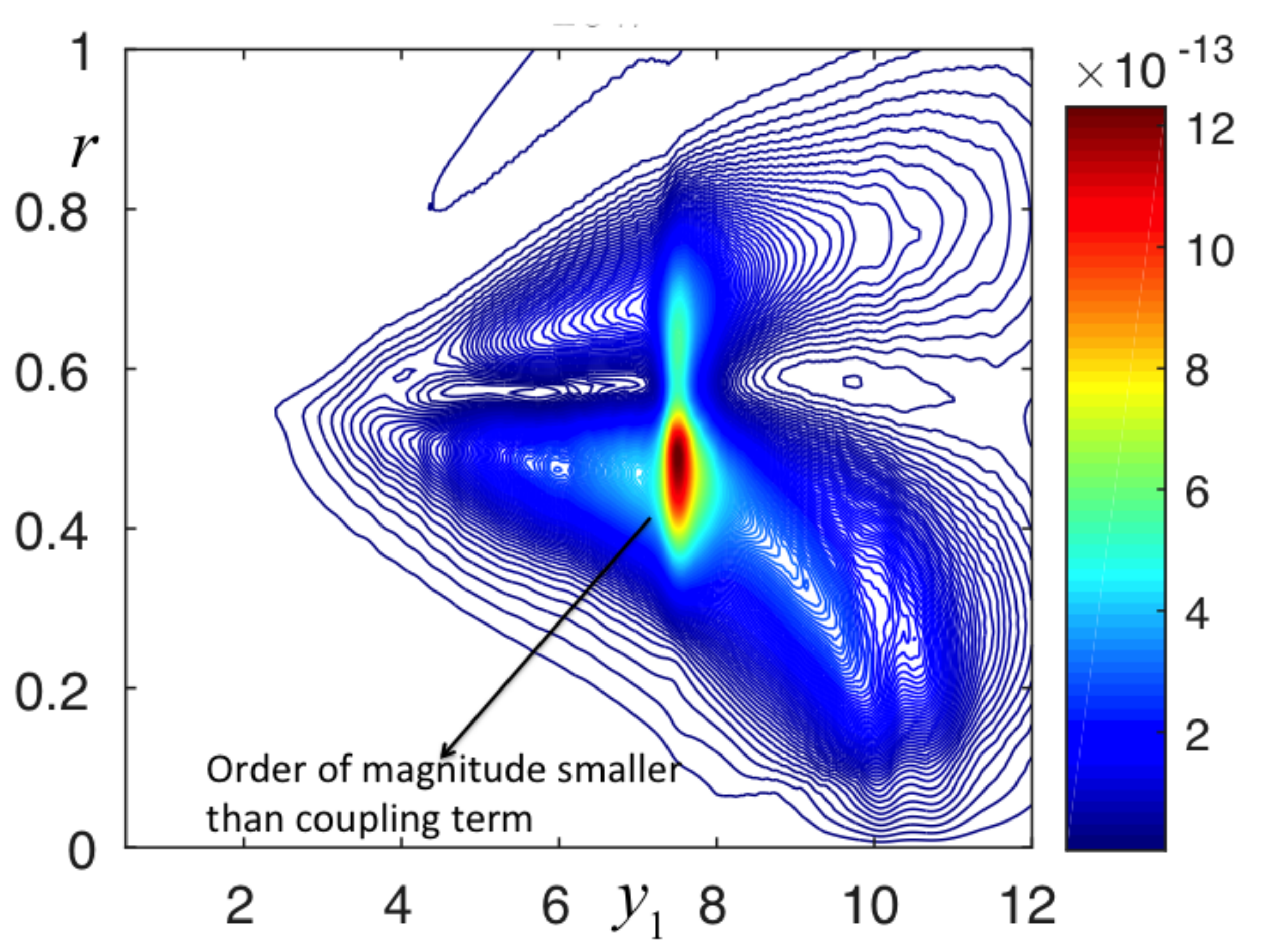}
        \caption{}
        \label{rILow_I3only}
    \end{subfigure}
    \begin{subfigure}[h]{0.4\textwidth}
        \centering
        \includegraphics[width=\textwidth]{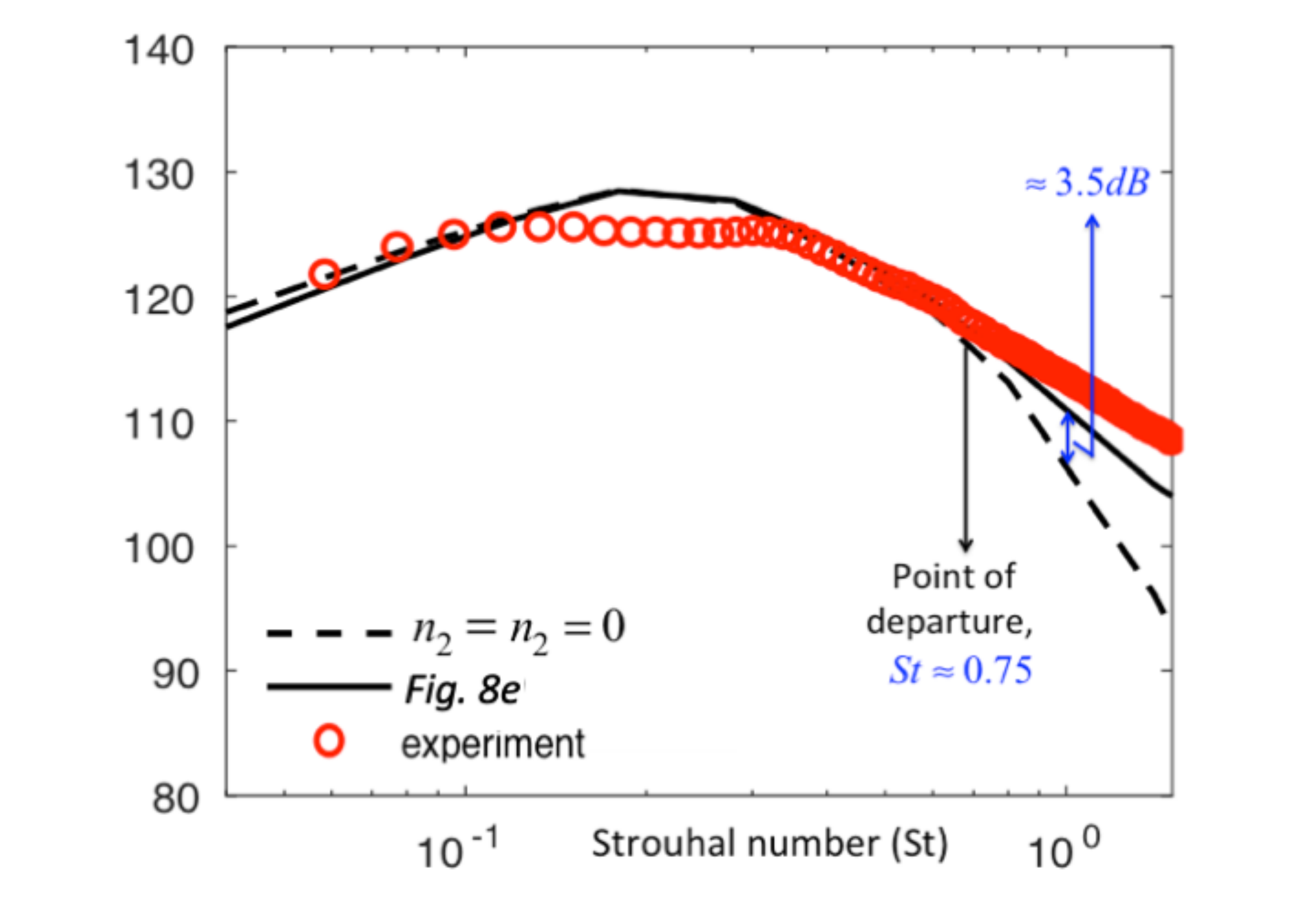}
        \caption{}
        \label{SPL_n2n3vary}
    \end{subfigure}
        \hfill
           \caption{Spatial distribution of each constituent propagator term in $r I({\boldsymbol x}, {\boldsymbol y};\omega)$, (\ref{I_low2}), for SP$49$. Figs. (a--c) computed at $(St,\theta)=(0.2,30^\circ)$ (see Fig. \ref{fig5_11_SPLpreds} for turbulence scales in (\ref{eq:SpecPhi1212_A4})$\&$ $(n_2,n_3)$ values):             %
           (a). $|G_{12}|^2$ only; (b). $Re\left\{ \Gamma_{41} G{}_{11}^* 
\right\}$ only; (c). $|\Gamma_{41}|^2 $ only; (d). sensitivity of $30^\circ$ spectrum to $(n_2,n_3)$. 
           }
    \label{SP49_sensitivity}
\end{figure}

We compared our model for $R_{1212}$ in the form of (\ref{eq:R1212_exp}), to $R_{1212}({\boldsymbol y},\eta_1, \boldsymbol{\eta}_T, \tau)$ data in Figs. \ref{fig_R1212_SP90} $\&$ \ref{fig_R1212_SP90} extracted from the LES solutions reported in Br\'es {\it et al}.\cite{Bres17} to determine the turbulence length scale ratio $c_1/c_0$ and anti-correlation parameters $(a_1, a_2)$. 
The numerical values for $(c_0, c_1, c_\perp, a_1, a_2)$ are summarized in Tables (\ref{Table_params}) $\&$ (\ref{LGscales}); see also Fig.\ref{fig_R1212+LG} and associated discussion.
We use the LES data\cite{Bres17} to estimate the value of $a_{1212}$  
at $r=0.5$. 
Encouragingly, we find that the values of $a_{1212} \approx (0.45, 0.5)$ remain largely constant throughout the jet between $2<y_1<20$ at $r=0.5$ for both the isothermal jet and heated supersonic cases respectively (see Afsar {\it et al}.\cite{Afsetal2017}).
These values are more-or-less consistent with Table 3 $\&$ Fig. 20 in Sharma $\&$ Lele\cite{SharmLele}
%
and (without anything else to go on) we use them for the constant $a_{1212}$ values of SP$90$ and SP$49$ in this paper. 
The values of the coefficients $(n_2, n_3)$ in (\ref{I_low2}) are defined in the caption of Fig. \ref{fig5_11_SPLpreds}
and, following Harper-Bourne\cite{HB10}, the convection Mach number in (\ref{eq:SpecPhi1212_A4}) is set at $U_c = 0.68$ for all predictions.

\begin{table}
\caption{Turbulence model parameters used in (\ref{eq:SpecPhi1212_A4}).  
}
\begin{ruledtabular}
\begin{tabular}{lcr}
 Tanna\cite{Tanna77} set point  &  $(c_0,c_1, c_\perp)$ & $(a_1, a_2)$ \\
\hline
SP$90$ & (0.1, 0.125, 0.022) & (0.19,0.01)\\
SP$49$ & (0.2, 0.17, 0.017) & (0.20, 0.01)\\
\end{tabular}
\end{ruledtabular}
\label{Table_params}
\end{table}

In Fig. \ref{fig5_11_SPLpreds} we show the acoustic predictions against data measured at the NASA Glenn Research Center\cite{BridgesWern}.
As mentioned in $\S.2$, the asymptotic theory finds greatest applicability at small $\theta$ where the peak noise occurs (typically at $\theta =30^\circ$).
We, therefore, show $2$ polar observation angles on either side of $\theta =30^\circ$; i.e., we consider $\theta =$ ($23.3^\circ$, $28.6^\circ$, $33.9^\circ$) for SP$90$ and ($25^\circ$, $30^\circ$, $35^\circ$) for SP$49$. 
The predictions remain accurate up to and beyond the peak noise.
For SP$90$, the agreement lies within $1$dB of the NASA data up to $St=0.5$ (this is almost at $St=0.6$ for $\theta = 23.3^\circ$ in Fig.\ref{fig5_11a_SP90theta=23}). 
Nevertheless, for SP$49$ the agreement is even better across the frequency spectrum; in particular, the low and high frequency decay remains within $1$dB of data up to $St \approx 0.8$. (The prediction does depart from the data near the peak frequency location $St\sim 0.2$ which we estimate at being $\lessapprox 2dB$ at $\theta = 35^\circ$ in Fig.\ref{fig5_11f_SP49theta=35}). 

We found a value of $c_\perp$ almost an order of magnitude smaller than that chosen for $c_1$ (which was found by comparison to turbulence data in Figs.\ref{fig_R1212_SP90} $\&$ \ref{fig_R1212_SP49}) gave the necessary amplitude scaling for the predictions to remain in agreement with acoustic data. That is, $(c_1, c_\perp) = (0.125, 0.022)$ for SP$90$ and $(c_1, c_\perp) = (0.17, 0.017)$ for SP$49$.
Note that experiments/simulations\cite{Pokora, M&Z, Karab2010} show that the transverse correlation length scale of $R_{1111}(\eta_1, \eta_\perp, \tau)$ reduced by almost an order of magnitude compared to streamwise (cf. Fig. 19b to 20b and 21b in Pokora $\&$ McGuirk\cite{Pokora}). 
As mentioned earlier, $R_{1111}$ is only relevant here inasmuch as its normalized space/time structure in $(\eta_1, \eta_\perp, \tau)$ was found to be similar to $R_{1212}$ in Semiletov $\&$ Karabasov\cite{Semil2015} (see their Fig. 1).

In Fig. \ref{SP49_sensitivity}, we show the contours of $r I({\boldsymbol x}, {\boldsymbol y};\omega)$  at the peak noise location $(St,\theta)=(0.2,30^\circ)$ when the momentum flux term $|G_{12}|^2$, the coupling term and the enthlapy flux terms are individually retained in (\ref{I_low2}).
From the contours of the coupling term propagator (\ref{G11_41}) $\&$ (\ref{G11_41b}) in Fig. \ref{fig:Prop_I2_I3}, it is clear that the coupling term makes a positive-definite contribution to the acoustic spectrum.
%
The peak value of radius weighted acoustic spectrum is two-orders of magnitude smaller (Fig. \ref{rILow_I2only} cf. \ref{rILow_G12sqrdonly}) than the contribution made by the momentum flux term $|\bar{G}_{12}|^2$ to (\ref{I_om_last}) at $(St,\theta)=(0.2,30^\circ)$.
Since the enthalpy flux term is even smaller (as Fig. \ref{rILow_I3only} indicates) we can legitimately approximate the integrand of (\ref{I_om_last}) by:
\begin{equation}
\label{I_approx}
I({\boldsymbol x}, {\boldsymbol y};\omega) 
\rightarrow
\left(\frac{\epsilon}{ c{}_\infty^2 |\boldsymbol{x}|}\right)^2
|\bar{G}_{12}|^2
 \Phi{}^*_{1212},
\end{equation}
as the lowest order term in the acoustic spectrum that captures the peak sound for a heated jet since Fig. \ref{SPL_n2n3vary} shows letting $(n_2, n_3) = 0$ in (\ref{I_low2}) gives predictions that are valid up to $St \approx 0.8$, which is well beyond the peak frequency.
This approximation therefore removes any influence of temperature-associated correlations in the acoustic spectrum formula.
Indeed, (\ref{I_approx}) is equivalent to taking $O(\epsilon)$ to be the error term in (\ref{Prop_Exp}), i.e. retaining the momentum flux propagator
\begin{equation}
\begin{split}
\label{Prop_Exp2}
\bar{\Gamma}_{\lambda, j}
(Y,r| {\boldsymbol x};\Omega)
=
&
\delta_{\lambda 1}
\delta_{j r}
\left(
\frac{\partial \bar{G}_1}{\partial r}
-
(\gamma-1)
\frac{\partial U}{\partial r}
\bar{G}_4
\right)
\\
+
&
\delta_{\lambda 4}
\delta_{j r}
\frac{\partial \bar{G}_4}{\partial r}
+
O(\epsilon),
\end{split}
\end{equation}
in $(Y,r;\Omega)$ co-ordinates and inserting the latter into (\ref{eq:IomWKB}) after using (\ref{eq:Scaled_G}).

It is an interesting artifact that only the second much smaller peaks in the coupling and enthalpy flux term propagators (circled in Figs. \ref{fig:I2_SP49_NP} $\&$ \ref{fig:I3_SP49_NP}) remain large when these terms are weighted with $\Phi{}^*_{1212}$ in (\ref{I_low2}). Both of these circled regions are centered on the shear layer of SP$49$ and although being significantly smaller than the larger peak region positioned on the jet axis near $y_1 \sim 10$ in Figs. \ref{fig:I2_SP49_NP} $\&$ \ref{fig:I3_SP49_NP} they still `produce noise' since $\Phi{}^*_{1212}$ remains large along the shear layer compared to that at the jet axis location, $r = 0$, at $y_1 \sim 10$ (see Fig. \ref{fig5_10d_Phi1212_SP49}). 

Fig. \ref{SP49_sensitivity} shows that the coupling and enthalpy flux terms in (\ref{I_low2}) have an impact on $I({\boldsymbol x}, {\boldsymbol y};\omega)$ for SP$49$ at $St>0.8$.
The spectra increases by $\leq3.5$ dB  above that predicted by retaining the momentum flux alone ($n_2=n_3 =0$) in (\ref{I_om_last}).
But this occurs between $0.8\leq St\leq 1.0$ and, therefore, is beyond the low frequency regime.
Essentially, then, both temperature-associated terms in (\ref{I_low}) \textit{remain acoustically silent for the entire low frequency sound regime of the supersonic jets we have considered here}. 
By this result, the sound prediction can be legitimately approximated by (\ref{I_approx}) for all $St< 0.8$.

\section{Discussion -- applicability of the asymptotic theory}
\label{S:6}

Our results confirm that non-parallelism has a pronounced effect on the spatial structure of the propagator causing both amplification in  its value and enlargement of the area in which the peak occurs relative to the locally parallel flow in contour plots of the integrand of the acoustic spectrum (\ref{I_low2}). For SP$90$ ($Ma,TR=1.5, 1.0$), Fig. \ref{fig:G12_SP90_NP} shows that this occurs between $2<y_1<10$. 
The amplification of the propagators, (\ref{G_12}) $\&$ (\ref{G11_41}), in non-parallel flow is clear from the fact that (\ref{Hyp3}), possesses the algebraic expansion $\bar{\nu}(Y,U)\sim \bar{\nu}_0(Y) + U\bar{\nu}_1(Y) +O(U^2\ln U)$, near the outer boundary $U\rightarrow 0$ where $\bar{\nu}_0(Y)$ is given by the outer boundary condition, (\ref{BC1}), and $\bar{\nu}_1(Y)$ by Eq. (5.42) in GSA. 
On the other hand, the locally parallel flow solution, obtained when $\bar{X}_1 = 0$ in Eq. (\ref{Hyp3}), can only behave like $\bar{\nu}\sim e^{-i\Omega Y\cos\theta/c_\infty}/(1-U\cos\theta/c_\infty)$.
%
In a parallel flow, the momentum flux propagator $|\bar{G}_{12}|^2$ is proportional to square of the local mean flow gradient, which peaks at the initial shear layers where $\partial U(y_1,r)/\partial r$ is maximum at fixed $y_1$ and not further downstream. Unless an appropriate uniformly valid solution is constructed (of Goldstein $\&$ Leib\cite{GanL} type), $|\bar{G}_{12}|^2$ will be singular for the parallel flow approximation to the Green's function in the thin critical layer at $\omega = O(1)$ frequencies. 

At the SP$49$ set point ($Ma,TR=1.5, 2.7$), the solution to ${ G_\sigma}
({\boldsymbol y}| {\boldsymbol x}; \omega)$ based on a locally parallel mean flow will also fail to capture the correct level of spectral amplification for a similar reason (see Fig. \ref{fig:G12_SP49_NP}). 
Moreover, the coupling term will fail to explain the reduction in sound because it will introduce cancellation owing to the odd power of inverse Doppler factor (see Eqs. 28 $\&$ 29 in AGF) when the mean flow is locally parallel.  
But we have shown that (\ref{G11_41}) $\&$ (\ref{G11_41b}) will always remain positive definite (Figs. \ref{fig:I2_SP49_NP} $\&$ \ref{fig:I3_SP49_NP} cf. Figs. \ref{fig:I2_SP49_P} $\&$ \ref{fig:I3_SP49_P} respectively) and are largely insignificant noise generators throughout the low frequency spectrum when the true non-parallel flow Green's function 
is used to determine it (see Fig. \ref{SPL_n2n3vary}). 
This is basically evident from the pre-factors in (\ref{G11_41}) $\&$ (\ref{G11_41b}), which show that both temperature-related terms are $O(\epsilon)$ and therefore their inclusion is not legitimately warranted at the lowest order expansion of the propagator (\ref{Prop_Exp}) and acoustic spectrum (\ref{I_low2}). 

The reason why a heated jet is quieter at fixed acoustic Mach number can be explained by the spatial localization and reduction in magnitude of the Favre-averaged turbulent kinetic energy ($k$) of SP$49$ in Fig. \ref{fig5_10b_TKE_SP49} compared to SP$90$ in Fig. \ref{fig5_10a_TKESP90}. The lower $k$ reduces the amplitude of $R_{1212}$ in (\ref{eq:R1212_model}) and, therefore, the acoustic spectrum (\ref{I_low2}) by a reduction in the spectral tensor component, $\Phi{}^*_{1212}$, in (\ref{eq:SpecPhi1212}).
Given that the Fluent simulations were run at the same turbulence intensity for both heated and isothermal jets, a physical explanation for this localization of $k$ at $TR>1$ is due to the heated jet carrying lower momentum owing to its reduced density by the equation of state.
In a sense, our results are consistent with the measurements of Ecker {\it et al}.\cite{Stuber} and Stuber {\it et al}.\cite{Stuber} whose experiments indicate that jet heating results in a reduction in convective amplification of noise sources contained within the turbulence. This can only come about through a reduction in the magnitude of the momentum flux term using our acoustic analogy model, (\ref{I_low2}).

Our results also show that the formula for the acoustic spectrum (\ref{I_approx}) remains accurate across most of $St$ range in Figs. \ref{fig5_11d_SP49theta=25} -- \ref{fig5_11f_SP49theta=35} $\&$ \ref{SPL_n2n3vary}, beyond the peak frequency.
This is probably because the non-parallel flow-based Green's function solution to (\ref{Hyp3}) prevents the formation of a critical layer at supersonic speeds (present in the locally parallel flow case), which ensures the amplification in propagator term (\ref{G_12}) is much greater than it would be in a subsonic flow (see Fig. 21 in GSA).

Further tests on the limit of applicability on the asymptotic theory we have developed in this paper can be assessed by considering what happens when we extend the range of prediction of (\ref{I_low2}) to higher polar angles ($44^\circ\leq\theta \leq 60^\circ$) with the turbulence scales kept the same as the peak sound predictions of Fig. \ref{fig5_11_SPLpreds}.
We find that (\ref{I_approx})
over predicts the acoustic data as $\theta$ increases. 
This is not surprising since the asymptotic theory 
applies only in the prediction region centered at ($St,\theta \approx 0.2, 30^\circ$).

\begin{figure}
  \centering
    \begin{subfigure}[h]{0.48\textwidth}
        \centering
        \includegraphics[width=\textwidth]{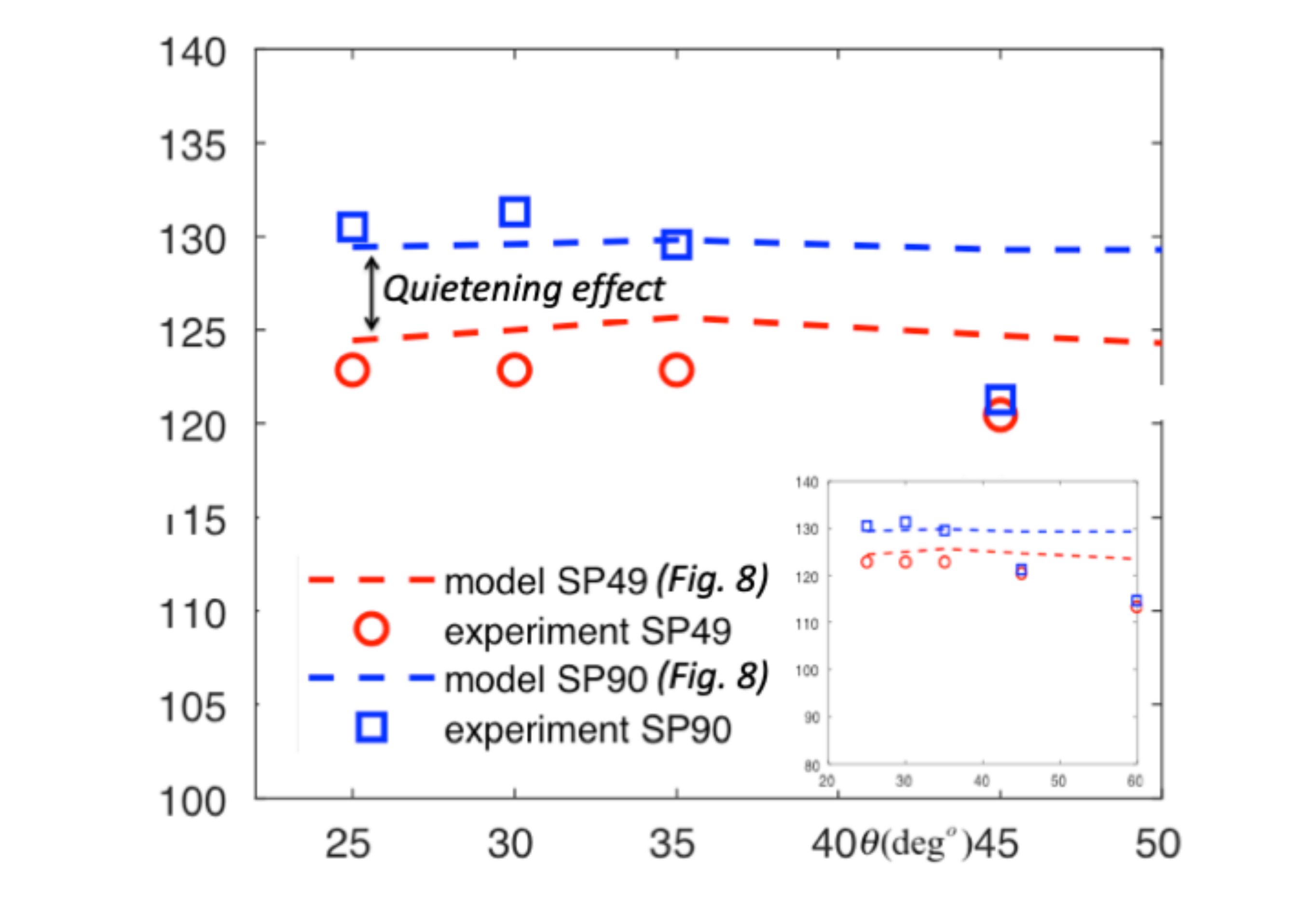}
        %
        \caption{ }
        \label{OASPL_Fig10values}
    \end{subfigure}
    \hfill
      \centering
    \begin{subfigure}[h]{0.48\textwidth}
        \centering
        \includegraphics[width=\textwidth]{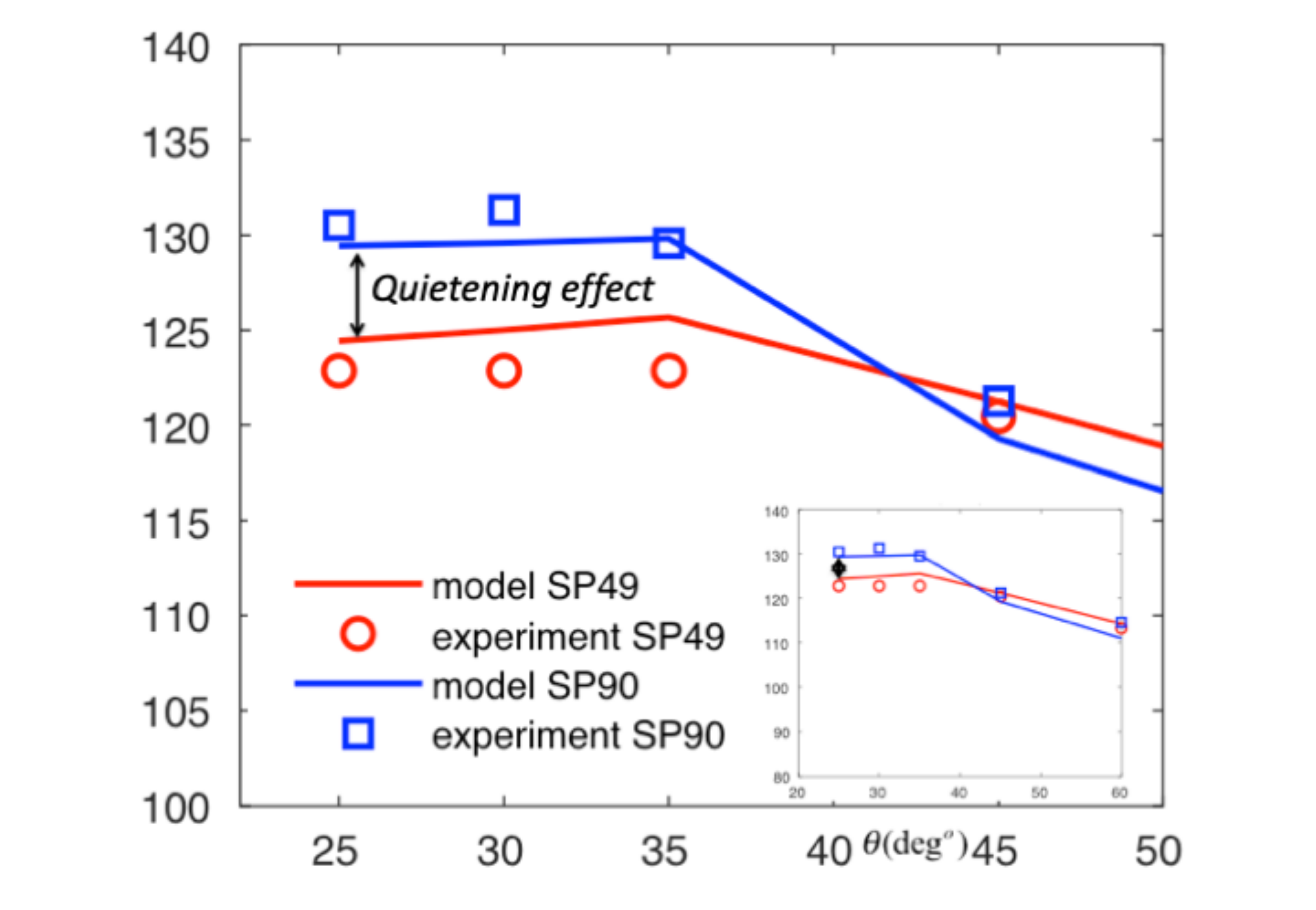}
        \caption{}
        \label{fig5_14_OASPL_cperptuned}
    \end{subfigure}
    \hfill
           \caption{Overall sound pressure level, OASPL (dB) 
         over frequencies $0.01<St<0.6$. See Table (\ref{Table_params}) for turbulence parameters in (\ref{eq:SpecPhi1212_A4}) and caption to Fig. \ref{fig5_11_SPLpreds} for $(n_2, n_3)$.  
        (a). Fig. (\ref{fig5_11_SPLpreds}) values for $c_\perp$;
        (b). $c_\perp$ tuned at each $\theta$.
           }
    \label{fig5_14_OASPL}
\end{figure}

The integrated effect of this over frequencies covering the peak sound regime ($0.01<St<0.6$) shown in Fig. \ref{fig5_14_OASPL} displays a similar conclusion. Namely that the peak sound predictions of Fig. \ref{fig5_11_SPLpreds} give accurate $OASPL$ values compared to the data between $25^\circ<\theta<35^\circ$, keeping $c_\perp$ fixed. Thereafter, (\ref{I_approx}) yields an over prediction for the $OASPL$. 
But since 
the predicted spectral shape remains a good match to the acoustic data (even though the amplitude is over predicted), one way to bring about an appropriate reduction in sound is to reduce the transverse length scale parameter  $c_\perp$ (defined by \ref{Eq:lengthscales}), given that the latter enters as the prefactor  $c{}_\perp^2$ in formula (\ref{eq:SpecPhi1212_A4}).
When doing this, the $OASPL$ predictions in Fig. \ref{fig5_14_OASPL_cperptuned} remain accurate for larger $\theta$ but this, of course, introduces more empiricism into the model.

The terms that we have neglected in deriving (\ref{I_low2}) involved the auto-covariance components (${R}_{4221}$, ${R}_{4242}$) $({\boldsymbol y}, {\boldsymbol \eta}; \tau)$. Gryazev {\it et al} \cite{Gryazev2019} showed these terms are negligible for a heated supersonic co-axial jet.
While Gryazev {\it et al}.\cite{Gryazev2019} found that {\it all} temperature-related correlations in $R_{\lambda j \mu l}$ were small, we retained the streamwise components  $R_{4 111}$ and $R_{4 141}$ to firstly, compare against AGF's locally parallel flow analysis and secondly, to assess the sensitivity of the acoustic spectrum to these terms. 
Note from Table \ref{Table_asym} the propagator terms associated with ${R}_{4221}$, ${R}_{4242}$ in the acoustic spectrum can be as large as $O(1)$, which means their outright exclusion must be more carefully assessed. 
%
However, we can bound the size of $|\bar{\Gamma}_{42}|$ in the propagator of these terms compared to $\bar{G}_{12}$ using (\ref{eq:Prop}) and the definition of ${\bar{\nu}}$ below (\ref{Hyp3}). That is, $\bar{\Gamma}_{42} = \partial \bar{G}_4/\partial r \approx (1/c{}_\infty^2) \partial \bar{\nu}/\partial r$ and, therefore, by the chain rule $|\bar{\Gamma}_{42}|\approx |(1/c{}_\infty^2)(\partial \bar{\nu}/\partial U) \partial U/\partial r|$, which  for a locally parallel flow will be less directive than $\bar{G}_{12}$ at low frequencies. 
This is clear using (7.1) $\&$ (7.2) in GSA, which show that $|\bar{\Gamma}_{42}|^2\sim \cos^2\theta /(1-U\cos\theta/c_\infty)^4$ whereas $|\bar{G}_{12}|^2\sim \cos^4\theta /(1-U\cos\theta/c_\infty)^6$. Just as the outer boundary conditions (\ref{BC1}) $\&$ (\ref{BC2}) determine the structure of the solution, $\bar{\nu}(Y,U)$, in (\ref{Hyp3}), the outer limit of any inner expansion of $\bar{\Gamma}_{42}(Y, U)$ and $\bar{\Gamma}_{12}(Y, U)$ must match onto the far-field $(r\rightarrow\infty)$ form of the above parallel flow results. Hence we can expect that the $\Omega =O(1)$ non-parallel solution to 
$|\bar{\Gamma}_{42}|^2$ should be less directive than $|\bar{G}_{12}|^2$.
So while the propagators associated with ${R}_{4221}$, ${R}_{4242}$ might be large as $O(1)$ in the acoustic spectrum, they are expected to be smaller in magnitude and less directive than the momentum flux term, $|\bar{G}_{12}|$, in (\ref{I_low2}).

\section{Conclusions}
\label{S:7}

Our main contributions in this paper involved extending the asymptotic theory in Ref. (\onlinecite{GSA12}, referred to here as GSA) 
to heated flows and using it within an acoustic analogy prediction model to explain, among other things, the observed spectral quietening of heated jets at supersonic acoustic Mach numbers, $Ma$.
We found that for an arbitrary axisymmetric jet flow with $O(1)$ spread rate, $\epsilon$, the adjoint linearized Euler equations (ALEE) in (\ref{eq:GAA}), can be transformed by taking $(y_1, U)$ as the two independent variables of choice.
Since the flow is heated, the Favre-averaged speed of sound ($\widetilde{c^2}$) in this case, satisfies the Crocco-Busseman relation, (\ref{CB_reln}).
The transformation results in the mixed partial differential equation given by (\ref{Hyp2}), where a hyperbolic operator $\mathcal{L}(y_1, U)$ appearing on the left hand side acts on combined Green's function variable, $\tilde{\nu}= \tilde{\nu}
(y_1, U) \equiv \widetilde{c^2} \tilde{G}_4 + \tilde{G}_5$.
As seen by using (\ref{S_r}), (\ref{S_funcs}), (\ref{S_1}) and also line below (\ref{nu_eqn}), the right hand side of (\ref{Hyp2}), $\mathcal{F}(\tilde{\boldsymbol S})$, couples the  $\tilde{\nu}$ solution  with the other components of the vector Green's function namely, $(\tilde{G}_4,\tilde{G}_r, \tilde{G}_\psi)$.
%
Here, ${G_\sigma} ({\boldsymbol y}| {\boldsymbol x}; \omega)$ is the Fourier transform, of the 5-dimensional adjoint vector Green's function ${g}{}_{\sigma4}^{a} ({\boldsymbol y}, \tau| {\boldsymbol x}, t)$ of the linearized Euler equations (Eq. 2.13 in Goldstein\cite{Gold03}, and  Eqs. 3.1--3.3 of Goldstein $\&$ Leib\cite{GanL}) where $\sigma = 1,...,5$.
But since high Reynolds number jets diverge slowly, i.e. $\epsilon \ll O(1)$, 
$\mathcal{F}(\tilde{\boldsymbol S})$ will always drop out of (\ref{Hyp2}) when 
non-parallel flow effects enter the lowest order Green's function solution, which Karabasov {\it et al}'s (2013) numerical simulations at high jet speeds indicate that they must. 
Therefore, ${g}{}_{\sigma4}^{a}
({\boldsymbol y}, \tau| {\boldsymbol x}, t)$ must evolve through a slow $O(1)$ time-scale, $\tilde{T}=\epsilon\tau$ in the low frequency regime.
Our analysis goes onto show that the richest dominant balance for $(\tilde{G}_4,\tilde{G}_r, \tilde{G}_\psi)$ (given by the Fourier transform of Eqs. 5.5 $\&$ 5.6 in GSA) will ensure that $\mathcal{F}(\tilde{\boldsymbol S})=0$. Interestingly, the inner equation for $\tilde{\nu}(Y,U)$ in (\ref{Hyp3}) at $r=\sqrt{y{}_2^2 + y{}_3^2} =O(1)$ distances from the jet center line,  is the same as that found by GSA in isothermal flows (their Eq. 5.31).
But (\ref{Hyp3}) now applies to heated jets where $TR\geq1$.

We used this extended asymptotic theory for the Green's function to determine the propagator, $\Gamma_{\lambda, j} ({\boldsymbol y}| {\boldsymbol x}; \omega)$, in the acoustic spectrum formula (\ref{I_low2}) at the  SP$90$ ($Ma=1.5$, $TR=1.0$) and SP$49$ ($Ma=1.5$, $TR=2.7$) set points that possesses a measureable region of low frequency spectral quietening (Tanna\cite{Tanna77}; Bridges\cite{Bridges06}; Bodony and Lele\cite{BodLel}).  
The acoustic spectrum model, (\ref{I_low2}), required only a single component of the generalized auto-covariance tensor ${R}_{\lambda j \mu l} ({\boldsymbol y}, {\boldsymbol \eta}; \tau)$ , ${R}_{1212}
$, to be modeled when the other streamwise momentum/temperature fluctuation-related components in (\ref{I_low2}) (${R}_{4111}
$ and ${R}_{4141}$) were assumed, as a first step, to be proportional to it.
We validated the model for ${R}_{1212}
({\boldsymbol y}, {\boldsymbol \eta}; \tau)$, given by (\ref{eq:R1212_model}), against LES data in 
Figs. \ref{fig_R1212_SP90} $\&$ \ref{fig_R1212_SP49} for jets that should have more-or-less the same space-time de-correlation.
This allowed determination of all relevant turbulence scales in the model function (\ref{eq:R1212_model}) apart from transverse correlation length scale. Although the latter was hand-tuned to give the correct level in the acoustic predictions in Fig. \ref{fig5_11_SPLpreds},  it was found to be an order of magnitude smaller than the streamwise length scale (see Table \ref{Table_params}).
This is consistent with axisymmetric turbulence approximation used in the paper and verified experimentally and computationally by, among others, Pokora $\&$ McGuirk (2015) and Karabasov {\it et al}. (2010) respectively.

The acoustic spectrum model (\ref{I_low2}) was derived for $\omega\ll O(\epsilon)$, but the predictions in Fig. \ref{fig5_11_SPLpreds} show excellent agreement over a wider frequency range (i.e. at $\omega \approx O(1)$) than the assumptions used to derive it. 

The sensitivity analysis in Fig. \ref{SPL_n2n3vary} indicates that both the (enthalpy flux/momentum flux) coupling term and enthalpy flux co-variance in (\ref{I_low2})
are largely silent at frequencies $St<0.8$ for the SP$49$ heated jet. 
Consequently, a Green's function based on locally parallel flow model will incorrectly estimate both the coupling and enthalpy flux propagators (\ref{G11_41}) $\&$ (\ref{G11_41b}) or result in an unbounded result due to the presence of a critical layer in the momentum flux propagator, (\ref{G_12}). 
%
We find that the inclusion of non-parallel flow effects does not allow propagator associated with the momentum/enthalpy flux coupling term to change sign at small observation angles. That is, the AGF conclusion of sign-change emerges only for a parallel mean flow using the low frequency asymptotic properties of the adjoint Rayleigh equation (i.e. equivalent to letting $\bar{X}_1 = 0$ in \ref{Hyp3}). 
As a result of this surprising discovery, we obtain an alternative explanation for the quietening of a supersonic heated jet that is more consistent with experimental data reported in Ecker {\it et al}.\cite{Ecker15}.
Namely that the reduction in sound with heating at fixed supersonic $Ma$ is due to the weakness of the momentum flux term in the acoustic spectrum, (\ref{I_approx}), as $TR$ increases. This conclusion emerges only when  the (true) non-parallel mean flow based Green's function is used to calculate the propagators in (\ref{I_approx}).

\vspace{-10pt}
\begin{acknowledgments}
Computational resources from HPC2, Mississippi State University, are appreciated.
MZA would like to thank Strathclyde University for financial support from the Chancellor's Fellowship.
We would also like to thank Dr. S. J. Leib (Ohio Aerospace Institute) for providing us with his spectral tensor routines.
\end{acknowledgments}

\appendix

\section{Generalization of the axisymmetric representation of $\Phi{}^*_{4jkl}$ for mirror-symmetry breaking}

\label{App:A}
AGF developed a model (Eq. C.5) for the axisymmetric representation of the spectral tensor associated with the coupling term, 
$\Phi{}^*_{4jkl}
({\boldsymbol y}, k_1, k{}_T^2 ; \omega)$, using Batchelor's\cite{Batchelor} (Eq. 3.3.10) formula of a rank-3 tensor that depends on two independent vectors $(\boldsymbol{k}, \boldsymbol{\lambda})$.
Batchelor's formulae in homogeneous axisymmetric turbulence (p.43 in Ref. \onlinecite{Batchelor}) can be used in the non-homogeneous setting of the jet flow problem because the field point, $\boldsymbol{y}$, is treated as being fixed for the stationary random functions $e{}_{\lambda j} 
$ and $e{}_{\mu l} 
$ in (\ref{eq:Rijkl}). Hence, the tensor ${R}_{\lambda j \mu l}
({\boldsymbol y}, {\boldsymbol \eta}; \tau)$ depends only on $\boldsymbol \eta$ with time delay $\tau$ acting as a parameter. In other words, as stated in Ref. \onlinecite{Afs2012}, the kinematic modeling  of  ${R}_{\lambda j \mu l}
({\boldsymbol y}, {\boldsymbol \eta}; \tau)$ is a locally-homogeneous field problem. 
In terms of the spectral tensor, at a given fixed field point $\boldsymbol{y}$, $\Phi{}^*_{4jkl}
({\boldsymbol y}, k_1, k{}_T^2 ; \omega)$ is a function of the wave number vector, $\boldsymbol{k}$ (defined by \ref{eq:Spec_Ten}), and $\boldsymbol{\lambda}$, which is a unit vector indicating the direction of symmetry (i.e. $\boldsymbol{\lambda} = \boldsymbol{e}_1$ in this case).

The axisymmetric model of $\Phi{}^*_{4jkl}
({\boldsymbol y}, k_1, k{}_T^2 ; \omega)$ is derived by determining all the basic invariants that can be formed from these two vector arguments and the unit tensor $\delta_{jk}$ in suffixes $(j,k)$ when the three-form $\Phi{}^*_{4jkl} a_j b_k c_l$ remains invariant to the full rotation group about the streamwise direction. 
But a tensor of odd suffixes (or parity) such as $\Phi{}^*_{4jkl}$  cannot remain invariant to improper rotations such as reflections of the transverse $(y_2-y_3)$ co-ordinate plane through the streamwise direction $y_1$ (Monin and Yaglom\cite{M&Y}, p.42; Batchelor\cite{Batchelor}, p.43 and p.2525 of AGF).
(This is sometimes referred to as parity breakage\cite{M&Y}).
Hence a more general representation requires that the three-form $\Phi{}^*_{4jkl} a_j b_k c_l$  is invariant to proper rotations only 
about $k_1$ axis with respect to the vector configuration formed by two field points: $\boldsymbol{y}$ and  $\boldsymbol{y}+\boldsymbol{k}$ separated by ${\boldsymbol k}$ with $\boldsymbol{\lambda}={\boldsymbol e}_1$ being the principal direction of symmetry (see Fig. A.1 in Ref. \onlinecite{Afs2012}).

The spectral tensor (\ref{eq:Spec_Ten}) is usually assumed to be a weak function of $\boldsymbol{k}_\perp$; this follows by Watson's lemma because the physical space tensor ${R}_{\lambda j \mu l}
({\boldsymbol y}, {\boldsymbol \eta}; \tau)$ that enters the Fourier transform integral, (\ref{eq:Spec_Ten}), through (\ref{eq:HFT}) and linear relation below (\ref{eq:Rijkl}), is a rapidly varying function of $|\boldsymbol{\eta}_\perp|$\cite{Pokora, M&Z}. 
%
In the limit of the infinitely long streamwise eddy,
$k_\perp =|\boldsymbol{k}_\perp|\rightarrow 0$ and $k_i = \delta_{i1} k_1$. But, as first argued by Afsar {\it et al}.\cite{Afsetal2010}, the irreducible invariant representation of $\Phi{}^*_{\lambda j \mu l}
({\boldsymbol y}, k_{_1},  k{}_{_\perp}^2 ; \omega)$ is identical to that of $\Phi{}^*_{\lambda j \mu l}
({\boldsymbol y}, k_{_1},  0 ; \omega)$ since $k{}_{_\perp}$ is a scalar argument.  When mirror invariance in $\Phi{}^*_{4jkl}$ is broken, on the other hand, a large number of additional basic invariants can be formed on top of those already contained in Batchelor (\cite{Batchelor}, Eq. $3.3.10$) for a rank-3 axisymmetric tensor.  More explicitly, 
\begin{equation}
\label{eq:Grassman}
\begin{split}
%
\Phi{}^*_{4jkl}
({\boldsymbol y}, k_1, k{}_T^2 ; \omega)
&
=
\delta_{j1}\delta_{kl}
A_1
+
\delta_{k1} \delta_{jl}
A_2
+
\delta_{l1} \delta_{jk}
A_3
 \\
&
+
\delta_{j1} \delta_{k1} \delta_{l1}A_4
+
\widetilde{\Phi}{}^*_{4jkl}
\end{split}
\end{equation}
where $A_{1,2,3,4} = A_{1,2,3,4}({\boldsymbol y}, k^2 , k_1 ; \omega)$ are arbitrary scalars that depend on invariants $k^2 = {\boldsymbol k}. {\boldsymbol k} = k{}_1^2 + k{}_T^2$ and $k_1 = {\boldsymbol e}_1.{\boldsymbol k}$ and $\widetilde{\Phi}{}^*_{4jkl}=\widetilde{\Phi}{}^*_{4jkl}({\boldsymbol y}, k_1, k^2 ; \omega)$ is given by the sum of permutations of the Grassman products (i.e. skew symmetric terms, see p. 92 of Bishop $\&$ Goldberg\cite{BishGold}) of the form: 
\begin{equation}
\label{eq:Grassman2}
%
\widetilde{\Phi}{}^*_{4jkl}
({\boldsymbol y}, k_1, k{}_T^2 ; \omega)
=
\sum_{\textnormal{perm}(j,k,l)}
\tilde{\Psi}{}_{jkl}^{(p,m,n)}
A_{4+I(p,m,n)}
({\boldsymbol y}, k^2 , k_1 ; \omega)
\end{equation}
where $A_{4+I(p,m,n)}
({\boldsymbol y}, k^2 , k_1 ; \omega)$ is a scalar field corresponding to a basic invariant formed by appropriate tensor multiplication of unit alternating tensor (Levi-Civita symbol), $\epsilon_{jkl}$, with vectors $(k_i, \lambda_i)$, for the $n$th permutation in suffixes $(j,k,l)$.

The notation $perm(j,k,l)$ denotes permutation over all possible combination of tensor suffixes $(j,k,l)$ where the function $I(p,m,n)$ is a mapping of a three-dimensional subspace $p= 1,2,3,...,P; m,n = 1,2,3$ of positive integers where the index, $p$, individuates the unique permutation of $perm(j,k,l)$ with $P$ being the total number of individual permutations possible.
That is (\ref{eq:Grassman2}) will include terms such as,
\begin{equation}
\label{eq:Grassman3}
\begin{split}
%
\widetilde{\Phi}{}^*_{4jkl}
&
=
A_5 \epsilon_{jkl}
+
A_6 \epsilon_{jkp} \epsilon_{lpq}k_q
+
A_7 \epsilon_{jkp} \epsilon_{lpq}\lambda_q
+
A_8 \epsilon_{jpq} \epsilon_{kpq}k_l
 \\
&
+
A_9 \epsilon_{jpq} \epsilon_{kpq}\lambda_l
+ 
A_{10} \epsilon_{jkq} k_q k_l +...+A_{13} \epsilon_{jkq} \lambda_q \lambda_l 
\\
&
+A_{14} \epsilon_{jpq} \epsilon_{jpq} k_j k_k k_l +...+A_{21} \epsilon_{jpq} \epsilon_{jpq} \lambda_j \lambda_k \lambda_l
\\
&
+A_{22} \epsilon_{jpq} k_p k_q k_k k_l +...+A_{30} \epsilon_{jpq}  \lambda_p \lambda_q \lambda_k \lambda_l
+...
\end{split}
\end{equation}
and so on. 

However, all but the leading term ($\epsilon_{jkl}$) in (\ref{eq:Grassman3}) are zero because they involve basic invariants that either reduce to those contained in (\ref{eq:Grassman}) using properties of Levi-Civita symbol (i.e. involve an even number of reflections which correspond to a rotation of the vector configuration in Fig. A.1 of Afsar\cite{Afs2012}) or are zero after the turbulence is assumed to have weak transverse correlation inasmuch as $\Phi{}^*_{\lambda j \mu l}
({\boldsymbol y}, k_{_1},  \boldsymbol{k}{}_{_\perp} ; \omega)\approx\Phi{}^*_{\lambda j \mu l}
({\boldsymbol y}, k_{_1},  0 ; \omega)$ and $k_i = \delta_{i1} k_1$ and $\lambda_i = \delta_{i1}$.
Thus, $A_{6+I(p,m,n)}
=0$ for all permutations in (\ref{eq:Grassman2}) and expansion (\ref{eq:Grassman3}) and we find that (\ref{eq:Grassman}) reduces to:
\begin{equation}
\label{eq:Grassman4}
\begin{split}
%
\Phi{}^*_{4jkl}
({\boldsymbol y}, k_1, k{}_T^2 ; \omega)
&
=
\delta_{j1}\delta_{kl}
A_1
+
\delta_{k1} \delta_{jl}
A_2
+
\delta_{l1} \delta_{jk}
A_3
 \\
&
+
\delta_{j1} \delta_{k1} \delta_{l1}
A_4 + \epsilon_{jkl}  A_5
\end{split}
\end{equation}

\section{Calculation of the spectral tensor component, (\ref{eq:SpecPhi1212})}
\label{App:B}
Inserting Eq. (\ref{eq:R1212_model}) into (\ref{eq:SpecPhi1212}) and re-writing terms algebraic in $(\tau,\eta_1)$ gives:
\begin{equation}
\label{eq:SpecPhi1212_p1}
 \begin{split}
\frac{2\pi \Phi{}^*_{1212}
({\boldsymbol y}, k_1, k{}_T^2 ; \omega)}{{R}_{1212}
({\boldsymbol y},{\boldsymbol 0}, 0)}
&
=
\int
\limits_{V_\infty({\boldsymbol \eta})}
\int\limits_{-\infty}^{\infty}
e^{i(\boldsymbol{k}.\boldsymbol{\eta} - \omega\tau)}
\biggl[
a_0
- 
\frac{a_1}{i}
\frac{\partial}{\partial \omega}
\frac{\partial}{\partial \tau}
+
\\
&
+
\frac{a_2}{i}
\frac{\partial}{\partial k_1}
\frac{\partial}{\partial \eta_1}
+
...
\biggr{]}
e^{\alpha-X(\eta_1, \eta_T, \tau)}
\,d\tau
\,d{\boldsymbol \eta}
\end{split}
\end{equation}
Introducing non-dimensional variables of integration: $\tilde{\eta}_1 = {\eta_i}/l_i$ ($l_i = (l_1, l_2, l_3)$) and $\tilde{\tau} = U_c\tau/l_0$ shows that the spectral function $X(\eta_1, \eta_T, \tau)$ can be written as $X(\eta_1, \eta_T, \tau) = \sqrt{\alpha^2+ \tilde{\eta}{}_1^2 + \tilde{\xi}^2+ f(\tilde{\eta}_T)}$ where $\tilde{\xi} = (\tilde{\eta}_1 l_1/l_0 - \tilde{\tau})$.
Integrating each term by parts gives and inserting this latter expression into (\ref{eq:SpecPhi1212_p1}) gives
\begin{equation}
\label{eq:SpecPhi1212_p2}
 \begin{split}
%
\frac{\Phi{}^*_{1212}
({\boldsymbol y}, k_1, k{}_T^2 ; \omega)}{{R}_{1212}
({\boldsymbol y},{\boldsymbol 0}, 0)}
=
&
\frac{l_0 l_1 l_2 l_3}{U_c}
\biggl{[}
(a_0 - a_1 - a_2)
- 
a_1 \tilde{\omega}
\frac{\partial}{\partial \tilde{ \omega}}
\\
&
-
a_2
\bar{k}_1
\frac{\partial}{\partial \bar{k}_1}
+
...
\biggr{]}
\Phi
({\boldsymbol y}, \tilde{\boldsymbol{k}} ; \omega)
\end{split}
\end{equation}
in which the unit-spectrum of turbulence, $\Phi
({\boldsymbol y}, \tilde{k}_1, \tilde{k}{}_T^2 ; \omega, \alpha)$, is defined by
\begin{equation}
\label{eq:SpecPhi_p3}
 \begin{split}
&
2\pi \Phi
({\boldsymbol y}, \tilde{k}_1, \tilde{k}{}_T^2 ; \omega, \alpha)
=
\int
\limits_{\boldsymbol{\eta}_T}
\,d\boldsymbol{\eta}_T
e^{i\tilde{{\boldsymbol k}}_T\tilde{\boldsymbol \eta}_T}
\int\limits_{-\infty}^{\infty}
\,d\tilde{\eta}_1
e^{i \bar{k}_1\tilde{\eta}_1}
\times
\\
&
\int\limits_{-\infty}^{\infty}
e^{i \tilde{\omega}\tilde{\xi}}
e^{\alpha - \sqrt{\alpha^2 + {\eta{}_1^2}/{l{}_1^2} + {(\eta_1 - U_c\tau)^2}/{l{}_0^2} + f(|\tilde{\eta}_T|) }}
\,d{\tilde{\xi}},
\end{split}
\end{equation}
where $\bar{k}_1 = \tilde{k}_1 - \tilde{\omega}(l_1/l_0)$, $\tilde{\omega} = \omega l_0/U_c$  and $\tilde{k}_i = l_i k_i$ (no sum on suffix $i = (1,2,3)$). Note that the square-brackets in (\ref{eq:SpecPhi1212_p2}) expand in the manner shown by G $\&$ L\cite{GanL} (Eq. 6.31): $a_0-a_1(1+\omega\partial/\partial\omega+...) - a_2(1+k_1\partial/\partial k_1+...)$ where further terms of type $a_n \omega^n \partial/\partial \omega_n$ and $a_m k{}_1^m \partial/\partial k{}_m$ (positive integer $(m,n)>1$) represent the oscillations of ${R}_{1212}
({\boldsymbol y},\eta_1, \eta_T, \tau)$ at $\tau = O(1)$.
Convergence of the space-time series (\ref{eq:R1212_model}) in  $\tau$ and/or $\boldsymbol{\eta} = (\eta_1, \eta_T)$ is guaranteed since each term will decay exponentially like, $e^{-X(\eta_1,\eta_T,\tau )}$, when $\tau\rightarrow \infty$ and   $|\boldsymbol{\eta}|=O(1)$ and vice versa. The spectral tensor component $\Phi_{1212}
({\boldsymbol y}, k_1, k{}_T^2 ; \omega)$, (\ref{eq:SpecPhi1212_p2}), on the other hand decays algebraically (see \ref{eq:SpecPhi1212_A4}) as $\tilde{\omega}^{-4}$ at infinity (which by the Abelian theorem in the theory of Fourier transforms, corresponds to the cusp of the auto-correlation of ${R}_{1212}
$ at $\tau = 0$ and $\boldsymbol{\eta}=O(1)$).

The $\tilde{\xi}-$integral can now be performed using result $\# 867$ in Campbell $\&$ Forster\cite{C&F}; thus giving:
\begin{equation}
\label{eq:SpecPhi_p4}
 \begin{split}
&
e^{-\alpha}
\Phi
({\boldsymbol y}, \tilde{k}_1, \tilde{k}{}_T^2 ; \omega, \alpha)
=
\int
\limits_{\boldsymbol{\eta}_T}
\,d\boldsymbol{\eta}_T
e^{i\tilde{{\boldsymbol k}}_T\tilde{\boldsymbol \eta}_T}
\int\limits_{-\infty}^{\infty}
\,d\tilde{\eta}_1
e^{i \bar{k}_1\tilde{\eta}_1}
\times
\\
&
\frac{\sqrt{\tilde{\eta}{}_1^2 +\alpha^2 + f(|\tilde{\eta}_T|)} }{\pi(\tilde{\omega}^2 + 1)^{1/2}}
K_1\left[ (\tilde{\omega}^2 + 1)^{1/2} \sqrt{\tilde{\eta}{}_1^2 +\alpha^2 + f(|\tilde{\eta}_T|)}
\right]
,
\end{split}
\end{equation}
where, $K_1[...]$ is the modified Bessel function of the second kind. Similar to above, the $\tilde{\eta}_1 -$integral is given by $2\pi$ times result $\# 917.8$ in Campbell $\&$ Forster\cite{C&F}. Whence:
\begin{equation}
\label{eq:SpecPhi_p5}
 \begin{split}
&
e^{-\alpha}
\Phi
({\boldsymbol y}, \tilde{k}_1, \tilde{k}{}_T^2 ; \omega, \alpha)
=
\int
\limits_{\boldsymbol{\eta}_T}
\,d\boldsymbol{\eta}_T
e^{i\tilde{{\boldsymbol k}}_T\tilde{\boldsymbol \eta}_T}
\times
\\
&
\frac{\left[1+ \chi^{1/2}\sqrt{f(|\tilde{\eta}_T|) +\alpha^2} \right]}{\chi^{3/2}}
e^{-\chi^{1/2} \sqrt{f(|\tilde{\eta}_T|) +\alpha^2} }
,
\end{split}
\end{equation}
where
\begin{equation}
\label{eq:Chi}
 \begin{split}
\chi
\equiv
\chi
(\bar{k}_1,\tilde{\omega})
=
\bar{k}{}_1^2
+
\tilde{\omega}^2
+
1
=
(\tilde{k}_1-(l_1/l_0)\tilde{\omega})^2 
+
\tilde{\omega}^2
+
1.
\end{split}
\end{equation}
Using the Leibnitz rule, we can re-write (\ref{eq:SpecPhi_p5}) into the remarkably simple result,
\begin{equation}
\label{eq:SpecPhi_p6}
 \begin{split}
-
\frac{e^{-\alpha}\Phi
({\boldsymbol y}, \tilde{k}_1, \tilde{k}{}_T^2 ; \omega, \alpha)}{2}
=
&
\frac{\partial}{\partial \chi}
\biggl{\{}
\frac{1}{\chi^{1/2}}
\\
&
\int
\limits_{\boldsymbol{\eta}_T}
\,d\boldsymbol{\tilde{\eta}}_T
e^{i\tilde{{\boldsymbol k}}_T\tilde{\boldsymbol \eta}_T}
e^{-\chi^{1/2} \sqrt{f(|\tilde{\eta}_T|) +\alpha^2} }
\biggr{\}}
,
\end{split}
\end{equation}
Changing variables to a cylindrical-polar co-ordinate system in which $\boldsymbol{\tilde{\eta}}_T = (\tilde{\eta}_T \cos\phi,\tilde{\eta}_T \sin\phi)$  where $\tilde{\eta}_T = |\tilde{\eta}_T|$ shows:
\begin{equation}
\label{eq:SpecPhi_p7}
\begin{split}
-\frac{e^{-\alpha} \Phi
({\boldsymbol y}, \tilde{k}_1, \tilde{k}{}_T^2 ; \omega, \alpha)}{4 \pi}
&
=
\frac{\partial}{\partial \chi}
\frac{1}{\chi^{1/2}}
\biggl{\{}
\\
&
\int
\limits_{0}^{\infty}
e^{-\chi^{1/2} \sqrt{f(|\tilde{\eta}_T|) +\alpha^2} }
\tilde{\eta}_T
J_0(\tilde{k}_T \tilde{\eta}_T)
\,d\tilde{\eta}_T
\biggr{\}}
,
\end{split}
\end{equation}
since $\int\limits_0^{2\pi} \,d\phi e^{i\tilde{k}_T \tilde{\eta}_T \cos\phi} = 2\pi J_0(\tilde{k}_T \tilde{\eta}_T)$ and where, $\tilde{k}_T = |\boldsymbol{\tilde{k}}_T|$ and $\boldsymbol{\tilde{k}}_T = \boldsymbol{l}_T \boldsymbol{k}_T = (l_2 k_2, l_3 k_3)$ (no sum on suffix $T = (2,3)$ here). But since this last integral in (\ref{eq:SpecPhi_p7}) is just the Hankel transform of
$F(\tilde{\eta}_T; \chi, \alpha ) = e^{-\chi^{1/2}\sqrt{f(\tilde{\eta}_T)+ \alpha^2 }}$, (\ref{eq:SpecPhi_p7}) takes the form of a purely algebraic formula, 
\begin{equation}
\label{eq:SpecPhi_p8}
%
\Phi
({\boldsymbol y}, \tilde{k}_1, \tilde{k}{}_T^2 ; \omega, \alpha)
=
-4 \pi
e^\alpha
\frac{\partial}{\partial \chi}
\frac{\mathcal{H}\left[F(\tilde{\eta}_T; \chi, \alpha )\right]}{\chi^{1/2}}
,
\end{equation}
in which the Hankel transform is defined by:
\begin{equation}
\label{eq:SpecPhi_p9}
%
\mathcal{H}\left[F(\tilde{\eta}_T; \chi, \alpha )\right]
=
\int
\limits_{0}^{\infty}
e^{-\chi^{1/2} \sqrt{f(|\tilde{\eta}_T|) +\alpha^2} }
\tilde{\eta}_T
J_0(\tilde{k}_T \tilde{\eta}_T)
\,d\tilde{\eta}_T
,
\end{equation}

The function $f (\tilde{\eta}_T )$, taken as a polynomial in $\tilde{\eta}_T$, such as: $f (\tilde{\eta}_T )\sim a\tilde{\eta}{}_T^m + b \tilde{\eta}{}_T^{m-1} + c\tilde{\eta}{}_T^{m-2} + ... $ (for positive integer $m$ ), ensures convergence of (\ref{eq:SpecPhi_p9}) since $\tilde{\eta}_T J_0(\tilde{k}_T\tilde{\eta}_T ) \rightarrow \tilde{\eta}{}_T^{1/2} \cos(\tilde{k}_T\tilde{\eta}_T )$ as $\tilde{\eta}_T \rightarrow \infty$. 
Eq. (\ref{eq:SpecPhi_p9}) can, therefore, be integrated numerically for any positive integral values of $m$ and inserted into (\ref{eq:SpecPhi_p8}).
This was partially demonstrated by Bassetti {\it et al}.\cite{Bass2007} who took $f (\tilde{\eta}_T ) = \tilde{\eta}{}_T^m$ and used values of $m$ between $2$ and $2.7$ in a model of the Reynolds
stress auto-covariance that involves only the first term of  (\ref{eq:R1212_model}) (i.e. without space-time anti-correlations). 
On the other hand, the Leib $\&$ Goldstein\cite{LG11} results indicate that taking $m = 4$ (as suggested by Harper-Bourne\cite{HB10}), and therefore letting $f (\tilde{\eta}_T ) \sim \tilde{\eta}{}_T^4$,
gives very good jet noise predictions across various subsonic and supersonic acoustic Mach numbers. 
They also showed that this model agreed with Harper-Bourne's measured data for ${R}_{1111}
({\boldsymbol y},\eta_1, \boldsymbol{\eta}_T, \tau)$ for a low Mach number jet ($Ma = 0.22$), which itself was remarkably close to an LES calculation of a round jet at $Ma = 0.75$ (see Fig. 6 in Karabasov {\it et al}.\cite{Karab2010}).
However, Figs. 4b $\&$ 4e in Semiltov {\it et al}.\cite{Semil2015} provide further evidence of fast decay of Reynolds stress auto-covariance with $\tilde{\eta}_T$ by showing that an exponential function in radial separation does not agree with either LES or Harper-Bourne's data over almost all radial locations and at two different frequencies.

Hence, inserting a transverse decay function of the form $f (\tilde{\eta}_T) = \tilde{\eta}{}_T^4 + 2\alpha\tilde{\eta}{}_T^2$ 
gives the following Hankel transform in the standard form of a zeroth-order Weber integral, 
\begin{equation}
\label{eq:Hank_eta4}
 \begin{split}
\mathcal{H}\left[F(\tilde{\eta}_{_T}; \chi, \alpha )\right]
=
&
e^{-\chi^{1/2} \alpha}
\int
\limits_{0}^{\infty}
e^{-\chi^{1/2} \tilde{\eta}{}_{_T}^2} 
\tilde{\eta}_{_T}
J_0(\tilde{k}_T \tilde{\eta}_T)
\,d\tilde{\eta}_T
\\
=
&
\frac{e^{-\chi^{1/2} \alpha}}{2\chi^{1/2}}
e^{-{\tilde{k}{}_{_T}^2}/{(4\chi^{1/2})}},
\end{split}
\end{equation}
after completing the square in argument of the exponential in $F(\tilde{\eta}_T; \chi, \alpha )$ defined above (\ref{eq:SpecPhi_p8}) and using the standard result in Lebedev\cite{Lebedev} (p.132). 
The unit spectrum, $\Phi
({\boldsymbol y}, \tilde{k}_1, \tilde{k}{}_T^2 ; \omega, \alpha)$, is found by substituting  
the above Hankel transform into (\ref{eq:SpecPhi_p8}); viz.: 
\begin{equation}
\label{eq:SpecPhi_eta4}
%
\Phi
({\boldsymbol y}, \tilde{k}_1, \tilde{k}{}_T^2 ; \omega, \alpha)
=
-2 \pi
e^\alpha
\frac{\partial}{\partial \chi}
\left\lbrace
\frac{e^{-\chi^{1/2}\alpha}}{\chi}
e^{-\tilde{k}{}_{_T}^2/(4 \chi^{1/2})}
\right\rbrace
,
\end{equation}
which reduces to $-2\pi \partial(e^{-\tilde{k}{}_{_T}^2/(4 \chi^{1/2})}/\chi)/\partial \chi$ at the cusp point $\alpha =0$ and finally to the algebraic formula, $-2\pi \partial (1/\chi)/\partial\chi$, when the transverse correlation lengths are relatively compact compared to that in the streamwise direction (i.e. $\tilde{k}_{_T} = 0$ in wave-number space).
Our calculations show that the acoustic prediction in the audible frequency range ($0.01<St<2.0$) obtained when ${k}{}_T=0$ is basically identical ($<0.25$dB) to that defined by taking ${k}{}_T\neq0$ for $\Phi{}^*_{1212}
$ in (\ref{eq:SpecPhi1212_A2}) $\&$ (\ref{eq:Spec_Pi_term}).

\section{\label{App:C} Algebraic formulae for $\Phi{}^*_{1212}
({\boldsymbol y}, k_1, k{}_T^2 ; \omega)$}
Since the unit turbulence spectrum $\Phi
({\boldsymbol y}, \tilde{k}_1, \tilde{k}{}_T^2 ; \omega, 0)$ depends on the streamwise wavenumber, $\tilde{k}_1$, through, $\bar{k}_1$, in the wavenumber-frequency function $\chi (\bar{k}_1,\tilde{\omega})$, (\ref{eq:Chi}), we re-write the independent variables  $(\tilde{k}_1,\tilde{\omega})$ in (\ref{eq:SpecPhi_eta4}) using the chain rule to transform the derivative $\partial/\partial\tilde{\omega}$ as
\begin{equation}
\label{eq:IVtrans}
%
\frac{\partial}{\partial \tilde{\omega}}
=
\left.\frac{\partial}{\partial \tilde{\omega}}
\right|_{\bar{k}_1}
+
\left(\frac{\partial \bar{k}_1}{\partial \tilde{\omega}}\right)
\left.\frac{\partial}{\partial\bar{k}_1}\right|_{\tilde{\omega}}
=
\frac{\partial}{\partial \tilde{\omega}}
-\left(
\frac{l_1}{l_0}
\right)
\frac{\partial}{\partial \bar{k}_1}
\end{equation}
so that $(\bar{k}_1,\tilde{\omega})$ are taken as independent variables where the modified wavenumber is $\bar{k}_1 = \tilde{k}_1 - \tilde{\omega}(l_1/l_0)$. Using (\ref{eq:IVtrans}) to re-write the derivative
with respect to $\tilde{k}_1$ in (\ref{eq:SpecPhi1212_p2}) and inserting the unit-spectrum $\Phi
({\boldsymbol y}, \tilde{\boldsymbol{k}} ; \omega)$, (\ref{eq:SpecPhi_eta4}), into this result shows that

\begin{equation}
\label{eq:SpecPhi1212_A1}
 \begin{split}
&
-\frac{
\Phi{}^*_{1212}
({\boldsymbol y}, k_1, k{}_T^2 ; \omega)}{2\pi{R}_{1212}
({\boldsymbol y},{\boldsymbol 0}, 0)}
=
\frac{l_0 l_1 l_2 l_3}{U_c}
\bigg{[}
(a_0 - a_1 - a_2)
\\
&
- 
a_1 \tilde{\omega}
\left(
\frac{\partial}{\partial \tilde{\omega}}
-
\frac{l_1}{l_0}
\frac{\partial}{\partial \bar{k}_1}
\right)
-
a_2
\left(\bar{k}_1
+\tilde{\omega}
\frac{l_1}{l_0}
\right)
\frac{\partial}{\partial \bar{k}_1}
+
...
\biggr{]}
\times
 \\
 &
 \frac{\partial}{\partial \chi}
 \left\lbrace
 \frac{e^{-\tilde{k}{}_{_T}^2/(4\chi^{1/2})}}{\chi}
 \right\rbrace
 \end{split}
\end{equation}
where  $\chi (\bar{k}_1,\tilde{\omega})$ is defined by (\ref{eq:Chi}).
%
The length scales in (\ref{eq:SpecPhi1212_A1}) are taken to be proportional to the local turbulent kinetic energy, $k(\boldsymbol{y})$, and the rate of energy dissipation, $\tilde{\epsilon}(\boldsymbol y)$ determined by RANS calculation through 
\begin{equation}
\label{Eq:lengthscales}
l_i = c_i (k^{3/2}/\tilde{\epsilon})(\boldsymbol y) 
\end{equation}
where suffix $i =(0,1,2,3)$ and $c_i$ are empirical parameters.

Performing the differentiation in (\ref{eq:SpecPhi1212_A1}) we obtain the final formula:
\begin{widetext}
\begin{eqnarray}
\label{eq:SpecPhi1212_A2}
%
&&-\frac{\Phi{}^*_{1212}
({\boldsymbol y}, k_1, k{}_T^2 ; \omega)}{2\pi
{R}_{1212}
({\boldsymbol y},{\boldsymbol 0}, 0)}  \\
&=& 
\frac{l_0 l_1 l_2 l_3}{U_c}
\biggr{[}
\frac{(a_0 - a_1 - a_2)}{\chi^2}
\left(
\frac{\tilde{k}{}_T^2}{8\chi^{1/2}}
-1
\right)
-
\big{(}
a_1 \tilde{\omega}^2
-
\bar{k}_1
(\tilde{\omega}
(a_1 
-
a_2
)l_1/l_0
-a_2
\bar{k}_1
)
\big{)}
\Pi(\tilde{k}_T, \chi)
\biggr{]}
e^{-\tilde{k}{}_T^2/(4\chi^{1/2})}   \nonumber
\end{eqnarray}
\end{widetext}
where $\Pi
(\tilde{k}{}_T, \chi)$
\begin{equation}
\label{eq:Spec_Pi_term}
%
\Pi
(\tilde{k}{}_T, \chi)
=
\frac{4}{\chi^3}
-
\frac{\tilde{k}{}_T^2}{4}
\left(
\frac{7}{2\chi^{7/2}}
-
\frac{\tilde{k}_T^2}{8 \chi^4}
\right)
\end{equation}
expands in powers of $\chi^{-(3+m)}$ and commensurately as powers of $\tilde{\omega}^{-2(m+3)}$ for $m=(0, 1/2,1)$ when $\tilde{\omega}\rightarrow \infty$. Thus, the spectral tensor component in (\ref{eq:SpecPhi1212_A1}) scales as a truncated power series in $\chi$: $\Phi{}^*_{1212}
({\boldsymbol y}, k_1, k{}_T^2 ; \omega)\sim \chi^{-(2+n)}$ and therefore is bounded as $\Phi{}^*_{1212} = O(\tilde{\omega}^{-2(n+2)})$ for $n=(0,1/2,1,3/2,2)$ at large frequencies.

When $\tilde{k}{}_T =0$ but $(a_1, a_2)\neq 0$, (\ref{eq:SpecPhi1212_A2}) reduces to the algebraic formula:
\begin{equation}
\label{eq:SpecPhi1212_A4}
 \begin{split}
&
\Phi{}^*_{1212}
({\boldsymbol y}, k_1, 0 ; \omega)
=
2\pi
{R}_{1212}
({\boldsymbol y},{\boldsymbol 0}, 0)
\frac{l_0 l_1 l{}_\perp^2}{\chi^2U_c}
\\
&
\left[
(1 - a_1 - a_2)
+
(a_1 \tilde{\omega}^2
-\bar{k}_1(\tilde{\omega}(a_1-a_2)l_1/l_0 -a_2\bar{k}_1))
\frac{4}{\chi}
\right]
 \end{split}
\end{equation}
where, consistent for an axisymmetric jet, we have put $ l_2 =l_3=l{}_\perp$ (or $c_2 =c_3 = c_\perp$ from \ref{Eq:lengthscales}) and $a_0=1$ so that ${R}_{1212}
({\boldsymbol y},0, 0, 0) /  {R}_{1212}
({\boldsymbol 0},\boldsymbol{0}, 0)= 1$. This model now depends on $5$ independent parameters that estimate the turbulence structure: that is, streamwise/temporal length scales, $(l_1, l_0)$; transverse length scale, $l_{\perp}$ and anti-correlation parameters: $(a_1, a_2)$. Note that (\ref{eq:SpecPhi1212_A4}) remains algebraically bounded at large frequencies $\tilde{\omega}$. This can be deduced by reducing the formula to the simple case where the anti-correlation (negative loops) region is negligible; i.e. at  $a_1=a_2=0$. 
The ratio
$\Phi_{1212}
({\boldsymbol y}, k_1, 0 ; \omega)/{R}_{1212}
({\boldsymbol y},{\boldsymbol 0}, 0)$ is then $\sim \chi^{-2}$ and remains $O(\tilde{\omega}^{-4})$ as $\tilde{\omega}\rightarrow \infty$  when $\tilde{k}_T \neq 0$ thus recovering the fact that the `spectral extent' of the peak jet noise producing region decays (algebraically in this case) at very high frequencies.

\subsection{Validation of turbulence model (\ref{eq:R1212_model}) and comparison to  Leib $\&$ Goldstein (\cite{LG11}, Eq. 54) }

Finally, in this appendix we present a validation of the turbulence model (\ref{eq:R1212_model}) against the post-processed turbulence data reported in Afsar {\textit{et al}.}\cite{Afsetal2017} of the Br\'es {\it et al}.\cite{Bres17} fixed $M_J>1$. We compare (\ref{eq:R1212_model}) used in the derivation of (\ref{eq:SpecPhi1212}) and predictions in Fig. \ref{fig5_11_SPLpreds}, to the Leib $\&$ Goldstein\cite{LG11} turbulence model (Eq. 54 in their paper) and subsequent predictions obtained by using that model as an alternative to (\ref{eq:R1212_model}) in  (\ref{I_om_last}) respectively.
Identical turbulence scales as given in Table \ref{Table_params} were used to determine $\Phi{}^*_{1212}
$ using the Leib $\&$ Goldstein model via Fourier transform (\ref{eq:SpecPhi1212}). Moreover, the same Green's function solution, (\ref{Hyp3})--(\ref{BC2}), was used to determine the propagators (\ref{G_12}), (\ref{G11_41}) $\&$ (\ref{G11_41b}) in (\ref{I_low2}). 

When $\eta_T=0$, the space-time structure of ${R}_{1212}
({\boldsymbol y},\eta_1, 0, \tau)$ in (\ref{eq:R1212_model}) expands as follows
\begin{equation}
\label{eq:R1212_exp}
  \begin{split}
\frac{{R}_{1212}
({\boldsymbol y},\eta_1, 0, \tau)}
{{R}_{1212}
({\boldsymbol y},{\boldsymbol 0}, 0)}
=
\biggl{[}
1
+
&
a_1
\frac{\tilde{\tau}}{X
}
 l{}_r^2(\tilde{\eta}_1-\tilde{\tau})
\\
&
-
a_2
\frac{\tilde{\eta}_1}{X
}
[\tilde{\eta}_1 +l{}_r^2(\tilde{\eta}_1 - \tilde{\tau}) ]
+
...
\biggr{]}
e^{-X}
 \end{split}
\end{equation}
where $X= X(\tilde{\eta}_1, \tilde{\eta}_T, \bar{\tau})
= \sqrt{ \tilde{\eta}_1^2 +|\tilde{\xi}|^2 }$. 
To simplify its graphical presentation, we use non-dimensional variables, $\tilde{\eta}_1$, defined below (\ref{eq:SpecPhi1212_p1}), and $\bar{\tau} = U_c \tau/l_1$, which allows the space-time dependence in (\ref{eq:R1212_exp}) to be determined through the scaled non-dimensional convected variable, $\tilde{\xi}= l_r(\tilde{\eta}_1 - \bar{\tau})$ where $l_r = l_1/l_0 = c_1/c_0$ by (\ref{Eq:lengthscales}).
Note by (\ref{eq:R1212_exp}) when $a_2 = 0$, ${R}_{1212}(\boldsymbol{y}, \tau)/{R}_{1212}(\boldsymbol{y})$ takes the simple exponential form $\sim (1-a_1\bar{\tau}l_r)e^{-l_r \bar{\tau}}$ at $\boldsymbol\eta=0$.

\begin{figure}
  \centering
    \begin{subfigure}[h]{0.4\textwidth}
        \centering
        \includegraphics[width=\textwidth]{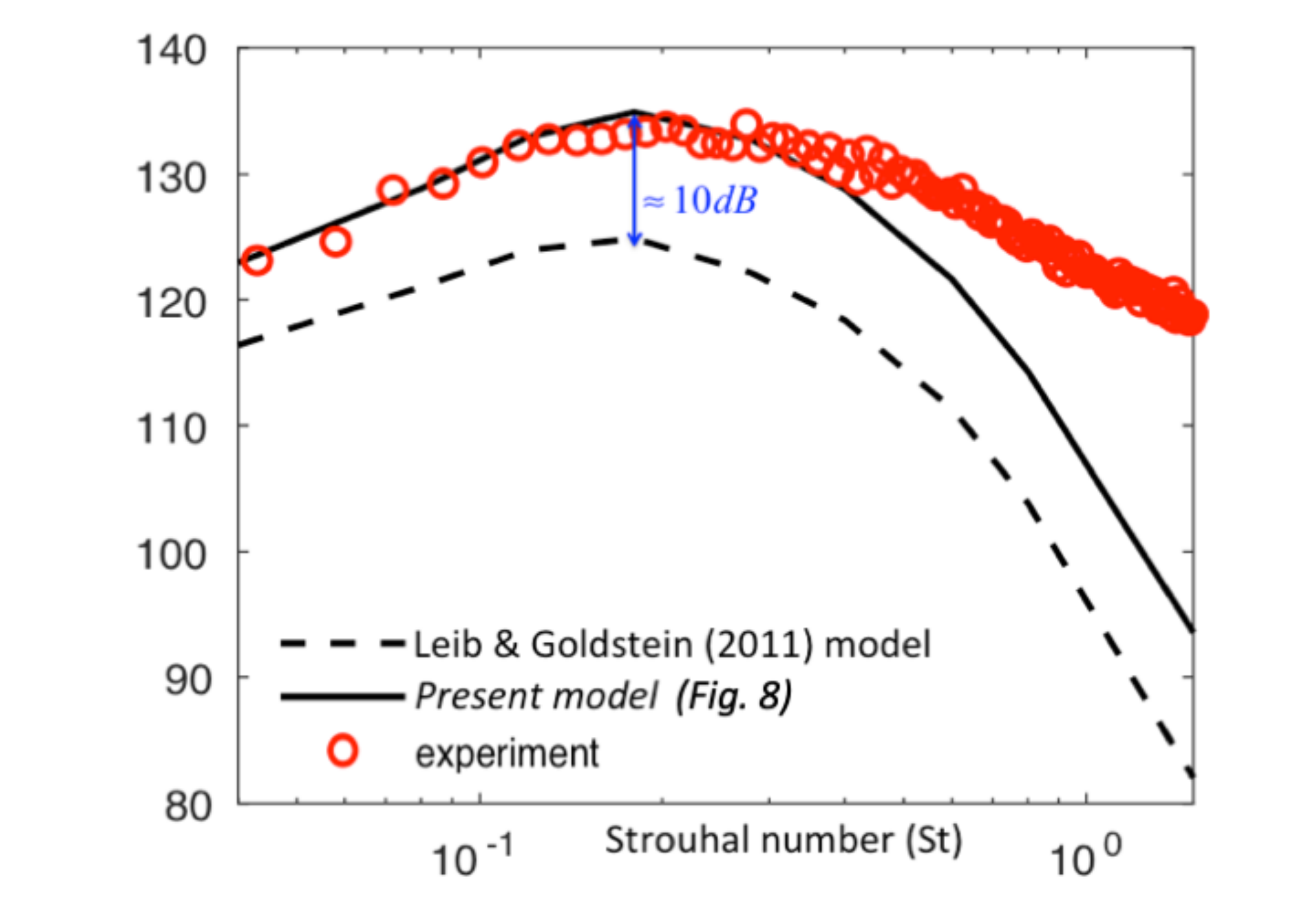}
        \caption{}
        \label{fig_LGASL_28degSP90}
    \end{subfigure}
      \centering
    \begin{subfigure}[h]{0.4\textwidth}
        \centering
        \includegraphics[width=\textwidth]{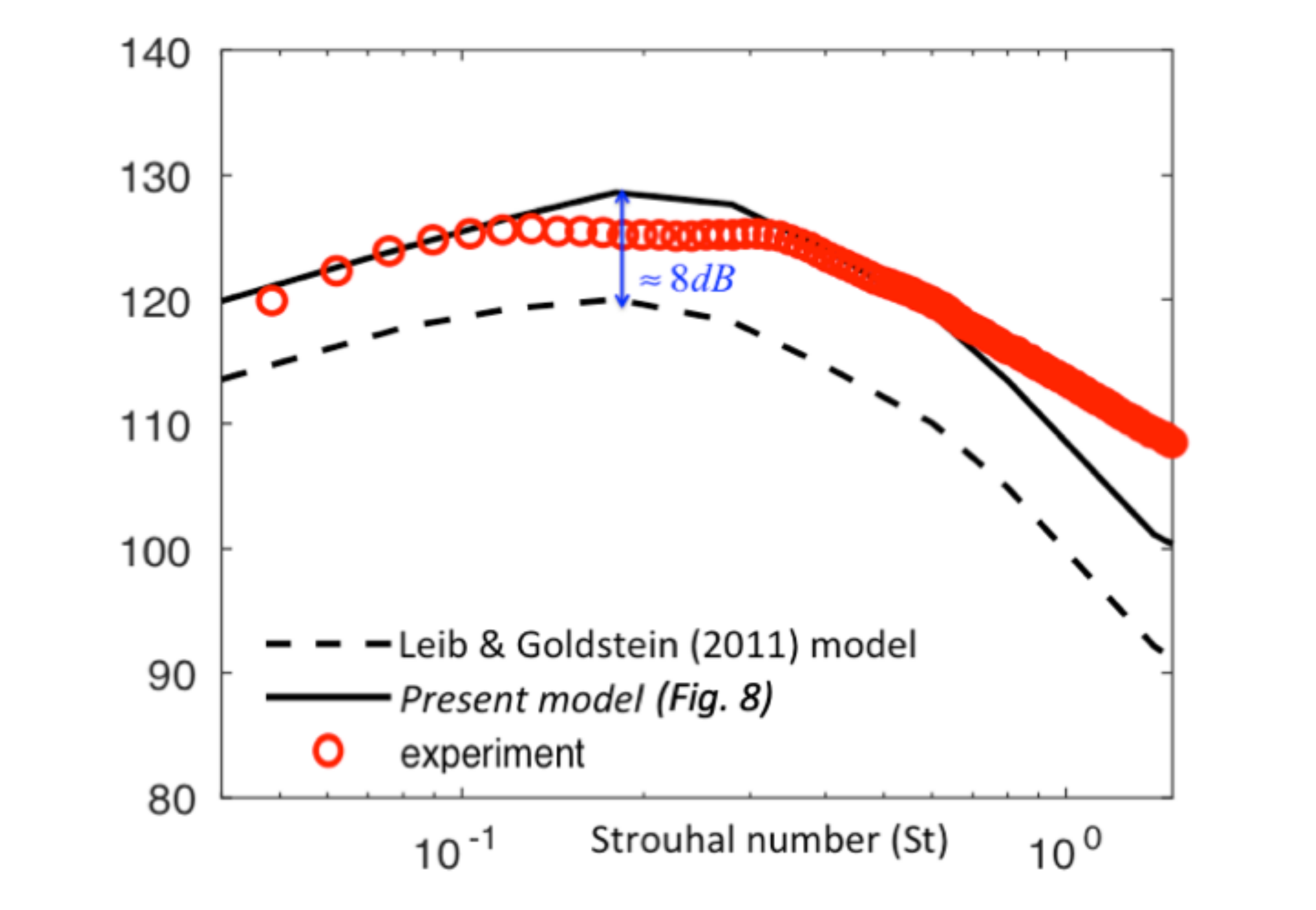}
        \caption{}
        \label{fig_LGASL_30SP49}
    \end{subfigure} \\
    \centering
    \begin{subfigure}[h]{0.4\textwidth}
        \centering
        \includegraphics[width=\textwidth]{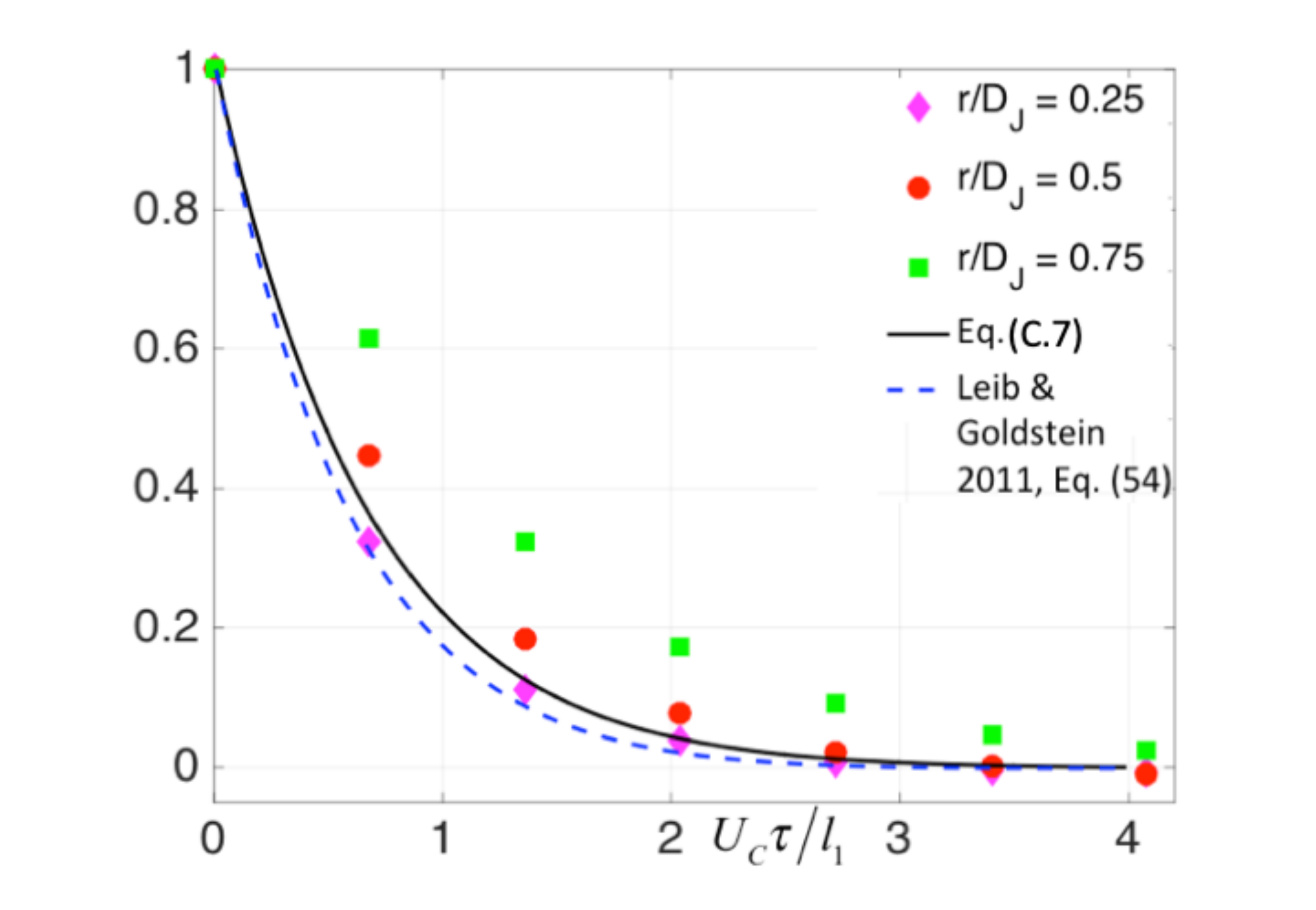}
  \caption{} 
        \label{fig_R1212_SP90}
    \end{subfigure}
    \begin{subfigure}[h]{0.4\textwidth}
        \centering
        \includegraphics[width=\textwidth] {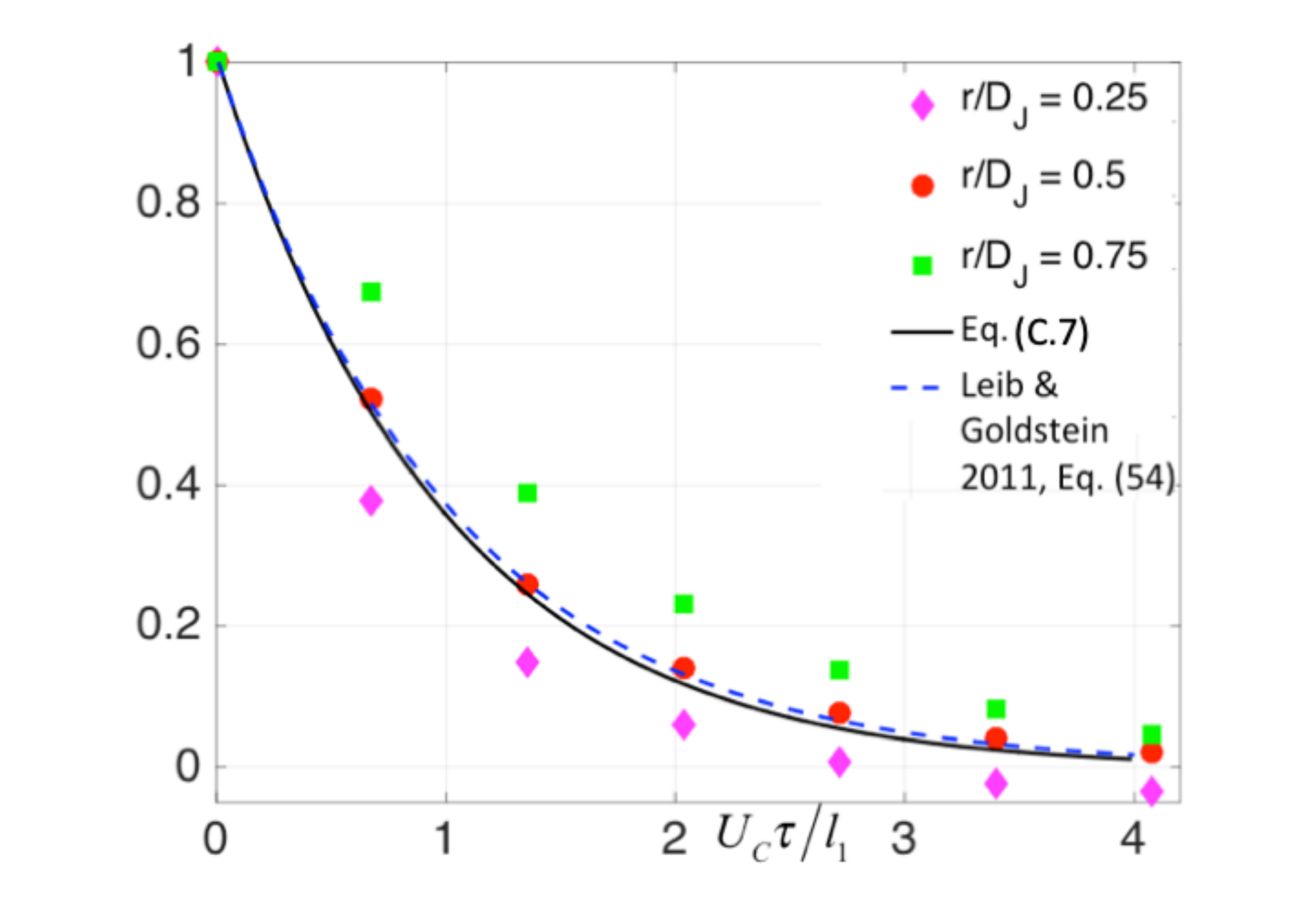}
\caption{}
        \label{fig_R1212_SP49}
    \end{subfigure}
           \caption{Comparison of Fig. \ref{fig5_11_SPLpreds} predictions with that obtained using Eq.54 in Leib $\&$ Goldstein\cite{LG11} to determine (\ref{eq:SpecPhi1212}) in (\ref{I_low2}).
           (a). SP90: $\theta = 28^\circ$; (b). SP49: $\theta = 30^\circ$; (c). SP$90$: $R_{1212}$; (d). SP$49$: $R_{1212}$.
           Figs. (c) $\&$ (d) compare (\ref{eq:R1212_exp}) $\&$ Eq. (54) in Leib $\&$ Goldstein\cite{LG11} to LES-extracted turbulence data\cite{Bres17,Afsetal2017} where SP$90$ is compared to a $(M_J, TR)=(1.5,1.0)$ jet and SP$49$ to a $(M_J,TR)=(1.5,1.7)$ one.
           %
           }
    \label{fig_R1212+LG}
\end{figure}

In Figs. \ref{fig_R1212_SP90} $\&$ \ref{fig_R1212_SP49} we set $\boldsymbol{\eta}=0$ for the comparison of (\ref{eq:R1212_exp}) and Eq. (54) in Leib $\&$ Goldstein (2011\cite{LG11}) against 
auto-correlation data of $R_{1212}$ with $\bar{\tau}$.
The LES data\cite{Bres17} was extracted at the streamwise position roughly at end of the jet potential core region, $y_1 = 8$, for three radial locations $r = (0.25, 0.5, 0.75)$ above/below the jet shear layer.
The agreement is slightly better for (\ref{eq:R1212_exp}) in the isothermal case ($M_J = Ma = 1.5$) which should have similar turbulence structure as SP$90$.   
In general, however, both models appear to compare favorably.

For the acoustic predictions on the other hand, using the Leib $\&$ Goldstein\cite{LG11} $R_{1212}$ model to determine (\ref{eq:SpecPhi1212}) in (\ref{I_low2}) under predicts the SPL by $(8-10)$dB in both SP$90$ and SP$49$ (Figs.\ref{fig_LGASL_28degSP90} and \ref{fig_LGASL_30SP49} respectively) when using the same turbulence parameters and non-parallel flow-based propagators as in the results of Fig. \ref{fig5_11_SPLpreds}.
Without altering $c_1$,   
this difference can be easily remedied by an appropriate increase in $c_\perp$ to $c_\perp=(0.063, 0,043)$ for SP$90$ and SP$49$ respectively.
In Table (\ref{LGscales}) we compare the corrected values of $c_\perp$ to those used in the Fig. \ref{fig5_11_SPLpreds} predictions.
It is clear that using Eq. 54 of Ref. \onlinecite{LG11} to determine $\Phi{}_{1212}^*$ results in a small increase of $c_\perp$ compared to using (\ref{eq:R1212_exp}) in (\ref{I_low2}). While the increase is modest, our choice of $c_\perp$ used in Fig. \ref{fig5_11_SPLpreds} remains more consistent with turbulence measurements\cite{Pokora, M&Z} since $c_\perp\ll c_1$ as Refs. \onlinecite{Pokora, M&Z, Karab2010} have found.

\begin{table}
\caption{ 
Streamwise/transverse parameters $(c_1, c_\perp)$ needed for accurate peak noise predictions in Figs. \ref{fig_LGASL_28degSP90} $\&$ \ref{fig_LGASL_30SP49}. 
}
\begin{ruledtabular}
\begin{tabular}{lcr}
 Tanna\cite{Tanna77} set point  &   Eq.(\ref{eq:R1212_exp})  & { L$\&$G (\cite{LG11}, Eq. 54)  } \\
\hline
SP$90$ & (0.125, 0.022) & (0.125,0.063)\\
SP$49$ & (0.17, 0.017) & (0.17, 0.0425)\\
\end{tabular}
\end{ruledtabular}
\label{LGscales}
\end{table}

\nocite{*}
\bibliography{aipsamp}

\end{document}